\documentclass[11pt,english]{article}
\usepackage[sort&compress,numbers]{natbib}
\usepackage{hyperref}
\usepackage{amsmath}
\usepackage{amsfonts}
\usepackage{amssymb}
\usepackage{graphicx}
\usepackage{enumitem}
\usepackage{eufrak}

\renewcommand{\P}{\mathcal{P}}
\newcommand{\res}{\mathcal{R}}
\renewcommand{\S}{{\rm S}}
\newcommand{\R}{{\rm R}}
\newcommand{\B}{\mathcal{B}}
\newcommand{\E}{\mathcal{E}}
\newcommand{\e}{\epsilon}
\newcommand{\tail}{{\rm tail}}
\newcommand{\lmax}{{\ell_{\rm max}}}
\newcommand{\qmax}{{q_{\rm max}}}

\newcommand{\nhat}{{\hat n}}
\newcommand{\man}{\mathcal{M}}
\newcommand{\manin}{\mathcal{M}_{\rm obj}}
\newcommand{\varLie}{\mathsterling}
\newcommand{\Lie}{\mathcal{L}}
\newcommand{\seed}{{\rm seed}}
\newcommand{\free}{{\rm free}}
\newcommand{\varext}{\mathfrak{g}}
\newcommand{\varh}{\mathfrak{h}}

\renewcommand{\O}{\mathcal{O}}
\newcommand{\D}{\mathcal{D}}
\newcommand{\gin}{g^{\rm obj}}

\newcommand{\beq}{\begin{equation}}
\newcommand{\eeq}{\end{equation}}

\begin{document}
\title{Motion of small objects in curved spacetimes: An introduction to gravitational self-force}

\author{Adam Pound\footnote{Email: a.pound@soton.ac.uk} \\
		Mathematical Sciences\\
		University of Southampton\\
		Southampton, SO17 1BJ, United Kingdom
        }

\date{\today}

\maketitle

\begin{abstract}
In recent years, asymptotic approximation schemes have been developed to describe the motion of a small compact object through a vacuum background to any order in perturbation theory. The schemes are based on rigorous methods of matched asymptotic expansions, which account for the object's finite size, require no ``regularization" of divergent quantities, and are valid for strong fields and relativistic speeds. Up to couplings of the object's multipole moments to the external background curvature, these schemes have established that at least through second order in perturbation theory, the object's motion satisfies a generalized equivalence principle: it moves on a geodesic of a certain smooth metric satisfying the vacuum Einstein equation. I describe the foundations of this result, particularly focusing on the fundamental notion of how a small object's motion is represented in perturbation theory. The three common representations of perturbed motion are (i) the ``Gralla-Wald" description in terms of small deviations from a reference geodesic, (ii) the ``self-consistent'' description in terms of a worldline that obeys a self-accelerated equation of motion, and (iii) the ``osculating geodesics" description, which utilizes both (i) and (ii). Because of the coordinate freedom in general relativity, any coordinate description of motion in perturbation theory is intimately related to the theory's gauge freedom. I describe asymptotic solutions of the Einstein equations adapted to each of the three representations of motion, and I discuss the gauge freedom associated with each. I conclude with a discussion of how gauge freedom must be refined in the context of long-term dynamics.
\end{abstract}

\tableofcontents

\section{Preamble and survey}

Consider a small object moving through a curved spacetime. What path does it follow? At the level of undergraduate physics, the answer is satisfyingly simple: if the object is sufficiently small and light, it can be idealized as a test particle, which does not affect the geometry around it and which moves on a geodesic of that geometry. But what if we do away with that idealization? The real object \emph{does} perturb the geometry around it; how, then, does the object move in that geometry, which it itself affects?

In Newtonian gravity, we may ask the analogous question---how does a massive body move when it contributes to the gravitational field around it?---and we may happily provide an answer without yet leaving undergraduate physics: if the body is sufficiently spherical, it can be treated as a point particle located at its center of mass, and the motion of that center of mass is governed by the \emph{external} gravitational fields produced by all \emph{other} masses in the system; the object does not feel its \emph{own} field, at least in so far as its center-of-mass motion is concerned. 

However, in general relativity, the situation becomes radically more complicated. Because of the nonlinearity of the Einstein equations, an extended object cannot, in general, be modelled as a point particle without invoking post-hoc regularization procedures~\cite{Geroch-Traschen:87,Steinbauer-Vickers:08}. Furthermore, in a curved background, an object, even an asymptotically small one, \emph{does} feel its own field, for reasons discussed below (and elsewhere in these proceedings). Hence, the field is said to exert a \emph{gravitational self-force} on the object. 

One might wonder if this effect is actually relevant in any realistic scenarios. The answer, unequivocally, is ``Yes''.  Suppose we are interested in a bound binary of widely separated compact objects of comparable masses $m_1$ and $m_2$ moving slowly in each other's weak mutual gravity. Each of the objects is small compared to the other scales in the system (for example, the typical orbital separation $R$, or the radius of curvature of $m_1$'s field at $m_2$'s position). In a Newtonian approximation, $m_1$ is subject only to $m_2$'s Newtonian field, which at the position of $m_1$ exerts a force (per unit mass) $F_N\sim m_2/R^2$. But if one requires anything more accurate than the Newtonian approximation, the moment one steps to the post-Newtonian level, self-forces can no longer be ignored: $m_1$'s field, which we can think of as scaling like $\Phi_N\sim m_1/R$, modifies the Newtonian fields, giving rise to a post-Newtonian force per unit mass that scales like $F_{PN}~\sim \Phi_N F_N\sim m_1m_2/R^3$~\cite{Einstein-Infeld-Hoffmann:38}, similar in magnitude to any other post-Newtonian effect.\footnote{In a certain sense, an object's mass affects its own motion even in a Newtonian binary. Each object follows a Keplerian orbit about the system's center of mass, not about the center of the other object. Since the center of mass is shifted by the object's own mass, the object affects its own motion; more plainly, $m_1$ influences its own motion by influencing that of $m_2$. This is a more indirect effect than the type described above, but in practice, distinguishing it from any other post-test-body effect is nontrivial. See Ref.~\cite{Detweiler:05} for a discussion.} 

In these systems just described, however, each object, while small compared to the radius of curvature of the ambient external field it finds itself in, is not small compared to the other object. What if we are interested in a case where there are truly only two scales? Take a binary system of compact objects of mass $m$ and $M$ satisfying  $m\ll M$, and specifically focus on the regime in which the orbital separation $R$ is of order $M$. In this case, it seems the gravity of $m$ must certainly have a very small effect on its own motion, and it must very nearly follow a geodesic of the metric of $M$. And yet, even in this scenario, $m$'s gravitational self-force cannot be neglected. Although the effect is very small over a few orbits, the system continually radiates energy in the form of gravitational waves (or equivalently, the self-force does negative work), causing the orbit to shrink, and eventually causing $m$ to collide with $M$ (or plunge into $M$, if $M$ is a black hole). In other words, the self-force has long-term, secular effects on the motion which make it impossible to ignore.

Both of these two types of systems---binaries of widely separated bodies whose mutual gravity is weak, and binaries of objects with very dissimilar masses, called extreme-mass-ratio inspirals (EMRIs)---are of increasing relevance to modern astrophysics. The prospect of directly detecting gravitational waves emitted from compact binaries, and extracting information about the binaries' strong-field dynamics from those waves, has spurred an international effort to study them. In the case of widely separated bodies of comparable mass, the main method of analysis has been post-Newtonian theory,\footnote{Of course, once the two bodies are sufficiently close to each other, they interact in a highly nonlinear, highly relativistic way. In that regime, one must use numerical relativity to solve the fully nonlinear Einstein equations for the system.} a historied subject modern overviews of which can be found in the review articles~\cite{Blanchet:14,Futamase-Itoh:07} and the recent textbook~\cite{Poisson-Will:14}. In the case of EMRIs,  the main method of analysis has been self-force theory; for summaries of efforts to model EMRIs, I refer the reader to the reviews~\cite{Barack:09,Poisson-Pound-Vega:11}, the more up-to-date but brief survey~\cite{Amaro-Seoane-etal:14}, and the contributions of Babak et al.~\cite{Babak-etal:15} and Wardell~\cite{Wardell:15} elsewhere in this book.

In this paper, I seek to provide a single, unified description of gravitational self-force theory. Roughly speaking, this formalism consists of a perturbative expansion in powers of the small object's mass $m$. Although I will refer to EMRIs to motivate many of the methods and problems I discuss, my focus will instead be on foundational issues related to the problem of motion of a small object. My aim is to complement extant reviews~\cite{Poisson-Pound-Vega:11,Barack:09} by concentrating on three themes given limited attention in those reviews: (i) self-force theory at arbitrary perturbative order, (ii) differing ways to represent perturbed motion, and the asymptotic expansions of the metric corresponding to each, and (iii) the relationship between perturbed motion and gauge freedom. My presentation mostly follows the methods and viewpoint of Refs.~\cite{Pound:10a,Pound:10b,Pound:10c,Pound:12a,Pound:12b,Pound:14a,Pound:15b}, though it takes additional inspiration from the work of Gralla and Wald~\cite{Gralla-Wald:08,Gralla:11,Gralla:12} and the classic papers of Mino, Sasaki, and Tanaka~\cite{Mino-Sasaki-Tanaka:97} and Detweiler and Whiting~\cite{Detweiler:01,Detweiler-Whiting:03}. To avoid excessive length, rather than striving for complete self-containment, I will often refer the reader to Refs.~\cite{Poisson-Pound-Vega:11,Pound:10a,Pound:12b,Pound:14a} for mathematical tools and technical details.

Most of what I discuss could be applied to objects carrying scalar or electric charges, but for simplicity I restrict my attention to the purely gravitational case. I refer the reader to Refs.~\cite{Quinn-Wald:97,Gralla-Harte-Wald:09,Harte:08,Harte:09,Harte:14,Poisson-Pound-Vega:11} for discussions of self-forces due to scalar and electromagnetic fields on fixed background geometries, and to Refs.~\cite{Linz-Friedman-Wiseman:14,Zimmerman-Poisson:14} for recent work on the coupled system in which the metric reacts to both the mass and the scalar or electromagnetic field.

In the remainder of this introduction, I make an extensive survey of the main concepts and results of self-force theory, beginning with the standard picture of a point mass in linearized perturbation theory and proceeding to describe the more robust framework now available for studying the motion of extended (but small) objects at any order in perturbation theory. 

\emph{Notation.} Throughout this paper, Greek indices run from 0 to 4 and are raised and lowered with a background metric $g_{\mu\nu}$. Latin indices run from 1 to 3 and are freely raised and lowered with the Euclidean metric $\delta_{ij}$. Sans-serif symbols such as ${\sf g}_{\mu\nu}$ refer to tensors on the perturbed spacetime rather than on the background. A semicolon and $\nabla$ refer to a covariant derivative compatible with $g_{\mu\nu}$,  ${}^{\sf g}\nabla_\mu$ to the covariant derivative compatible with ${\sf g}_{\mu\nu}$, and  $\tilde\nabla_\mu$ to the covariant derivative compatible with an ``effective metric'' $\tilde g_{\mu\nu}$. $\e\equiv 1$ is used to count powers of $m$, and labels such as the  ``$n$'' in $h^n_{\mu\nu}$ refer to perturbative order $\e^n$; I freely write these labels as either super- or subscripts. I work in geometric units with $G=c=1$.




\subsection{A point particle picture: the MiSaTaQuWa equation and its interpretations}\label{point_particle_picture}
\subsubsection{Self-interaction with the tail of the perturbation}\label{tail_picture}
As mentioned above, a point particle stress-energy tensor is not a well-defined source in the nonlinear Einstein equations. However, it \emph{is} a fine source in the \emph{linearized} theory. For simplicity, assume our object of mass $m$ is isolated, such that in some large region, we can take it to be the sole source of stress-energy in the system. Now consider a metric ${\sf g}_{\mu\nu} = g_{\mu\nu}+\e h^1_{\mu\nu}+\O(\e^2)$, where the background metric  $g_{\mu\nu}$ is a vacuum solution to the nonlinear Einstein equations, and $h^1_{\mu\nu}$ is the leading-order perturbation due to our small object. Linearizing the Einstein equations in $h^1_{\mu\nu}$, we obtain 
\beq\label{dG=T}
\delta G_{\mu\nu}[h^1]=8\pi T^1_{\mu\nu},
\eeq
where $\delta G_{\mu\nu}[h]\equiv \frac{dG_{\mu\nu}[g+\lambda h]}{d\lambda}\big|_{\lambda=0}$ is the linearized Einstein tensor, and $T_1^{\mu\nu}$ is the leading-order approximation to the object's stress-energy tensor. Now suppose that the the object's stress-energy can be approximated by 
\beq\label{pp_stress-energy}
T_1^{\mu\nu}(x;z)=\int_\gamma mu^\mu u^\nu \delta(x,z(\tau))d\tau,
\eeq
which is the stress-energy of a point mass moving on a worldline $\gamma$ with coordinates $z^\mu$ in the background spacetime. Here $u^\mu\equiv \frac{dz^\mu}{d\tau}$ is the particle's four-velocity, $\tau$ is its proper time (as measured in $g_{\mu\nu}$), and $\delta(x,z)\equiv \frac{\delta^4(x^\alpha-z^\alpha)}{\sqrt{-g}}$ is a covariant delta distribution, with $g$ being the determinant of $g_{\mu\nu}$. 

Unlike the nonlinear Einstein equations with a point particle source, Eq.~\eqref{dG=T} has a perfectly well-defined solution. If we introduce the trace-reversed perturbation $\bar h^1_{\mu\nu}\equiv h^1_{\mu\nu}-\frac{1}{2}g_{\mu\nu}g^{\alpha\beta}h^1_{\alpha\beta}$ and impose the Lorenz gauge condition $\nabla^{\nu}\bar h^1_{\mu\nu}=\O(\e)$, then the linearized Einstein equation takes the form of a wave equation,
\beq
E_{\mu\nu}[\bar h^1] = -16\pi T^1_{\mu\nu},
\eeq
where 
\beq\label{E}
E_{\mu\nu}[\bar h^1]\equiv\Box \bar h^1_{\mu\nu}+2R_{\mu}{}^\alpha{}_\nu{}^\beta\bar h^1_{\alpha\beta},
\eeq
$\Box\equiv g^{\mu\nu}\nabla_\mu\nabla_\nu$, and $R_{\mu\alpha\nu\beta}$ is the Riemann tensor of the background. Assuming the existence of a global retarded Green's function $G_{\mu\nu\mu'\nu'}(x,x')$\footnote{My conventions for the Green's function are those of Ref.~\cite{Poisson-Pound-Vega:11}; Ref.~\cite{Poisson-Pound-Vega:11} also contains a pedagogical introduction to bitensors, objects which live in the tangent spaces of two different points $x$ and $x'$.} for this wave equation, we can write the retarded solution as
\beq
\bar h^1_{\mu\nu}(x;z) = 4\int G_{\mu\nu\mu'\nu'}T_1^{\mu'\nu'}dV = 4\int mG_{\mu\nu\mu'\nu'}u^{\mu'}u^{\nu'}d\tau,
\eeq
where a primed index refers to the tangent space at the source point $x'$. 

Now, if the background were flat, waves would propagate precisely on the light cone, and the retarded Green's function would be supported only on points connected by a null curve: $G^{\mu\nu}{}_{\mu'\nu'}=\delta^\mu_{\mu'}\delta^\nu_{\nu'}\frac{\delta(t'-[t-|x^a-x^{a'}|])}{|x^a-x^{a'}|}$, where $(t,x^a)$ is a Cartesian coordinate system, $t-|x^a-x^{a'}|$ is the retarded time, and $|x^a-x^{a'}|\equiv\sqrt{\delta_{ab}(x^a-x^{a'})(x^b-x^{b'})}$ is the spatial distance between the points $x$ and $x'$. 
However, in a curved spacetime, Huygen's principle no longer holds. Waves scatter off the spacetime curvature, causing them to propagate from a source point $x'$ both on the future null cone of $x'$ and \emph{within} that cone. Correspondingly, the Green's function at a point $x$ has support both on the past null cone of $x$ and within it. This implies that if we look at the field $\bar h^1_{\mu\nu}$ at a point $x$ near the worldline, we can split it into two pieces: a \emph{direct} piece, corresponding to the portion of the field that propagated to $x$ from a point $z^\mu(\tau_{\rm ret})$ along a null curve; and a so-called \emph{tail} piece,  $h^\tail_{\mu\nu}$, corresponding to the portion of the field that propagated from all \emph{earlier} points $z^\mu(\tau<\tau_{\rm ret})$ on the worldline. The direct piece diverges like a Coulomb field, behaving as $1/r$ (where $r$ is a geodesic distance from the particle). The tail piece, on the other hand, is finite.

A detailed analysis (such as the one presented in the bulk of this paper) reveals that at leading order, the mass $m$ is constant, and the force on the particle vanishes; in other words, it behaves as a test particle in $g_{\mu\nu}$. The same analysis applied at subleading order reveals that the direct piece of the field exerts no force on the particle, but the tail piece \emph{does} exert a force, and the equation of motion is found to be
\beq\label{MiSaTaQuWa}
\frac{D^2z^\mu}{d\tau^2}  = -\frac{1}{2}\e P^{\mu\nu} \left(2h^\tail_{\nu\alpha\beta}-h^\tail_{\alpha\beta\nu}\right)u^\alpha u^\beta+\O(\e^2),
\eeq
where $\frac{D}{d\tau}=u^\mu \nabla_{\mu}$, $P^{\mu\nu}=g^{\mu\nu}+u^\mu u^\nu$ projects orthogonally to the worldline, and the tail term is given by
\beq\label{tail}
h^\tail_{\mu\nu\rho}(z(\tau)) = 4m\int_{-\infty}^{\tau-0^+} \nabla_\rho\bar G_{\mu\nu\mu'\nu'}u^{\mu'}u^{\nu'}d\tau';
\eeq
the integral covers all of the worldline earlier than the point $z^\mu(\tau)$ at which the force is evaluated. The bar atop $G^{\mu\nu}{}_{\mu'\nu'}$ again denotes a trace reversal.

Equation~\eqref{MiSaTaQuWa} is termed the MiSaTaQuWa equation, after Mino, Sasaki, and Tanaka~\cite{Mino-Sasaki-Tanaka:97}, who first derived it, and Quinn and Wald \cite{Quinn-Wald:97}, who re-derived it very shortly thereafter using a very different, independent method. The intuitive picture to glean from the MiSaTaQuWa result is that the direct piece of the field is analogous to a Coulomb field, moving with the particle and exerting no force on it, in the same way the self-field exerts no force on a body in Newtonian gravity. Very loosely speaking, from the perspective of the particle, the tail, consisting as it does of backscattered radiation, is indistinguishable from any other incoming radiation. In other loose words, it is effectively an external field, and like an external field, it exerts a force.

Much of this paper is devoted to showing how the MiSaTaQuWa result can be robustly justified, and higher-order corrections to it can be found,  within a systematic expansion of the Einstein equations. As a byproduct of that analysis, in Sec.~\ref{skeleton} I will show that the setup of the linearized system with a point particle source in this section is rigorously justified---even for a non-material object such as a black hole, and even though at nonlinear orders the field equations cannot be written in terms of such a source. However, before moving on to those matters, we may say significantly more on the basis of the point-particle result.

\subsubsection{The Detweiler-Whiting description: a generalized equivalence principle}\label{Detweiler-Whiting}
In the original MiSaTaQuWa papers~\cite{Mino-Sasaki-Tanaka:97,Quinn-Wald:97}, the authors noted that Eq.~\eqref{MiSaTaQuWa} appears to have the form of the geodesic equation in a metric $g_{\mu\nu}+h^\tail_{\mu\nu}$, when that geodesic equation is expanded to linear order in $h^\tail_{\mu\nu}$.\footnote{I refer the reader to Appendix~\ref{geodesic_equation_expansion} for the expansion of the geodesic equation in powers of a metric perturbation.} However, $h^\tail_{\mu\nu}$ is not in any way a nice field. It does not satisfy any particularly meaningful field equation, nor is it even differentiable at the particle~\cite{Poisson-Pound-Vega:11} [despite superficial appearances in Eq.~\eqref{MiSaTaQuWa}]. 

Detweiler and Whiting provided a more compelling form of the MiSaTaQuWa result~\cite{Detweiler:01,Detweiler-Whiting:03}. Rather than splitting the retarded field into a direct piece and a tail, they split it as
\beq
h^1_{\mu\nu} = h^{\S1}_{\mu\nu}+h^{\R1}_{\mu\nu}.
\eeq
The \emph{singular field} $h^{\S1}_{\mu\nu}$ can be interpreted as the bound self-field of the particle. Like the direct piece of the field, it exhibits a $1/r$, Coulomb divergence at the particle; but unlike the direct piece, it satisfies the inhomogeneous linearized Einstein equation $E_{\mu\nu}[\bar h^{\S1}]=-16\pi T^1_{\mu\nu}$, bolstering its interpretation as a self-field. Similarly, the \emph{regular field} $h^{\R1}_{\mu\nu}$ improves on the interpretation of the tail: it includes all the backscattered radiation in the tail, but it is a smooth solution to the homogeneous wave equation $E_{\mu\nu}[\bar h^{\R1}]=0$. Hence, more than we could of the tail, we can think of $h^{\R1}_{\mu\nu}$ as an effectively external field, propagating independently of the particle. From it we can define what I will variously call an \emph{effective metric} or \emph{effectively external metric} $\tilde g_{\mu\nu}=g_{\mu\nu}+\e h^{\R1}_{\mu\nu}$.

Fittingly, given this interpretation of $h^{\R1}_{\mu\nu}$, Detweiler and Whiting showed that the MiSaTaQuWa equation~\eqref{MiSaTaQuWa} can be equivalently written as
\beq\label{Detweiler-Whiting-form1}
\frac{D^2z^\mu}{d\tau^2} = -\frac{1}{2}\e P^{\mu\nu} \left(2h^{\R1}_{\nu\alpha;\beta}-h^{\R1}_{\alpha\beta;\nu}\right)u^\alpha u^\beta+\O(\e^2),
\eeq
or (following Appendix~\ref{geodesic_equation_expansion}) explicitly as the geodesic equation in the metric $\tilde g_{\mu\nu}$,
\beq\label{Detweiler-Whiting-form2}
\frac{\tilde D^2z^\mu}{d\tilde\tau^2} = \O(\e^2),
\eeq 
where $\frac{\tilde D}{d\tilde\tau}\equiv\tilde u^\mu\tilde \nabla_{\mu}$ is a covariant derivative compatible with $\tilde g_{\mu\nu}$, $\tilde u^\mu=\frac{dz^\mu}{d\tilde\tau}$ is the four-velocity normalized in $\tilde g_{\mu\nu}$, and $\tilde\tau$ is the proper time along $z^\mu$ as measured in $\tilde g_{\mu\nu}$. Equation~\eqref{Detweiler-Whiting-form1} is equivalent to Eq.~\eqref{MiSaTaQuWa} because on the worldline, $h^{\R1}_{\mu\nu}$ differs from $h^\tail_{\mu\nu}$ only by (i) background Riemann terms that cancel in Eq.~\eqref{Detweiler-Whiting-form1} and (ii) terms proportional to the worldline's acceleration, which can be treated as effectively higher order because the acceleration is already $\sim\e$. 

 
Allow me to dwell longer on the interpretation of the regular field. Because $\tilde g_{\mu\nu}$ is a smooth vacuum solution, at the particle's position an observer cannot distinguish it from $g_{\mu\nu}$. Although a portion of $\tilde g_{\mu\nu}$ comes from the retarded field sourced by the particle, to the observer on the worldline, it appears just as would any metric sourced by a distant object. However, this interpretation of the effective metric as an effectively \emph{external} metric is delicate. To effect the desired split into $h^{\S1}_{\mu\nu}$ and $h^{\R1}_{\mu\nu}$, both fields must be made acausal when evaluated off the worldline~\cite{Detweiler-Whiting:03,Poisson-Pound-Vega:11}. More precisely, in addition to depending on the particle's causal past, $h^{\R1}_{\mu\nu}(x)$ depends on the particle at \emph{spatially} related points $x'$. So in that sense its interpretability as a physical external field is limited. Yet when evaluated \emph{on} the worldline, $h^{\R1}_{\mu\nu}$ and its derivatives \emph{are} causal, and that is the sense in which $\tilde g_{\mu\nu}$ appears as a physical metric on the worldline.

I impress upon the reader the significance of these properties of $h^{\R1}_{\mu\nu}$; they are what makes Eq.~\eqref{Detweiler-Whiting-form2} a meaningful result. Although it may not be an obvious fact at first glance, \emph{any equation of motion can be written as the geodesic equation in} some \emph{smooth piece of the metric.} This is most easily seen by writing the equation of motion in a frame that comoves with the particle. In locally Cartesian coordinates $(t,x^i)$ adapted to that frame, such as Fermi-Walker coordinates~\cite{Poisson-Pound-Vega:11}, the particle's equation of motion reads $a_i = F_i$, where $a^\mu\equiv \frac{D^2z^\mu}{d\tau^2}$ is the particle's covariant acceleration and $F^\mu$ is the force (per unit mass) acting on it. Now suppose we were to define some smooth field $h^{r1}_{\mu\nu}$ and a corresponding singular field $h^{s1}_{\mu\nu}\equiv h^1_{\mu\nu}-h^{r1}_{\mu\nu}$. In the comoving coordinates, the  linearized geodesic equation in the regular metric $g_{\mu\nu}+h^{r1}_{\mu\nu}$, in the form analogous to Eq.~\eqref{Detweiler-Whiting-form1}, reads $a_i = -h^{r1}_{ti,t}+\frac{1}{2}h^{r1}_{tt,i}$. No matter what force $F^\mu$ acts on the particle, the equation of motion $a_i=F_i$ could be written as the geodesic equation in $g_{\mu\nu}+h^{r1}_{\mu\nu}$ simply by choosing $h^{r1}_{ti,t}\big|_\gamma=0$ and $h^{r1}_{tt,i}\big|_\gamma=2F_i$, for example. Besides those two conditions, the regular field $h^{r1}_{\mu\nu}$ could be entirely freely specified, and regardless of the specification we made, we would have defined a split $h^1_{\mu\nu}=h^{s1}_{\mu\nu}+h^{r1}_{\mu\nu}$ in which $h^{s1}_{\mu\nu}$ is singular and exerts no force, and $g_{\mu\nu}+h^{r1}_{\mu\nu}$ is a regular metric in which the motion is geodesic. 

Given this fact, it is of no special significance that the MiSaTaQuWa equation is equivalent to geodesic motion in \emph{some} effective metric. But it \emph{is} significant that the MiSaTaQuWa equation is equivalent to geodesic motion in the \emph{particular} effective metric $\tilde g_{\mu\nu}=g_{\mu\nu}+h^{\R1}_{\mu\nu}$ identified by Detweiler and Whiting, because of the particular, `physical' properties of that metric: $\tilde g_{\mu\nu}$ is a smooth vacuum solution that is causal on the particle's worldline. Because of those properties, we may think of the MiSaTaQuWa equation as a generalized equivalence principle: any object, if it is sufficiently compact and slowly spinning, regardless of its internal composition, falls freely in a gravitational field $\tilde g_{\mu\nu}$ that can be thought of (loosely if not precisely) as the physical `external' gravitational field at its `position'. I stress that this is a derived result, not an assumption, as will be shown in Sec.~\ref{algorithm}. By that stage in the paper, the principle's reference to extended, ``sufficiently compact and slowly spinning'' objects will have become clear. 

Another aspect that will have become clearer is the non-uniqueness of the effective metric and the limitations of interpreting it in a strongly physical way. Nonetheless, the split of the metric into a self-field and an effective metric will be a recurring theme. In many ways, the generalized equivalence principle just described is both the central tool and the core result of self-force theory. It is conceptually more compelling than the expression for the force in terms of a tail,\footnote{Another compelling physical interpretation is provided by Quinn and Wald~\cite{Quinn-Wald:97}. They showed that the MiSaTaQuWa equation follows from the assumption that the net force is equal to an average over a certain ``bare'' force over a sphere around the particle. (This assumption was later proved to be true in a large class of gauges~\cite{Gralla-Wald:08,Gralla:11,Pound-Merlin-Barack:14}.) In the language of Detweiler and Whiting fields, the force lines of the singular field are symmetric around the particle and vanish upon averaging, while the force lines of the regular field are asymmetric and add up to a net force on the particle.} it is less tied to any particular choice of gauge~\cite{Pound:15b}, and it is often easier to use as a starting point from which to derive formal results. Perhaps most importantly, it is easily carried to nonlinear orders: as I describe in later sections, at least through second order in $\e$, the generalized equivalence principle stated above holds true~\cite{Pound:12a,Pound:14a}.

\subsection{Extended bodies and the trouble with point particles}\label{extended_bodies}
Although the point-particle picture seemingly works well within linearized theory, it is not obvious \emph{a priori} how it fits within a systematic asymptotic expansion that goes to higher perturbative orders. For as I have reiterated above, one cannot use a point particle in the full nonlinear theory. Let me now expound on that point. 

Suppose we attempt to model the extended object as a point particle in the exact spacetime, with a stress-energy tensor 
\beq
{\sf T}^{\mu\nu} = \int m\frac{dz^\mu}{d{\sf t}}\frac{dz^\nu}{d{\sf t}}\frac{\delta^4(x^\alpha-z^\alpha)}{\sqrt{-{\sf g}}}d{\sf t},
\eeq
 where ${\sf t}$ is proper time on $z^\mu$ as measured in ${\sf g}_{\mu\nu}$. If we expand the metric as ${\sf g}_{\mu\nu}=g_{\mu\nu}+\sum_{n>0} \e^n h^n_{\mu\nu}$, the linearized Einstein equation is exactly as described in the previous section. But at second order, severe problems arise. The second-order term in the Einstein equation $G^{\mu\nu}[\sf{g}]=8\pi \sf{T}^{\mu\nu}$ reads
\beq\label{pp_second-order}
\delta G^{\mu\nu}[h^2] = -\delta^2 G^{\mu\nu}[h^1,h^1] +8\pi T_2^{\mu\nu},
\eeq
where $\delta^2 G^{\mu\nu}[h,h]\equiv \frac{1}{2}\frac{d^2G^{\mu\nu}[g+\lambda h]}{d\lambda^2}\big|_{\lambda=0}$ and $T_2^{\mu\nu}$ is the second-order term in ${\sf T}^{\mu\nu}$. There are two problems with Eq.~\eqref{pp_second-order}. Most obviously, $T_2^{\mu\nu}$ contains terms like 
\begin{equation*}
\int mu^\mu u^\nu \left(-\frac{1}{2}g^{\rho\delta}h^1_{\rho\delta}\right)\frac{\delta^4(x^\alpha-z^\alpha)}{\sqrt{-g}}d\tau. 
\end{equation*}
As described in the preceding section, $h^1_{\mu\nu}$ diverges as $1/r$ near the particle; hence, the stress-energy diverges in the distributionally ill-defined manner $\frac{1}{r}\delta(r)$. Even if we could somehow mollify the ill behavior of this piece of the source, the other source term in Eq.~\eqref{pp_second-order} would still present a problem. $\delta^2 G^{\mu\nu}[h^1,h^1]$ behaves schematically as $(\partial h^1)^2+h^1\partial^2 h^1$. Therefore, it diverges as $1/r^4$. Such a divergence is non-integrable, meaning it is well defined as a distribution only if it can be expressed as a linear operator acting on an integrable function. In the present case, there does not appear to be a unique way of so expressing it.

These arguments make clear that a point particle poses increasingly worsening difficulties at nonlinear orders in perturbation theory. And it is well known that in any well-behaved space of functions \emph{there exists no solution} to the original, fully nonlinear equation $G^{\mu\nu}[\sf{g}]=8\pi \sf{T}^{\mu\nu}$ with a point particle source~\cite{Geroch-Traschen:87,Steinbauer-Vickers:08}. 

Despite these obstacles, one might suppose that the point-particle model could be adopted and the divergences resolved using post-hoc regularization methods. Evidence for this reasoning is given by the successful use of dimensional regularization in post-Newtonian theory~\cite{Blanchet:14}. In the fully relativistic context, the most promising route to such regularization appears to be effective field theory~\cite{Galley-Hu:08,Galley:12b}.

However, at a fundamental level, there should be no need for such regularization in general relativity. Outside of curvature singularities inside black holes, everything in the problem should be perfectly finite. For that reason, in this paper I will be interested only in formalisms that deal with finite, well-defined quantities throughout.

So let us do away with the fiction of a point particle and think of an extended object. Perhaps the most obvious route, at least at first glance, to determining this object's motion is to go to the opposite extreme from the linearized point particle model, by looking instead at a generic extended object in fully nonlinear general relativity; if one is interested in the case of a small object, one can always examine an appropriate limit of one's generic results. This line of attack has been most famously pursued by Dixon~\cite{Dixon:74} (inspired by early work by Mathisson~\cite{Mathisson:37}) and Harte~\cite{Harte:12}. Specializing to material bodies, Dixon showed that all information in the stress-energy ${\sf T}^{\mu\nu}$ can be encoded in a set of multipole moments, and all information in the conservation law ${}^{\sf g}\nabla_\mu {\sf T}^{\mu\nu}=0$ can be encoded in laws of motion for the body. These laws take the form of evolution equations for some suitable representative worldline in the object's interior and for the object's spin about that worldline. A ``good'' choice of representative worldline may be made by, for example, defining the object's mass dipole moment relative to any given worldline and then choosing the worldline for which the mass dipole moment vanishes, establishing the worldline as a center of mass~\cite{Ehlers-Rudolph:77}. However, in order to transform these general results into practical equations of motion, Dixon's method requires an assumption that the metric and its derivatives do not vary much in the body's interior. Given that assumption, the force and torque appearing on the right-hand side of the equations of motion can be written as a simple expansion composed of couplings of the metric's curvature to the object's higher multipole moments; the higher moments, beginning with the quadrupole, may be freely specified, and their specification is entirely equivalent to a specification of the object's stress-energy tensor. Equations of motion of this form can be viewed in Eqs.~\eqref{TH-pdot}--\eqref{TH-Sdot} below.

Unfortunately, for a reasonably compact, strongly gravitating body, the physical metric does \emph{not} vary slowly in its interior, due to the body's own contribution to the metric. To make progress, one must treat the object as a test body, an extended object that is non-gravitating but is equipped with a multipole structure---or, as discussed in Sec. 13 of Ref.~\cite{Dixon:74}, one must find an efficacious means of   separating the object's self-field from the ``external" field, analogous to the trivial split in Newtonian theory and to the Detweiler-Whiting split in the linearized point particle model. A well-chosen ``external'' field will vary little in the body's interior, a well-chosen ``self-field'' will have minimal influence on the motion, and one can hope to arrive at equations of motion expressed in terms of couplings to the curvature of the external field alone. 

In the fully nonlinear theory, finding such a split would seem to be highly nontrivial. Nevertheless, in a series of papers~\cite{Harte:08,Harte:09,Harte:10} culminating in Ref.~\cite{Harte:12} (see also Harte's contribution to these proceedings~\cite{Harte:14}), Harte has succeeded in finding a suitable split by directly generalizing the Detweiler-Whiting decomposition. He has shown that the ``self-field" he defines modifies the equation of motion only by shifting the values of the object's multipole moments, and the object behaves as a test body moving in the effectively external metric he defines. This extends the Detweiler-Whiting result from the linearized model to the fully nonlinear problem. However, there is one caveat to this generalization: beyond linear order, Harte's effective metric loses one of the compelling properties of the Detweiler-Whiting field: it is not a solution to the vacuum Einstein equation. Despite this feature, Harte's work is a tour de force in the problem of motion.

The approach I take in this paper is complementary to Harte's. Rather than beginning with the fully nonlinear problem, I will proceed directly to perturbation theory. There are several advantages to this. The perturbative approach naturally applies to black holes, while Harte's formalism, because it is based on  integrals over the object's interior, is restricted to material bodies. The perturbative approach also naturally leads to a split into self-field and effective metric in which the effective metric satisfies the vacuum Einstein equation at all orders and is causal on the worldline. Most importantly, the perturbative approach provides a practical means of solving the Einstein equations. In the fully nonlinear approach, one arrives at equations of motion given the metric, but not a practical way to find that metric.

In the next section, I will begin to discuss the perturbative formalism. First, I note that with the work of Dixon and Harte, a new theme has been introduced: an object's bulk motion can be expressed in terms of a set of multipole moments, and the $\ell\geq2$ moments are freely specifiable. Like the decomposition of the metric into a self-field and effectively external field, this second theme will appear prominently in the remainder of this paper.

\subsection{When perturbation theory fails near a submanifold: the method of matched asymptotic expansions}\label{matched_expansions_intro}
As soon as we seek a perturbative description of the problem, we run into a new challenge. Say we assume an expansion of the exact spacetime $({\sf g}^\e_{\mu\nu},\man_\e)$ about a background spacetime $(g_{\mu\nu},\man_0)$, as in ${\sf g}^\e_{\mu\nu} = g_{\mu\nu}+\e h^1_{\mu\nu}+\O(\e^2)$, where all $\e$-dependent terms are created by the small object (or by nonlinear interactions of its field with itself). This expansion assumes the object has only a small effect on the metric. Clearly, if the object is compact, this cannot be true everywhere: sufficiently near the object, where $r\sim \e$, the object's own gravitational field will contain a Coulomb term $\sim m/r\sim\e^0$---just as large as the background $g_{\mu\nu}$, and because it varies on the spatial scale $\e$ rather than $\e^0$, having much stronger curvature than $g_{\mu\nu}$. 

This fact motivates the use of \emph{matched asymptotic expansions}. At distances $r\sim\e^0$ from the object (for example, $r\sim M$ in an EMRI), we expand the exact spacetime around $(g_{\mu\nu},\man_0)$, as above; I will call this the \emph{outer expansion}. At distances $r\sim\e$ from the object (or $r\sim m$ in dimensionful units), we introduce a second expansion,  ${\sf g}_{\mu\nu} = \gin_{\mu\nu}+\e H^1_{\mu\nu}+\O(\e^2)$, where $(\gin_{\mu\nu},\man_{\rm obj})$ is the spacetime of the object were it isolated, and the perturbations  $H^n_{\mu\nu}$ are due to the fields of external objects (and to nonlinear interactions); I call this the \emph{inner expansion}. In a \emph{buffer region} around the object, defined by $\e\ll r\ll\e^0$, we assume a \emph{matching condition} is satisfied: if the outer expansion is re-expanded in the limit $r\ll\e^0$ and the inner expansion is re-expanded in the limit $r\gg\e$, the two expansions must agree term by term in powers of $r$ and $\e$, since they both began as expansions of the same metric. The relationship between the various regions and expansions is shown schematically in Fig.~\ref{Fig:matching}, and it will be described precisely in Sec.~\ref{matched expansions}.

Historically, matched asymptotic expansions have been a highly successful way of treating singular perturbation problems in which the behavior of the solution rapidly changes in a localized region. For general discussions of singular perturbation theory in applied mathematics, I refer readers to the textbook~\cite{Kevorkian-Cole:96}, and for more rigorous treatments, to  Refs.~\cite{Fenichel:79,Eckhaus:79}. For general discussions in the context of general relativity, I refer them to~\cite{Kates:81, Pound:10b}. 

In the context of spacetimes containing small objects, matched expansions have been the standard method of tackling the problem; Refs.~\cite{DEath:75, DEath:96, Kates:80, Thorne-Hartle:85, Mino-Sasaki-Tanaka:97, Detweiler:01,Gralla-Wald:08,Pound:10a,Pound:12a,Gralla:12} are but a small sample. When it comes to obtaining equations of motion, the method hearkens back to an early insight of Einstein and others~\cite{Weyl:21,Einstein-Grommer:27,Einstein-Infeld-Hoffmann:38,Einstein-Infeld:49}: an object's equations of motion can be determined from the Einstein equations in a region \emph{outside} the object. Specifically, it can be determined from the field equations in the buffer region defined above. Relative to the object's scale $r\sim\e$, the buffer region $\e\ll r\ll \e^0$ is at asymptotically large distances, allowing one to make an asympotic characterization of how well ``centered'' the buffer region is around the object. 
One such characterization is based on a definition of multipole moments in the buffer region. Again, because the buffer region is at asymptotic infinity in the object's metric $g^{\rm obj}_{\mu\nu}$, we can define the object's multipole moments by examining the form of the metric there, rather than having to refer to the object's stress-energy. We can then use the metric's mass dipole moment in the buffer region as a measure of centeredness: if we install a timelike curve $\gamma$ in the background spacetime and define the mass dipole moment in coordinates centered on that worldline, then $\gamma$ is a good representative worldline if the mass dipole moment vanishes. Centeredness conditions along these lines will be described in more detail in Secs.~\ref{matched expansions} and \ref{gauge and motion}; for a schematic preview, see Figs.~\ref{Fig:matching} and \ref{Fig:GW-vs-SC}.

Prior to its application to self-force analyses~\cite{Mino-Sasaki-Tanaka:97, Detweiler:01,Gralla-Wald:08,Pound:10a,Pound:12a,Gralla:12}, this program was pursued furthest by Thorne and Hartle~\cite{Thorne-Hartle:85}. Where Dixon stood relative to the later non-perturbative work of Harte, Thorne and Hartle stand in the same position relative to perturbative self-force constructions. As did Dixon, they derived general laws of motion and precession for compact objects. They considered an object immersed in some external spacetime, say $(\varext_{\mu\nu},\mathfrak{M})$, and found forces and torques made up of couplings between the external curvature and the object's multipole moments; specifically, in Cartesian coordinates $(\mathfrak{t},x^i)$ that are at rest relative to a geodesic $z^\mu$ of the external spacetime (and in which the mass dipole moment vanishes), they found
\begin{align}
\frac{dp^i}{d\mathfrak{t}}&= -\mathfrak{B}^i{}_j S^j+\O(\e^3)\label{TH-pdot},\\
\frac{dS^i}{d\mathfrak{t}}&= -\epsilon^i{}_{ab}(\mathfrak{E}^b{}_c Q^{ac}+\tfrac{4}{3}\mathfrak{B}^b{}_c \mathcal{Q}^{ac}) +\O(\e^4),\label{TH-Sdot}
\end{align}
where $p^\mu$ and $S^\mu$ are the object's linear and angular momentum relative to $z^\mu$, $Q^{ab}$ and $\mathcal{Q}^{ab}$ are its ``mass and current" quadrupole moments (see the ends of Secs.~\ref{seeds} and \ref{imposing_gauge}), $\mathfrak{t}$ is proper time (as measured in $\varext_{\mu\nu}$) on $z^\mu$, and $\epsilon_{ijk}$ is the flat-space, Cartesian Levi-Civita tensor. The quantities $\mathfrak{E}_{ab}$ and $\mathfrak{B}_{ab}$ are the electric-type and magnetic-type quadrupole tidal moments of the external universe, which describe the tidal environment the object is placed in; they are related to the Riemann tensor of $\varext_{\mu\nu}$ according to $\mathfrak{E}_{ab} \equiv \mathfrak{R}_{\mathfrak{t}a\mathfrak{t}b}$ and $\mathfrak{B}_{ab} \equiv -\frac{1}{2}\epsilon^{pq}{}_{(a}\mathfrak{R}_{b)\mathfrak{t}pq}$. 

Physically, Eqs.~\eqref{TH-pdot}--\eqref{TH-Sdot} say that \emph{the object moves as a test body in the external metric}; the equations have the same structure as the  test-body equations of motion mentioned above in the context of Dixon's work~\cite{Dixon:74}. Up to the coupling of the object's moments to the tidal moments of the external universe, the motion is geodesic, and the spin parallel-propagated, in the external metric. But just as in Dixon's work, \emph{these results become useful only once one knows how to split the full metric into a self-field and an effectively external metric}. What Thorne and Hartle call the ``external metric" is not $g_{\mu\nu}$, but rather it is essentially defined to be whatever creates tidal fields across the object. As Thorne and Hartle themselves remarked, this ``external'' field includes a contribution from the object's own field. (Likewise, the ``object's" multipole moments can be altered by the external field.) 

In this sense, the self-force game is played by finding a useful split in which Thorne and Hartle's general laws are valid. Doing so requires developing a systematic theory of matched expansions for spacetimes containing small objects, as described in the body of this paper. Using those matched expansions, we will find that at linear order, we can circle back to the point particle picture: as first shown by D'Eath~\cite{DEath:75,DEath:96}, the linearized perturbation $h^1_{\mu\nu}$ in the outer expansion \emph{is identical to the linearized field sourced by a point particle} (see also Refs.~ \cite{Gralla-Wald:08,Pound:10a,Pound:12b} for more refined derivations).  And at that linearized level, we will find that the effectively external metric $\varext_{\mu\nu}$ of Thorne and Hartle can be taken to be the Detweiler-Whiting effective metric $\tilde g_{\mu\nu}=g_{\mu\nu}+\e h^{\R1}_{\mu\nu}$.

\subsection{Gauge, motion, and long-term dynamics}\label{long_term_dynamics}
Despite all the above preparation, we are still not ready to broach the problem of motion in perturbation theory. Two additional facts must first be understood: in perturbation theory, motion is intimately related to gauge freedom~\cite{Barack-Ori:01,Gralla:11,Pound:15b}; and in problems of astrophysical interest, the most important dynamical effects occur on the very long time scale $\sim1/\e$~\cite{Hinderer-Flanagan:08}. 

To the first point. At leading order, the object's motion is geodesic in the background metric $g_{\mu\nu}$; all deviation from that motion is driven by an order-$\e$ force. Suppose the self-accelerated worldline $\gamma$ is a smooth function of $\e$. Then we can write its coordinates as an expansion 
\beq\label{z_linear_expansion}
z^\mu( s,\e)=z^\mu_0( s)+\e z^\mu_1( s)+\O(\e^2), 
\eeq
where $ s$ is a parameter on the worldline, and the zeroth-order term $z_0^\mu$ is a geodesic of $g_{\mu\nu}$. Now consider the effect of a gauge transformation. Under a transformation generated by a vector $\e\xi^\mu$, a curve $z^\mu$ is shifted to a curve $z'^{\mu}=z^\mu-\e \xi^\mu+\O(\e^2)$. Nothing prevents us from choosing $\xi^\mu=z^\mu_1$, which leaves us with $z'^\mu=z_0^\mu+\O(\e^2)$, entirely eliminating the first-order deviation from $z^\mu_0$. This same idea can be carried to arbitrary order, meaning we can precisely set $z'^\mu=z_0^\mu$. In other words, the effect of the self-force appears to be pure gauge.

In one sense, this result is true. If we look at any finite region of spacetime and consider the limit $\e\to0$ in that region, the deviation from a background geodesic is, indeed, pure gauge. This does not mean it is irrelevant: in any given gauge, it must be accounted for to obtain the correct metric in that gauge. But it need not be accounted for in the \emph{linearized} metric. We can always substitute the expansion~\eqref{z_linear_expansion} into $h^1_{\mu\nu}(x;z)$ to obtain 
\beq\label{h1_expansion}
\e h^1_{\mu\nu}(x;z)=\e h^1_{\mu\nu}(x;z_0)+\e^2 \delta h^1_{\mu\nu}(x;z_0,z_1)+\O(\e^3),
\eeq
and we can then transfer the term $\delta h^1_{\mu\nu}(x;z_0,z_1)$ into the second-order perturbation, $\e^2 h^2_{\mu\nu}$. (An explicit expression for $\delta h^1_{\mu\nu}(x;z_0,z_1)$ is given in Eq.~\ref{dh_zperp}.)

However, this analysis assumes we work in a fixed, finite domain---and as mentioned in the first paragraph of this section, we do \emph{not} typically work in such a domain in problems of interest. Consider an EMRI. Gravitational waves carry away orbital energy from the EMRI at a rate $\dot E/E\sim \e$. It follows that the inspiral occurs on the time scale $t_{rr}\sim E/\dot E\sim 1/\e$, which is called the \emph{radiation-reaction time}. So in practice, we are not looking at the limit $\e\to 0$ on a finite interval of time $[0,T]$, where $T$ is independent of $\e$; instead, we are looking at the limit $\e\to 0$ on a time interval $[0,T/\e]$ that blows up. 

This consideration forces us to adjust our thinking about motion and gauge. Loosely speaking, the deviation from geodesic motion, $\e z^\mu_1$, is governed by an equation of the form $\frac{d^2z^\mu_1}{dt^2}\sim F^\mu_1$\footnote{More precisely, it is governed by Eq.~\eqref{GW_ddotz1}.}. On the radiation-reaction time scale, it therefore behaves as  $\e z^\mu_1\sim \e F^\mu_1 t_{rr}^2\sim 1/\e$. In other words, it blows up in the limit $\e\to0$. So on this domain, one cannot rightly write the worldline as a geodesic plus a self-forced correction, and one cannot use a small gauge transformation to shift the perturbed worldline onto a background geodesic; the gauge transformation would have to blow up in the limit $\e\to0$. 

Because we are solving partial differential equations and not ordinary ones, these arguments about time scales translate into arguments about spatial scales. For example, if we seek a solution in a Schwarzschild background spacetime, fields (including inaccuracies in them) propagates outward toward null infinity along curves of constant $t-r^*$, or inward toward the future horizon along curves of constant $t+r^*$, where $r^*$ is the tortoise coordinate. If one's accuracy is limited to a time span $\Delta t$, then it is also limited to a spatial region of similar size.

To organize our thinking, let us denote by $\D_{\varsigma(\e)}$ a spacetime region roughly of size $\varsigma(\e)$ (both temporal and spatial). I call an asymptotic solution to the Einstein equations a ``good'' solution in $\D_{\varsigma(\e)}$ if it is uniform in $\D_{\varsigma(\e)}$. That is, the asymptotic expansion ${\sf g}_{\mu\nu}=g_{\mu\nu}+\sum_{n>0} \e^n h^n_{\mu\nu}$ must satisfy $\lim_{\e\to0} \frac{\e h^1_{\mu\nu}}{g_{\mu\nu}}=0$ and $\lim_{\e\to0} \frac{\e^n h^n_{\mu\nu}}{\e^{n-1} h^{n-1}_{\mu\nu}}=0$ \emph{uniformly} (e.g., in a sup norm).

For the EMRI problem, we are interested in obtaining a good solution in a domain $\D_{1/\e}$. Suppose we use an asymptotic expansion of the form~\eqref{h1_expansion} and incorporate $\delta h^1_{\mu\nu}(x;z_0,z_1)$ into $h^2_{\mu\nu}$. In a gauge such as the Lorenz gauge, $z_1^\mu$ grows as $\sim F_1^\mu t^2$, and so $\delta h^1_{\mu\nu}(x;z_0,z_1)$ likewise grows as $t^2$. Hence, on $\D_{1/\e}$, its contribution to $\e^2 h^2_{\mu\nu}$ behaves at best as $\e^0$, comparable to $g_{\mu\nu}$. Clearly, this is not a good approximation. Suppose we instead eliminated $z_1^\mu$ using a gauge transformation generated by $\xi^\mu=z^\mu_1$. This removes the offending growth in $h^2_{\mu\nu}$, but it commits a worse offense: it alters $\e h^1_{\mu\nu}$ by an amount $2\e\xi_{(\mu;\nu)}$, which behaves at best as $\e t$, or as $\e^0$ on $\D_{1/\e}$. Hence, if we are in a gauge where the self-force is nonvanishing, $h^2_{\mu\nu}$ behaves poorly; if we are in a gauge where the self-force is vanishing, even $h^1_{\mu\nu}$ behaves poorly.

Let us chase the consequences of this. To obtain a good approximation in $\D_{1/\e}$, we need to work in a class of gauges compatible with uniformity in $\D_{1/\e}$. This means, in particular, that if we obtain a good approximation in a particular gauge---call it a \emph{good gauge}---we must confine ourselves to a class of gauges related to the good gauge by uniformly small gauge transformations. In turn, this means that the effects of the self-force are \emph{not} pure gauge on $\D_{1/\e}$. Due to dissipation, $z^\mu$ will deviate from any given geodesic $z_0^\mu$ by a very large amount in $\D_{1/\e}$, but by using an allowed gauge transformation we may shift it only by a very small amount, of order $\e$, on that domain. In other words, \emph{although the self-forced deviation from $z_0^\mu$ is pure gauge on a domain like $\D_{\e^0}$, it is no longer pure gauge in the domain $\D_{1/\e}$}.


\subsection{Self-consistent, Gralla-Wald, and osculating-geodesics approximations}\label{SC_GW_osculating}

In the preceding sections, we encountered several core concepts pertaining to the motion of a small object: (i) the metric can be usefully split into a self-field and an effectively external metric in which the object behaves as a test body, (ii)  the bulk motion of the object can be described in terms of forces and torques generated by freely specifiable multipole moments, (iii) by using matched expansions, all of this can be done in a vacuum region \emph{outside} the object, where multipole moments can be defined from the metric and laws of motion can be derived from the vacuum Einstein equation, and (iv) the representation of the object's bulk motion, the asymptotic expansion of the metric, and notions of gauge must all be tailored to suit the long timescales on which self-force effects accumulate.

Let me now combine this information into a cohesive framework, mostly following Ref.~\cite{Pound:10a} (but see also Refs.~\cite{Pound:12a, Pound:12b, Pound:14a, Pound:15b, Pound-Miller:14}). The overarching method I describe consists of solving the Einstein equation with an ``outer expansion'' in some vacuum region $\D_{\varsigma(\e)}$ outside the object, using only minimal information from the ``inner expansion" to determine the behavior of the solution very near the object.

Since we work in the vacuum region $\D_{\varsigma(\e)}$, we seek an asymptotic solution to the vacuum Einstein equation
\beq\label{vacuum_EFE}
R_{\mu\nu}[{\sf g}]=0.
\eeq
Following the lessons of post-Newtonian theory, I write this equation in a ``relaxed'' form~\cite{Blanchet:14,Futamase-Itoh:07,Poisson-Will:14}. I define $h_{\mu\nu}\equiv {\sf g}_{\mu\nu}-g_{\mu\nu}$, impose the Lorenz gauge condition\footnote{Reference~\cite{Pound:12b} describes how the entirety of this section can be performed in any gauge in which the linearized Einstein tensor is hyperbolic. Section~\ref{gauge} below offers a more general discussion of gauge.}
\beq\label{Lorenz_gauge_condition}
L_\nu[h]\equiv\nabla^\mu (h_{\mu\nu}-\frac{1}{2}g_{\mu\nu}g^{\alpha\beta}h_{\alpha\beta})=0,
\eeq
and write Eq.~\eqref{vacuum_EFE} with that condition imposed, making it read 
\beq\label{vacuum_relaxed_EFE}
E_{\mu\nu}[h] = S_{\mu\nu}[h],
\eeq
where $E_{\mu\nu}$ is the wave operator introduced in Eq.~\eqref{E} (but here acting on the metric perturbation rather than on its trace reverse),\footnote{In the Lorenz gauge in a vacuum background, the linearized curvature tensors are related by $E_{\mu\nu}[h]=-2\delta R_{\mu\nu}[h]=-2\overline{\delta G_{\mu\nu}[h]}=\overline{E_{\mu\nu}[\bar h]}$.} and  the ``source" $S_{\mu\nu}\equiv R_{\mu\nu}[g+h]-E_{\mu\nu}[h]$ is a nonlinear functional of $h_{\mu\nu}$. The background $g_{\mu\nu}$ is chosen to be a smooth solution to $R_{\mu\nu}[g]=0$ (if matter exists outside the object, it is assumed to be sufficiently far away to lie outside $\D_{\varsigma(\e)}$). This makes Eq.~\eqref{vacuum_relaxed_EFE} a weakly nonlinear hyperbolic equation for the perturbation $h_{\mu\nu}$; at this stage, that equation is still exact. 

I wish to solve Eq.~\eqref{vacuum_relaxed_EFE} in $\D_{\varsigma(\e)}$ subject to two types of boundary conditions:
\begin{itemize}[leftmargin=0.65in]
\item[{ (BC1)}] \emph{Global boundary conditions}. Examples of these are retarded boundary conditions or specified Cauchy data.
\item[{ (BC2)}] \emph{The matching condition}. In the buffer region, the solution must be compatible with an inner expansion. 
\end{itemize}
Together these conditions ensure we are describing the correct physical situation. But they do not yet uniquely determine the solution. Equation~\eqref{vacuum_relaxed_EFE} is called ``relaxed" because, unlike Eq.~\eqref{vacuum_EFE}, it can be solved no matter how the object moves; this relaxation arises because Eq.~\eqref{vacuum_relaxed_EFE} is not constrained by the Bianchi identity, unlike Eq.~\eqref{vacuum_EFE}. The object's motion is determined only once the gauge condition is also imposed, thereby making the solution to the relaxed equation also a solution to the unrelaxed one. As discussed earlier, in the present problem, the motion of the object is defined by the mass dipole moment of the metric in the buffer region, and we will find that the evolution of that mass dipole moment is determined by the gauge condition. 

Beyond these broad ideas, the specifics of the method, and the size of the region $\D_{\varsigma(\e)}$ in which it applies, depend crucially on how one represents the object's perturbed motion, which determines how one formulates the asymptotic solution to the Einstein equation. Here I describe three representations and corresponding asymptotic solutions: what I call ``self-consistent'', ``Gralla-Wald", and ``osculating geodesics" approximations.

\subsubsection{Self-consistent approximation}\label{SC-approx}

In the self-consistent approximation, to avoid the secularly growing errors described in Sec.~\ref{long_term_dynamics}, I seek to directly determine an accelerated worldline $z_\e^\mu(s)$ that represents the object's bulk motion; I do \emph{not} wish to expand  that worldline in powers of $\e$. To accommodate this, I write the perturbation $h_{\mu\nu}$ as $h_{\mu\nu}(x,\e;z_\e,\dot z_\e)$, where $\dot z_\e^\mu\equiv \frac{dz_\e^\mu}{ds}$ and the quantities after the semicolon denote a functional dependence. I expand this functional as
\beq\label{SC_expansion}
h_{\mu\nu}(x,\e;z_\e,\dot z_\e) = \sum_{n>0} \e^n h^n_{\mu\nu}(x;z_\e,\dot z_\e).
\eeq
Despite the fact that $z_\e^\mu$ and $\dot z_\e^\mu$ depend on $\e$, they are not expanded; in other words, I hold them fixed while taking the limit $\e\to0$. Later, I will suppress the functional dependence on $\dot z_\e^\mu$ and simply write $h^n_{\mu\nu}(x;z_\e)$. 

The self-consistent representation of motion is given its name because $z_\e^\mu$ must be determined simultaneously with $h^n_{\mu\nu}(x;z_\e)$. It was the representation adopted in the original derivations of the MiSaTaQuWa equation~\cite{Mino-Sasaki-Tanaka:97,Quinn-Wald:97,Tanaka:private}, and it is the one I used in describing that equation in Sec.~\ref{point_particle_picture}. It was first put on a sound and robust basis, as part of a systematic expansion of the Einstein equation, in Ref.~\cite{Pound:10a}. In this section, I outline that expansion.

I first refine the region in which I seek a solution to the relaxed Einstein equation~\eqref{vacuum_relaxed_EFE}. Install the timelike curve $\gamma_\e$ (with coordinates $z_\e^\mu)$ in the background spacetime, and let $D_{\varsigma(\e)}\supset\gamma_\e$ be a region of size $\varsigma(\e)$. I define $\B_{\gamma_\e}$ to be a region of proper radius $r_B$ centered on $\gamma_\e$, with $\e\ll r_B\ll\e^0$. The region I seek a solution in is then $\D_{\gamma_\e,\varsigma(\e)}=D_{\varsigma(\e)}\setminus\B_{\gamma_\e}$. The inner boundary of this region lies in the buffer region, and there the solution must satisfy the matching condition (BC2).

\subsubsection{Field equations}
Since the coefficients $h^n_{\mu\nu}(x;z_\e)$ in the expansion \eqref{SC_expansion} depend on $\e$, it may seem they are not uniquely determined. However, here I define the functional $h_{\mu\nu}(x,\e;z)$ to be the solution to the \emph{relaxed} Einstein equation~\eqref{vacuum_relaxed_EFE}; the relaxed equation places no constraints on $z^\mu$, and each function $z^\mu:\mathbb{R}\to\mathbb{R}^4$ yields a different solution. (In the present context this means there are no constraints on the motion of the region $\B_{\gamma}$.) The coefficients $h^n_{\mu\nu}(x;z)$ are then uniquely determined to be the solution to the $n$th-order term in an ordinary power-series expansion of the relaxed Einstein equation. That $n$th-order term has the form $E_{\mu\nu}[h^n]=S^n_{\mu\nu}[h^1,\ldots,h^{n-1}]$. Up to $n=3$, it reads
\begin{subequations}\label{SC_EFEs}%
\begin{align}
E_{\mu\nu}[h^1] &= 0\qquad x\in \D_{\gamma,\varsigma(\e)},\label{SC_EFE1} \\
E_{\mu\nu}[h^2] &= 2\delta^2R_{\mu\nu}[h^1,h^1] \qquad x\in \D_{\gamma,\varsigma(\e)},\label{SC_EFE2}\\
E_{\mu\nu}[h^3] &= 2\delta^3R_{\mu\nu}[h^1,h^1,h^1]+4\delta^2R_{\mu\nu}[h^1,h^2] \qquad x\in \D_{\gamma,\varsigma(\e)},\label{SC_EFE3}
\end{align}
\end{subequations}
where I have defined the ``$n$th-order Ricci tensor'' to be the $n$th variation\footnote{Cross terms like $\delta^2R_{\mu\nu}[h^1,h^2]$ are as defined in Eq.~\eqref{dnA-gen}.}
\beq
\delta^n R_{\mu\nu}[h,\ldots,h] = \frac{1}{n!}\frac{d^n}{d\lambda^n}R_{\mu\nu}[g+\lambda h]\big|_{\lambda=0}.
\eeq
In the concrete calculations in this paper, I will require only $\delta^2 R_{\mu\nu}$, which is given explicitly by
\begin{align}
\delta^2R_{\alpha\beta}[h,h] &=
		-\tfrac{1}{2}\bar h^{\mu\nu}{}_{;\nu}\left(2h_{\mu(\alpha;\beta)}
		-h_{\alpha\beta;\mu}\right)  
		+\tfrac{1}{2}h^{\mu}{}_{\beta}{}^{;\nu}\left(h_{\mu\alpha;\nu} -h_{\nu\alpha;\mu}\right)\nonumber\\
		&\quad  +\tfrac{1}{4}h^{\mu\nu}{}_{;\alpha}h_{\mu\nu;\beta}-\tfrac{1}{2}h^{\mu\nu}\left(2h_{\mu(\alpha;\beta)\nu}-h_{\alpha\beta;\mu\nu}-h_{\mu\nu;\alpha\beta}\right).
		\label{second-order_Ricci}
\end{align}

Solving Eqs.~\eqref{SC_EFEs} subject to the boundary conditions (BC1)--(BC2) yields a functional-valued asymptotic solution to the relaxed Einstein equation~\eqref{vacuum_relaxed_EFE}. Now we must find a particular function $z^\mu=z_\e^\mu$ for which $h_{\mu\nu}(x,\e;z)$ is also an asymptotic solution to the unrelaxed equation~\eqref{vacuum_EFE}. To do this, we must ensure $h_{\mu\nu}(x,\e;z)$ also satisfies the gauge condition~\eqref{Lorenz_gauge_condition} to the same order.
\footnote{Equation~\eqref{gauge_condition_SC} illustrates more explicitly, using a point-particle field, how the gauge condition implies an equation of motion.} 

I accomplish this in a systematic way by writing the accelerated equation of motion of $z_\e^\mu$ as
\beq\label{SC_eq_mot}
a_\e^\mu\equiv\frac{D^2z_\e^\mu}{d\tau^2} =F^\mu(\tau,\e),
\eeq
and then assuming that like $h_{\mu\nu}$, the force (per unit mass) appearing on the right-hand side can be expanded as
\beq\label{SC_F_expansion}
F^\mu(\tau,\e) = \sum_{n\geq0} \e^n F_n^\mu(\tau;z_\e,\dot z_\e).
\eeq
I substitute this expansion, together with the one in Eq.~\eqref{SC_expansion}, into the gauge condition~\eqref{Lorenz_gauge_condition} and solve order by order in $\e$ \emph{while holding $(z^\mu_\e,\dot z^\mu_\e)$ fixed}. By holding $(z^\mu_\e,\dot z^\mu_\e)$ fixed during this procedure, rather than expanding their $\e$ dependence, I preserve the particular accelerated worldline that satisfies some appropriate mass-centeredness condition, such as the vanishing of a suitably defined mass dipole moment in the buffer region centered on $\gamma_\e$. Solving the sequence of gauge conditions to higher and higher order yields a better and better approximation to the equation of motion of that particular worldline, \emph{without ever expanding the worldline itself}. The first few of these gauge conditions are
\begin{subequations}\label{gauge_conditions_SC}%
\begin{align}
L_\mu[h^1,F_0] &= 0,\label{SC_gauge1}\\
\delta L_\mu[h^1,F_1] &= -L_\mu[h^2,F_0],\label{SC_gauge2}\\
\delta L_\mu[h^1,F_2] &= -L_\mu[h^3,F_0]-\delta L_\mu[h^2,F_1]-\delta^2L_\mu[h^1,F_1,F_1],
\end{align}
\end{subequations}
where I have defined $L_\mu[h,F]=L_\mu[h]\big|_{a=F}$ and 
\beq
\delta^n L_\mu[h,F,\ldots,F]\equiv \frac{1}{n!}\frac{d^n}{d\lambda^n}L_\mu[h,F_0+\lambda F]\big|_{\lambda=0}.
\eeq
This sequence determines the forces $F_n^\mu$,\footnote{It also constrains other quantities in $h_{\mu\nu}$, particularly determining the evolution of the object's mass and spin.} thereby determining the equation of motion~\eqref{SC_eq_mot}.

\subsubsection{Solution method}
We solve Eqs.~\eqref{SC_EFEs} and \eqref{gauge_conditions_SC} by working outward from the buffer region. Solving them in the buffer region using a local expansion, subject to (BC2), yields two things: the local form of the metric outside the object, and an equation of motion for the object. 

The local form of the metric is described in Sec.~\ref{algorithm}. In line with the themes of the earlier sections, it allows a natural split into an effectively external metric $\tilde g_{\mu\nu}=g_{\mu\nu}+h^\R_{\mu\nu}$ and a self-field  $h^\S_{\mu\nu}\equiv h_{\mu\nu}-h^\R_{\mu\nu}$, where $\tilde g_{\mu\nu}$ satisfies all the ``nice'' properties of the Detweiler-Whiting regular field. 

Derivations of the equations of motion at first and second order are sketched in Secs.~\ref{algorithm} and \ref{gauge and motion}. If at leading order the object's spin and quadrupole moment both vanish, then the equation of motion is~\cite{Pound:12a,Pound:14a}
\beq
\frac{D^2z_\e^\mu}{d\tau^2} = -\frac{1}{2}P^{\mu\gamma}(g_\gamma^\nu-h^{\rm R}_\gamma{}^\nu)(2h^{{\rm R}}_{\rho\nu;\sigma}-h^{{\rm R}}_{\rho\sigma;\nu})u^\rho u^\sigma 
											+\O(\e^3),\label{SC_eq_mot_2nd}
\eeq
where $h^\R_{\mu\nu}=\e h^{\R1}_{\mu\nu}+\e^2 h^{\R2}_{\mu\nu}+\O(\e^3)$. Following the steps of Appendix~\ref{geodesic_expansion_in_h}, this equation of motion can also be written as $\frac{\tilde D^2 z_\e^\mu}{d\tilde\tau^2}=\O(\e^3)$, the geodesic equation in the effective (smooth, vacuum) metric $\tilde g_{\mu\nu}$. In other words, at least through second order in $\e$, the generalized equivalence principle described in Sec.~\ref{Detweiler-Whiting} holds.

After obtaining the local results in the buffer region, one might think to solve Eqs.~\eqref{SC_EFEs} and \eqref{gauge_conditions_SC} globally in $\D_{\gamma_\e,\varsigma(\e)}$ by imposing agreement with the local results on the inner boundary $\partial\B_{\gamma_\e}$ and then moving $\partial\B_{\gamma_\e}$ using the equation of motion. However, in practice, a global solution is instead obtained by analytically extending the buffer-region results into $\B_{\gamma_\e}\setminus\gamma_\e$, replacing the physical metric there with the fictitious, analytically extended metric, while insisting that outside $\B_{\gamma_\e}$, the metric is unaltered. This procedure (described in detail in Sec.~\ref{puncture}) allows us to work with field equations on the whole of $\D_{\varsigma(\e)}=\D_{\gamma_\e,\varsigma(\e)}\cup\B_{\gamma_\e}$. At order $\e$, the procedure reveals that $h^1_{\mu\nu}(x;z_\e)$ in $\D_{\gamma_\e,\varsigma(\e)}$ is precisely equal to the perturbation produced by a point mass moving on $z_\e^\mu$, as promised in Sec.~\ref{tail_picture}. More generally, at all orders, it leads to a practical \emph{puncture scheme}~\cite{Barack-Golbourn:07,Vega-Detweiler:07,Dolan-Barack:11,Vega-Wardell-Diener:11,Pound:12b,Pound-Miller:14}, in which the \emph{puncture} $h^\P_{\mu\nu}$, a local approximation to $h^\S_{\mu\nu}$, moves on $\gamma_\e$, and the field equations in $\B_{\gamma_\e}$ are recast as equations for a \emph{residual field} $h^\res_{\mu\nu}$ that locally approximates $h^{\R}_{\mu\nu}$.\footnote{Note that although a puncture scheme utilizes approximations to $h^\S_{\mu\nu}$ and $h^\R_{\mu\nu}$, it is designed to \emph{exactly} obtain $h^\R_{\mu\nu}$ (and any finite number of its derivatives) on the worldline, meaning it does not introduce any approximation into the motion of $\gamma_\e$. Nor does it introduce approximations into the physical field $h_{\mu\nu}=h^\R_{\mu\nu}+h^\S_{\mu\nu}=h^\res_{\mu\nu}+h^\P_{\mu\nu}$.} 

The setup of a puncture scheme is compactly summarized in Eqs.~\eqref{h1_SC}--\eqref{motion_SC} below.  Using this scheme, one can directly solve for the effective metric on the worldline and use it to evolve that worldline via the equation of motion, and at the same time one can obtain the physical metric outside $\B_{\gamma_\e}$. Although we begin with a potentially complicated extended object, this scheme illuminates the fact that in self-force theory, we do not need to know anything about the particularities of that object: \emph{at the end of the day, all necessary physical information about it is absorbed into the puncture and the motion of that puncture}.  

\subsubsection{Accuracy estimates}\label{accuracy}
How accurate will this self-consistent approximation be on a domain $\D_{\gamma,\varsigma(\e)}$? Let us make the reasonable assumption that the largest secularly growing error in the approximation arises from truncating the expansion~\eqref{SC_F_expansion} at some order $\e^n$, leading to an error in $z^\mu$ of order $\delta z^\mu\sim \e^{n+1}F^\mu_{n+1}t^2$.\footnote{Here I return to what will become my common practice of dropping the subscript $\e$ on $z^\mu$ for simplicity, though I refer to the self-consistently determined center-of-mass worldline, not the freely specifiable worldline for which the relaxed Einstein equation can be solved.} The largest error in $h_{\mu\nu}$ is then 
\beq
\text{Error in } h_{\mu\nu}\sim\e h^1_{\mu\nu}\cdot \delta z^\alpha =\O(\e^{n+2}t^2). 
\eeq
The order of accuracy depends on the size of the domain we work in. Suppose we work in  $\mathcal{D}_{\gamma,1/\e}$, corresponding to the radiation-reaction time. On that domain, the error from neglecting $F_{n+1}^\mu$ is $\O(\e^{n+2}\cdot 1/\e^2)=\O(\e^n)$. Therefore, if we include only $F^\mu_1$ in the equation of motion, solving (via a puncture scheme) the coupled system comprising Eqs.~ \eqref{SC_EFE1} and \eqref{Detweiler-Whiting-form1}, then the result contains errors  $\O(\e^3 \cdot 1/\e^2)=\O(\e)$---which is as large as our first-order perturbation. In other words, this approximation fails on $\mathcal{D}_{1/\e}$. If in addition we include $F^\mu_2$, solving the coupled system comprising Eq.~\eqref{SC_EFE1}, \eqref{SC_EFE2}, and \eqref{SC_eq_mot_2nd}, then the error is $\O(\e^4 \cdot 1/\e^2)=\O(\e^2)$; hence, with this approximation, we can have faith in our field $h^1_{\mu\nu}(x;z)$.

In the smaller domain $\mathcal{D}_{1/\sqrt\e}$, corresponding to the so-called \emph{dephasing time} $t_{\rm dph}\sim 1/\sqrt{\e}$,  the approximations are more accurate. Including only $F_1^\mu$, by solving Eqs.~ \eqref{SC_EFE1}  and \eqref{Detweiler-Whiting-form1}, yields a first-order-accurate solution with errors $\O(\e^3 \cdot 1/\e)=\O(\e^2)$. Including $F_2^\mu$, by solving Eqs.~ \eqref{SC_EFE1}, \eqref{SC_EFE2}, and \eqref{SC_eq_mot_2nd}, yields a second-order-accurate solution with errors $\O(\e^4 \cdot 1/\e)=\O(\e^3)$.

\subsubsection{Gralla-Wald approximation}\label{GW-approx}
I next consider the Gralla-Wald representation of perturbative motion, named after the authors of Refs.~\cite{Gralla-Wald:08,Gralla:12}. In this approximation, as in Eqs.~\eqref{z_linear_expansion}--\eqref{h1_expansion}, one considers the effect of the self-force to be a small perturbation of the worldline, and one expands the worldline as
\beq\label{z_expansion}
z^\mu( s,\e) = \sum_{n\geq0}\e^n z_n^\mu(s),
\eeq
where the terms $z^\mu_{n>0}$ measure the deviation of $z^\mu$ from the zeroth-order worldline $z_0^\mu$. Substituting this expansion into Eq.~\eqref{SC_expansion}, one obtains a new expansion of the metric perturbation:
\beq\label{GW_h_expansion}
h_{\mu\nu}(x,\e;z,\dot z) = \sum_{n>0} \e^n \check{h}^n_{\mu\nu}(x;z_0,\ldots,z_{n-1}).
\eeq
Section~\ref{worldline} describes this expansion in some detail. Explicitly, at first and second order, $h^1_{\mu\nu}(x;z_0)=h^1_{\mu\nu}(x;z_0)$ and $\check{h}^2_{\mu\nu}(x;z_0,z_1)=h^2_{\mu\nu}(x;z_0)+\delta h^1_{\mu\nu}(x;z_0,z_1)$, as described below Eq.~\eqref{h1_expansion}. Since the individual terms $z_n^\mu(s)$ are independent of $\e$ in Eq.~\eqref{z_expansion}, Eq.~\eqref{GW_h_expansion} is an ordinary expansion in which the coefficients $\check{h}^n_{\mu\nu}$ do not depend on $\e$, unlike in Eq.~\eqref{SC_expansion}. Rather than starting from the self-consistent representation and then expanding the worldline, one could instead simply start with this ordinary expansion, as was done by Gralla and Wald. In this paper, to explicate the relationship between the two, I will instead almost always derive Gralla-Wald results from self-consistent results.

In the Gralla-Wald approximation, the role of the domain $\D_{\gamma,\varsigma(\e)}$ is played by $\D_{\gamma_0,\varsigma(\e)}$, which excludes a small region around $\gamma_0$ (the worldline with coordinates $z^\mu_0)$. In that region Eqs.~\eqref{SC_EFEs} become 
\begin{subequations}\label{GW_EFEs}%
\begin{align}
E_{\mu\nu}[ \check{h}^1] &= 0 \qquad x\in \D_{\gamma_0,\varsigma(\e)},\label{GW_EFE1}\\
E_{\mu\nu}[ \check{h}^2] &= 2\delta^2R_{\mu\nu}[ \check{h}^1, \check{h}^1] \qquad x\in \D_{\gamma_0,\varsigma(\e)},\label{GW_EFE2}\\
E_{\mu\nu}[\check{h}^3] &= 2\delta^3R_{\mu\nu}[\check{h}^1,\check{h}^1,\check{h}^1]+4\delta^2R_{\mu\nu}[\check{h}^1,\check{h}^2] \qquad x\in \D_{\gamma_0,\varsigma(\e)}.\label{GW_EFE3}
\end{align}
\end{subequations}
The gauge conditions~\eqref{gauge_conditions_SC} become simply
\beq
L_\mu[\check{h}^n]=0.
\eeq
The equation of motion~\eqref{SC_eq_mot_2nd} becomes a sequence of equations for $z_n^\mu$:
\begin{align}
\frac{D^2z^\mu_0}{d\tau_0^2} &= 0,\label{GW_ddotz0}\\
\frac{D^2z^\mu_1}{d\tau_0^2} &= \check{F}_1^\mu(\tau_0;z_0)-R^\mu{}_{\alpha\beta\gamma}u^\alpha_0 z^\beta_1 u^\gamma_0 ,\label{GW_ddotz1}\\
\frac{D^2z^\mu_{2F}}{d\tau_0^2} &= \check{F}^\mu_2(\tau_0;z_0,z_1) 
							-R^\mu{}_{\alpha\beta\gamma}\left(u^\alpha_0 z_{2F}^\beta u_0^\gamma+2u_1^\alpha z_1^\beta u_0^\gamma\right)\nonumber\\
				&\quad			+2R^\mu{}_{\alpha\beta\gamma;\delta}z_1^{(\alpha} u_0^{\beta)} z_1^{[\gamma} u_0^{\delta]},\label{GW_ddotz2}
\end{align}
where $\tau_0$ is proper time on $\gamma_0$ as measured in $g_{\mu\nu}$, $u_0^\mu\equiv \frac{dz_0^\mu}{d\tau_0}$, $u_1^\mu\equiv \frac{Dz_1^\mu}{d\tau_0}$, and  the forces $\check{F}_1^\mu$ and $\check{F}^\mu_2$ are constructed from $\check{h}^\R_{\mu\nu}$ according to Eqs.~\eqref{F1_generic}--\eqref{F2_generic}. The Riemann terms in these equations of motion are geodesic-deviation terms; they correspond to the fact that even in the absence of a force, two neighbouring curves $z^\mu$ and $z^\mu_0$ will deviate from one another due to the background curvature. Appendix~\ref{geodesic_expansion_in_h_and_dz} describes how Eqs.~\eqref{GW_ddotz0}--\eqref{GW_ddotz2} are derived from Eq.~\eqref{SC_eq_mot_2nd}. As explained more thoroughly there, the quantities $z_1^\mu$ and $z_{2F}^\mu$ are vectors that live on $\gamma_0$. $z_{2F}^\mu$ is defined by applying the expansion~\eqref{z_expansion} in a normal coordinate system centered on $\gamma_0$; it is related to $z_2^\mu$ in any other coordinate system by the coordinate-dependent relation $z^\mu_{2F} = z^\mu_2 + \frac{1}{2}\Gamma^\mu_{\nu\rho}(z_0)z_1^\nu z_1^\rho$. 



Just as in the self-consistent case, we can work with field equations on $\D_{\varsigma(\e)}=\D_{\gamma_0,\varsigma(\e)}\cup\B_{\gamma_0}$ by replacing the physical metric in $\B_{\gamma_0}$ with the analytical extension of the buffer-region metric. Re-expanding the results from the self-consistent case, we find that $\check{h}^1_{\mu\nu}(x;z_0)$ in $\D_{\gamma_0,\varsigma(\e)}$ is identical to the perturbation sourced by a point particle moving on $z_0^\mu$; the expansion of $h^1_{\mu\nu}(x;z)$ around $\check{h}^1_{\mu\nu}(x;z_0)$ is derived in Sec.~\ref{pp_quantities_expansion}. We also arrive at a substantially simplified practical puncture scheme: rather than having to solve for $z^\mu$ and $h_{\mu\nu}$ together, as a coupled system, one can first specify a geodesic $z_0^\mu$ and then calculate in sequence (i) the perturbations $\check{h}^1_{\mu\nu}(x;z_0)$ and $\check{h}^{\R1}_{\mu\nu}(x;z_0)$, (ii) the deviation $z_1^\mu$ driven by $\check{h}^{\R1}_{\mu\nu}$, (iii)  the perturbations $\check{h}^2_{\mu\nu}(x;z_0,z_1)$ and $\check{h}^{\R2}_{\mu\nu}(x;z_0,z_1)$, (iv) the deviation $z^\mu_2$, and so on. \emph{At all orders, the puncture moves on $z_0^\mu$}; the deviations $z_{n>0}^\mu$, through their appearance in $\check{h}^{n>1}_{\mu\nu}$, merely alter the singularity structure of the puncture. This puncture scheme is compactly summarized in Eqs.~\eqref{z0_GW}--\eqref{h2_GW} below.

On what domain is the Gralla-Wald approximation valid? 
There is no obvious estimate for the rate of growth of the terms $z_{n>1}^\mu$, but as in Sec.~\ref{long_term_dynamics}, we can easily estimate the growth of $\e z_1^\mu$ to be of order $\e t^2$.\footnote{See Ref.~\cite{Pound:14c} for an explicit solution to Eq.~\eqref{GW_ddotz1} in a particular scenario.} If we assume that the dominant error in $h_{\mu\nu}$ arises from the $z^\mu_1$ term in $\check{h}_{\mu\nu}^2$, then we have
\beq
\text{Error in } \check{h}_{\mu\nu}\sim \e^2 \check{h}^1_{\mu\nu} z_1^\alpha =\O(\e^2t^2). 
\eeq
If this estimate is valid, we can make the approximation valid to any order on domains $\D_{\gamma_0,\varsigma(\e)}$ with sufficiently small $\varsigma(\e)\ll t_{dph}=1/\sqrt{\e}$. On domains comparable to the dephasing time, $\D_{\gamma_0,1/\sqrt\e}$, the deviation vector $\e z_1^\mu$ becomes of order $1$, the second-order metric perturbation becomes as large as the first, and the expansion of the worldline ceases to be sensible. If higher-order deviations $z_{n>1}^\mu$ grow large much more quickly than $z_1^\mu$, as we might surmise from Eq.~\eqref{GW_ddotz2}, then the domain of validity of the expansion may be substantially smaller than $\D_{\gamma_0,1/\sqrt\e}$ even for first-order accuracy; if at each higher order the deviations grow more rapidly than the last, it may be the case that the domain of validity cannot be extended beyond $\D_{\gamma_0,1}$.

\subsubsection{Osculating geodesics}\label{osc-approx}
Finally, I consider an approximation intermediate between the self-consistent and Gralla-Wald expansions, one which makes use of both the expanded and unexpanded representations of the worldline. Starting from the self-consistent representation, the idea is at each instant $\tau$ on $\gamma$, to perform a Gralla-Wald expansion in a region $\D_{\gamma_{0(\tau)},\varsigma(\e)}$, where $\gamma_{0(\tau)}$ is a geodesic of $g_{\mu\nu}$ that is instantaneously tangential to $\gamma$ at time $\tau$; $\gamma_{0(\tau)}$ is called an \emph{osculating geodesic}~\cite{Pound-Poisson:08a}.\footnote{The scheme I describe here should not be confused with the general method of osculating geodesics, which is simply a way of using instantaneously tangential geodesics to rewrite an equation of motion $\frac{D^2z^\mu}{d\tau^2}=F^\mu$ in terms of more convenient variables; that general method, inherited from celestial mechanics, is exact and does not inherently involve an expansion of $z^\mu$, although it is particularly well suited to the osculating-geodesic approximation discussed here~\cite{Pound-Poisson:08a}.}  By solving the field equations of the Gralla-Wald approximation, one may calculate the self-force at time $\tau$, and then use that force to evolve $z^\mu$ to the next time step. By following this procedure at each time step, one eventually obtains $\gamma$ over the entire timespan of interest. The terms in the self-consistent approximation $h_{\mu\nu}(x;z)=\sum_\e \e^nh^n_{\mu\nu}(x;z)$ can then be found simply by solving Eqs.~\eqref{SC_EFE1}--\eqref{SC_EFE3} (and higher-order analogues) with $\gamma$ already pre-determined. 

More concretely, at each instant $\tau$ on the worldline, one perform a Gralla-Wald expansion $z^\mu(\tau',\e) = \sum \e^n z_{n(\tau)}^\mu(\tau')$, where $\tau$ is the specific instant of interest and $\tau'$ is variable, and one substitutes this expansion into the right-hand side of Eq.~\eqref{SC_F_expansion}---\emph{but not into the left-hand side}---to obain
\beq\label{eq_mot_osculating}
\frac{D^2z^\mu}{d\tau^2} = \e \check{F}^\mu_1(\tau;z_{0(\tau)}) + \e^2 \check{F}^\mu_2(\tau;z_{0(\tau)},z_{1(\tau)})+\O(\e^3).
\eeq
The forces on the right-hand side are constructed from fields $\check{h}^{\R1}_{\mu\nu}(x;z_{0(\tau)})$ and $\check{h}^{\R2}_{\mu\nu}(x;z_{0(\tau)},z_{1(\tau)})$ (and higher-order fields for higher order forces) according to Eqs.~\eqref{F1_generic}--\eqref{F2_generic}. For example, in terms of the tail of the perturbation, $\check{F}^\mu_1(\tau;z_{0(\tau)})$ is given by the order-$\e$ term on the right-hand side of Eq.~\eqref{MiSaTaQuWa} but with the tail integral~\eqref{tail} evaluated over $z_{0(\tau)}^\mu$ rather than over $z^\mu$.  These fields are found by solving the following sequence of equations at each value of $\tau$: \eqref{GW_ddotz0} for the osculating geodesic $z_{0(\tau)}^\mu$; \eqref{GW_EFE1} for $\check{h}^1_{\mu\nu}(x;z_{0(\tau)})$ in $\D_{\gamma_{0(\tau)},\varsigma(\e)}$; \eqref{GW_ddotz1} for $z^\mu_{1(\tau)}$; and \eqref{GW_EFE2} for $\check{h}^2_{\mu\nu}(x;z_{0(\tau)},z_{1(\tau)})$ in $\D_{\gamma_{0(\tau)},\varsigma(\e)}$. This suffices to compute $\check{F}^\mu_1(\tau;z_{0(\tau)})$ and $\check{F}^\mu_2(\tau;z_{0(\tau)},z_{1(\tau)})$, but in principle the calculations could proceed to higher order. All of these equations are to be solved subject to the ``osculation conditions'' $z_{0(\tau)}^\mu=z^\mu(\tau)$, $u^\mu_{0(\tau)}(\tau)=u^\mu(\tau)$, $z_{n>0(\tau)}^\mu=0$, and $u^\mu_{n>0(\tau)}(\tau)=0$, which state that $\gamma_{0(\tau)}$ is tangential to $\gamma$ at time $\tau$. They must also be solved subject to boundary conditions that ensure the sum $\sum_{n>0}\e^n\check{h}^n_{\mu\nu}(x;z_{0(\tau)})$ agrees with the full field $h_{\mu\nu}$ in $\D_{\gamma_{0(\tau)},\varsigma(\e)}$; at nonlinear orders, finding those boundary conditions may be highly nontrivial.

At linear order, the osculating-geodesic approximation has already been concretely implemented to find $z^\mu$ using the force $ \check{F}^\mu_1(\tau;z_{0(\tau)})$~\cite{Warburton-etal:12,Lackeos-Burko:12,Burko-Khanna:13} (building on the framework in Ref.~\cite{Pound-Poisson:08a}) and to compute $h^1_{\mu\nu}(x;z)$~\cite{Lackeos-Burko:12,Burko-Khanna:13}. However, to my knowledge, the brief sketch above is the first time it has been described at nonlinear orders (although possibly equivalent ideas have been presented by Mino~\cite{Mino:05}). Considerably more work must be done to establish that the scheme is viable beyond linear order. If it is, then one may naively estimate that it is valid to the same order on the same domain as the self-consistent expansion: including only $\check{F}^\mu_1(\tau;z_{0(\tau)})$ in Eq.~\eqref{eq_mot_osculating} yields an approximation valid up to $\O(\e^2)$ errors in a domain $\D_{\gamma,1/\sqrt\e}$, and including both $\check{F}^\mu_1(\tau;z_{0(\tau)})$ and $\check{F}^\mu_2(\tau;z_{0(\tau)},z_{1(\tau)})$ yields an approximation valid up to $\O(\e^2)$ errors in a domain $\D_{\gamma,1/\e}$.

\subsection{Outline of this paper}

As discussed above, the bulk of this paper focuses on the formalism presented in Ref.~\cite{Pound:10a} and further developed in the series of papers~\cite{Pound:10b,Pound:12a,Pound:12b,Pound-Miller:14,Pound:14a,Pound:15b}.

I begin in Sec.~\ref{matched expansions} with a more complete description of matched asymptotic expansions, the concepts of which underly most of self-force theory. I focus on the importance of the buffer region and formulating appropriate definitions of the small object's representative worldline.

Sections \ref{algorithm} and \ref{puncture} present an algorithm for constructing an $n$th-order expansion in the self-consistent approximation. In Sec.~\ref{algorithm}, I describe how to obtain the general solution in the buffer region. This algorithm determines the equation of motion as well as a natural split of the general solution into a self-field and an effectively external field. 
In Sec.~\ref{puncture}, I describe how to generate the global solution in $\D_{\gamma,\varsigma(\e)}$, using as input the self-field obtained in the buffer region. Along the way, I show how at linear order, the point-particle picture is recovered, and how at nonlinear orders, a certain point-particle ``skeleton'' can be defined to characterize the object's multipole structure.

In Sec.~\ref{worldline}, I show how to recover a Gralla-Wald expansion or an osculating-geodesics expansion from the results of the self-consistent expansion.

In Sec.~\ref{gauge}, I discuss the gauge freedom in each of the three types of expansions---self-consistent, Gralla-Wald, and osculating-geodesics---and how that freedom relates to the representation of perturbative motion in each.

In Sec.~\ref{gauge and motion}, I sketch how one can derive equations of motion by obtaining further information about the inner expansion and then utilizing the relationship between gauge and motion. This is an alternative to the algorithmic approach in the buffer region, and it is the method that has been used in practice to derive second-order equations of motion~\cite{Pound:12a, Pound:14a, Gralla:12}.

I conclude in Sec.~\ref{conclusion} with a summary and a discussion of future directions.

The appendices contain more general results: expansions of the geodesic equation in a perturbed spacetime, expansions of point particle functionals of an accelerated worldline around a geodesic, and some identities pertaining to gauge transformations of curvature tensors. 

Some portions of this paper first appeared in slightly different forms in Ref.~\cite{Pound:15b}: specifically, Appendices~\ref{geodesic_equation_expansion}, \ref{pp_quantities_expansion}, and~\ref{gauge-identities}, and parts of Sec.~\ref{gauge}.


								\section{Matched asymptotic expansions}\label{matched expansions}


\subsection{Outer, inner, and buffer expansions}

Traditionally in applied mathematics, the method of matched asymptotic expansions has been used to find both inner and outer expansions to some desired degree of accuracy and then combine them to obtain a uniformly accurate solution in the entire domain. But for my purposes here, I will not be interested in obtaining an accurate solution in the region $r\sim\e$; instead, I will be interested in the inner expansion only insofar as it constrains the outer expansion. Finding an accurate inner expansion would require specifying the type of compact object we are examining, while here I am interested in generic results that apply for any compact object. As it turns out, finding those generic results requires only minimal knowledge of the inner expansion---in fact, just its general form in the buffer region suffices. 

To build toward that conclusion, let me describe the formalism more geometrically. I focus for the moment on the self-consistent case. In the outer expansion, we wish to approximate the exact spacetime $({\sf g}_{\mu\nu},\man_\e)$ in the domain $\D_{\gamma,\varsigma(\e)}$ outside a small region of size $r\ll\e^0$ around the object. I expand the exact spacetime around a background $(g_{\mu\nu},\man_0)$ by adopting some identification $\varphi:\D_{\gamma,\varsigma(\e)}\subset\man_0\to\man_\e$ between the spacetimes (in the region outside the inner region). In a given coordinate system $x^\mu:\man_0\to\mathbb{R}^4$, the identification map assigns points $p\in\man_0$ and $\varphi(p)\in\man_\e$ the same coordinate values $x^\mu(p)$. I next write ${\sf g}_{\mu\nu}$ as a functional of a worldline $\gamma\subset\man_0$ as ${\sf g}_{\mu\nu}(x,\e;z)$ in those coordinates, where $z^\mu(s)=x^\mu(\gamma(s))$, and I then expand for small $\e$ while holding both $x^\mu$ and $z^\mu$ fixed. I thence arrive at Eq.~\eqref{SC_expansion}.\footnote{In truth, it is unlikely that any of the expansions I consider, whether self-consistent, Gralla-Wald, or osculating-geodesics, is convergent. More likely, they are asymptotic approximations only. So when performing the self-consistent expansion, I actually assume that $|{\sf g}_{\mu\nu}(x,\e)-{\sf g}^N_{\mu\nu}(x,\e;z)|=o(\e^N)$, where ${\sf g}^N_{\mu\nu}(x,\e;z)=g_{\mu\nu}(x)+\sum_{n=1}^N\e^n h^n_{\mu\nu}(x;z)$. The notation $o(k(\e))$ means ``goes to zero faster than $k(\e)$''. } Note that $z^\mu$ is defined in the background, not in the perturbed spacetime; the identification map is assumed to exist only in $\D_{\gamma,\varsigma(\e)}$, not in the inner region where there may lie a black hole rather than an identifiable worldline, for example. Later, I will replace the physical metric in the interior of the object with the effective metric $\tilde g_{\mu\nu}$, and via the identification map, the worldline will have identical coordinate values in the effective spacetime $(\tilde g_{\mu\nu},\tilde\man)$ as in the background.

Now, the inner expansion is constructed by choosing some coordinates $(t,x^i)$ centered on $\gamma$, and then rescaling spatial distances according to $\bar x^i\equiv x^i/\e$. The inner expansion is performed by expanding for $\e\to0$ while holding the scaled coordinates $(t,\bar x^i)$ fixed, as in\footnote{Here indices refer to the unscaled coordinates $(t,x^i)$. If components are written in the scaled coordinates, overall factors $\e$ and $\e^2$ appear in front of $ta$ and $ab$ components, respectively. These overall factors have no practical impact.}  
\begin{equation}\label{inner_expansion}
{\sf g}_{\mu\nu}(t,\bar x^a,\e)=\gin_{\mu\nu}(t,\bar x^a)+\sum_{n\geq1}\e^n H^n_{\mu\nu}(t,\bar x^a;z).
\end{equation}
Tensors in this expansion live on a manifold $\manin$, where $(\gin_{\mu\nu},\manin)$ is identified as the object's spacetime were it isolated. The perturbations $H^n_{\mu\nu}(t,\bar x^a;z)$ describe the effect of interaction with the external spacetime. What is the meaning of this expansion? The scaled coordinates serve to keep distances fixed relative to the object's mass in the limit $\e\ll1$, effectively zooming in on the object by sending all distances much larger than the mass off toward infinity. The use of a single scaling factor makes the approximation most appropriate for compact objects, whose linear dimension is comparable to their mass. Scaling only distances, not $t$, is equivalent to assuming the object possesses no fast internal dynamics; that is, there is no evolution on the short timescale of the object's mass and size. 

Note that the treatment of the ``region around the object'' depends strongly on whether one considers a self-consistent or Gralla-Wald expansion. In the self-consistent case, the outer expansion takes the limit as the object shrinks toward zero size around the self-consistently determined, accelerated worldline $\gamma\subset\man_0$, and the inner expansion blows up a region around that accelerated worldline; in the Gralla-Wald case, the outer expansion takes the limit as the object shrinks to zero size around the zeroth-order, background geodesic $\gamma_0$, and the inner expansion blows up a region around that background geodesic. 

\begin{figure}[t]
\begin{center}
\includegraphics[width=\textwidth]{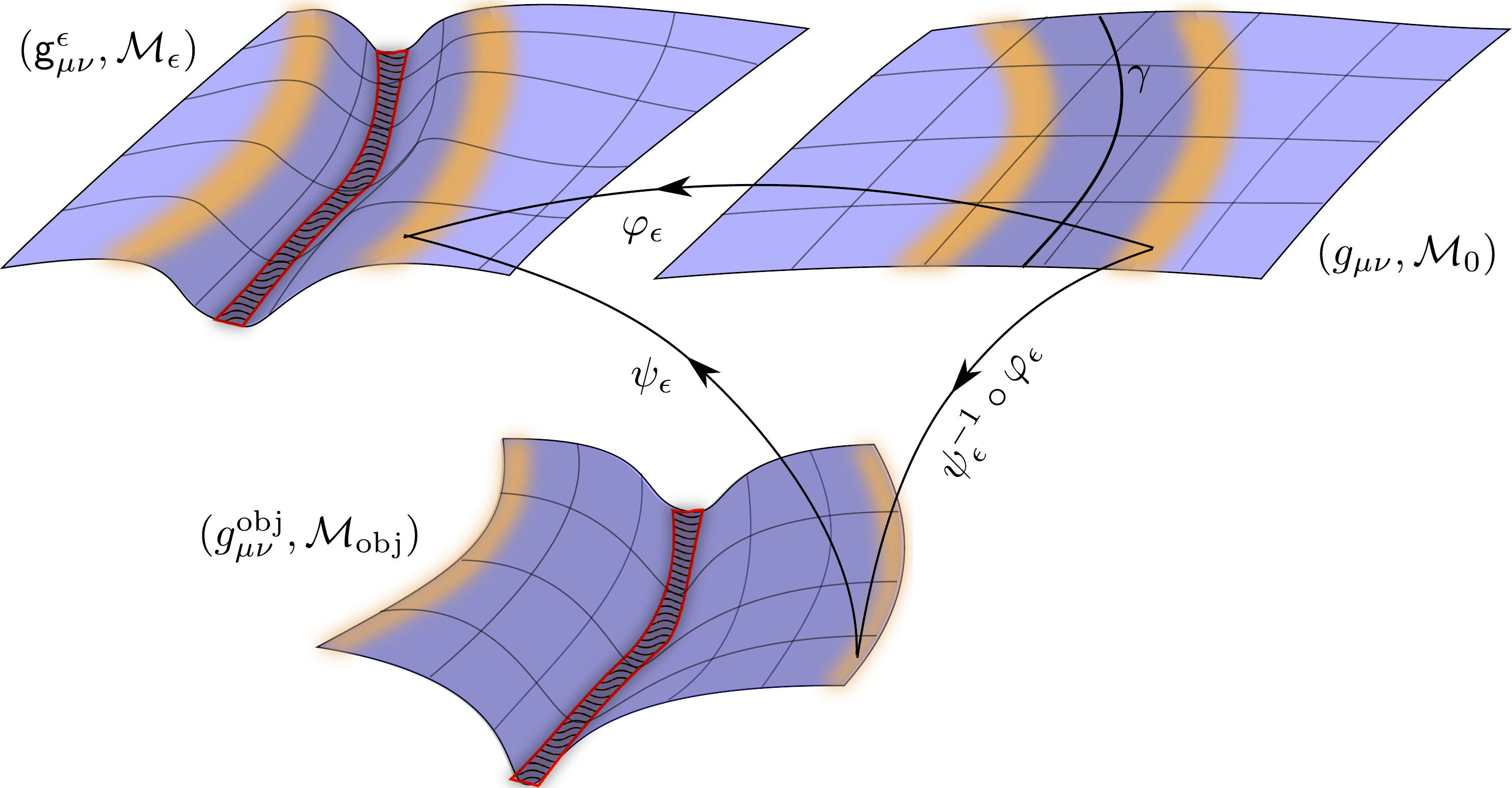}
\caption{\label{Fig:matching} The exact spacetime (top left), external background spacetime (top right), and inner background spacetime (bottom), with maps between them. In the exact spacetime, we have some small compact object, here shown as a material body (in dark red and black), but possibly a black hole or more exotic object. Around the object is a buffer region, shown in orange. Points outside the object can be identified (via the identification map $\varphi_\e$) with points in the external background manifold $\man_0$. In a chart $(t,x^i):\man_0\to\mathbb{R}^4$ centered on a worldline $\gamma\in\man_0$, the points $p\in\man_0$ and $\varphi_\e(p)\in\man_\e$ are assigned the same coordinate values $(t(p),x^i(p))$. The map $\psi^{-1}_\e\circ\varphi_\e$ zooms in on the object by rescaling spatial coordinates; in the same chart $(t,x^i)$, the point $q=\psi^{-1}_\e(\varphi_\e(p))$ is assigned coordinate values $(\bar t(q),\bar x^i(q))=(t(p),x^i(p)/\e)$.}
\end{center}
\end{figure}

In either case, the relationship between the two expansions is illustrated in Fig.~\ref{Fig:matching}. I now use this relationship to feed information from the inner expansion out to the outer expansion. This exchange of information is done in the buffer region around the object. From the perspective of the inner expansion, the buffer region lies at asymptotic infinity in $\manin$. In that region, the inner expansion can be expressed in unscaled coordinates as 
\begin{equation}
{\sf g}_{\mu\nu}(t,x^a/\e,\e)=\gin_{\mu\nu}(t,x^a/\e)+\sum_{n\geq1}\e^n H^n_{\mu\nu}(t,x^a/\e;z)
\end{equation}
and then re-expanded for small $\e$ (or equivalently, expanded for $r \gg \e$; i.e., for distances that are large on the scale of the inner expansion). Conversely, from the perspective of the outer expansion, the buffer region lies in a tiny region around the worldline. Hence, in that region the outer expansion can be expanded for $r\ll\e^0$ (i.e., for distances that are small on the scale of the outer expansion). Since the inner and outer expansions are assumed to approximate the same metric, it is assumed that the results of these re-expansions in the buffer region must match order by order in both $r$ and $\e$.\footnote{This matching condition amounts to the assumption that nothing too ``funny'' happens in the buffer region. It can instead be replaced by more explicit assumptions on the behavior of the full metric ${\sf g}_{\mu\nu}$, such as the conditions assumed in Ref.~\cite{Gralla-Wald:08} or various others discussed in Ref.~\cite{Eckhaus:79}.}

Now, say the $n$th-order outer perturbation is expanded as $h^n_{\mu\nu}(t,x^i)=\sum_p r^p h^{np}_{\mu\nu}(t,n^i)$ in the buffer region, where $n^i=x^i/r$.\footnote{$\ln r$ terms also generically arise. For simplicity, I incorporate those terms into $h^{np}_{\mu\nu}$ for the moment. Their presence does not spoil the well-orderedness of the expansion, since $r^p(\ln r)^q\ll r^{p'}(\ln r)^{q'}$ for $p>p'$. Similarly, $\ln\e$ terms can occur in solving the relaxed Einstein equation~\cite{Pound:12b}, and I absorb them into the coefficients $h^n_{\mu\nu}(x;z)$.} One must allow negative powers of $r$, since at least part of the field will fall off with distance from the body. But there is a bound on the most negative power at a given order in $\e$: Since the inner expansion is assumed to be well behaved, it must include no negative powers of $\e$. And since any term in $\e^n h^n_{\mu\nu}$ must coprrespond to a term in the inner expansion, if $\e^n h^n_{\mu\nu}$ is written as a function of the scaled distance $\bar r=r/\e$, it must likewise have no negative powers of $\e$. From this it follows that 
\begin{equation}
\e^n h^n_{\mu\nu}(t,x^i) = \frac{\e^n}{r^n}h^{n,-n}_{\mu\nu}(t,n^i)+\O(\e^n r^{-n+1}).\label{most_singular_part}
\end{equation}
Any higher power of $1/r$ would induce illegal powers of $\e$; for example, $\frac{\e^n}{r^{n+1}}=\frac{1}{\e\bar r^{n+1}}$. 

We can go one step further with this general analysis. Since $\e^n/r^n$ is independent of $\e$ in the scaled coordinates, and $\gin_{\mu\nu}$ is the only term in the inner expansion~\eqref{inner_expansion} that does not depend on $\e$, it must be that $h^{n,-n}_{\mu\nu}$ is equal to a term in $\gin_{\mu\nu}$. If we write  $\gin_{\mu\nu}(t,\bar x^i)$ in terms of the unscaled coordinates and expand for $r\gg\e$, we find its form in the buffer region is\footnote{The fact that the inner background must be asymptotically flat, containing no positive powers of $r$, follows from the assumption that the outer expansion contains no negative powers of $\e$, in the same manner as the cutoff on powers of $1/r$ in Eq.~\eqref{most_singular_part}.}
\begin{equation}\label{gin_buffer}
\gin_{\mu\nu}(t,\bar x^i/\e)=\sum_{n\geq0}\frac{\e^n}{r^n}g^{{\rm obj},n}_{\mu\nu}(t,n^i).
\end{equation}
From the matching condition, we then have 
\beq
h^{n,-n}_{\mu\nu}=g^{{\rm obj},n}_{\mu\nu}. 
\eeq
Therefore, at each order in $\e$, the most singular (as a function of $r$) piece of the metric perturbation $h^n_{\mu\nu}$ in the buffer region is determined by the $r\gg \e$ asymptotic behavior of the object's unperturbed metric. In addition, note that because quantities in the inner expansion vary slowly in time relative to their variation in space, they are quasistationary: after changing to scaled coordinates, a derivative with respect to $t$ effectively increases the power of $\e$ relative to a derivative with respect to $\bar x^i$. Hence, on any short time, we can think of $\gin_{\mu\nu}$ being stationary, and we can write its asymptotic form~\eqref{gin_buffer} in terms of a canonical set of multipole moments~\cite{Geroch:70,Hansen:74}. So our final statement is that the inner expansion constrains the outer expansion to have the form~\eqref{most_singular_part}, and the coefficients $h^{n,-n}_{\mu\nu}$ can be expressed in terms of multipole moments of the small object's unperturbed spacetime $(\gin_{\mu\nu},\manin)$. This is all the information that will be required from the inner expansion (until we get to Sec.~\ref{gauge and motion}).

\subsection{Defining the worldline}

Let us now use the above formalism to define what we mean by the object's worldline. First consider the self-consistent case. In the coordinates $(t,x^i)$ centered on $\gamma$, calculate the mass dipole moment $M^i$ of the spacetime $\gin_{\mu\nu}$. A mass dipole moment indicates the position of the center of mass relative to the origin of the coordinates. If the coordinates are mass-centered, a Coulomb-like piece of the field behaves as $m/r$; if the origin lies slightly away from the center of mass, by an amount $\xi^i$, then the Coulomb-like field behaves as $m/|x^i-\xi^i|$. Expanding this around $\xi^i=0$, we find $m/r + m\xi_i n^i/r^2+\O(|\xi|^2)$. The quantity $m\xi^i$ is the mass dipole moment $M^i$. Ergo, if this mass dipole moment vanishes in the coordinates centered on $\gamma$, then the object is appropriately centered ``on" $\gamma$, and we identify $\gamma$ as a good representative worldline.\footnote{This notion of mass-centeredness based on the mass dipole moment of  $\gin_{\mu\nu}$ applies only to order-$\e$ deviations from $z^\mu$. For higher-order deviations, mass-dipole-moment terms in the perturbations $H^n_{\mu\nu}$ must also be considered, or some other copacetic centeredness condition must be imposed, as discussed in Secs.~\ref{algorithm} and \ref{gauge and motion}.} Because the definition only utilizes quantities in the buffer region, this definition makes sense even if the object is a black hole or contains topological oddities such as a wormhole: even if there exists no identification map between the background and the exact spacetime in the region inside the object, at least in the buffer region the coordinates $(t,x^i)$ can be used to chart the manifolds $\man_\e$ and $\manin$, and in those coordinates the metrics ${\sf g}_{\mu\nu}$ and $\gin_{\mu\nu}$ have reference to the worldline $\gamma$ that is defined only in $\man_0$. I refer the reader again to Fig.~\ref{Fig:matching} to illuminate this.

\begin{figure}[t]
\begin{center}
\includegraphics[width=\textwidth]{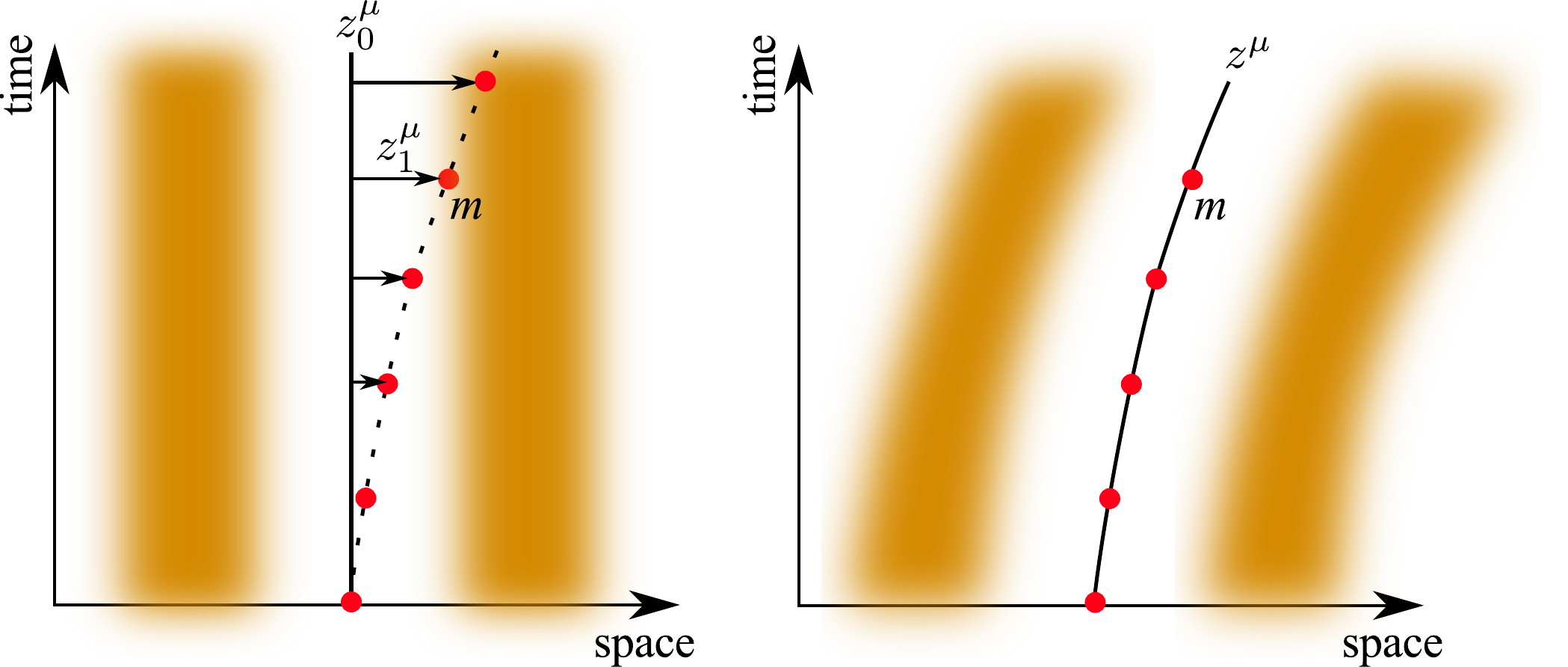}
\caption{\label{Fig:GW-vs-SC}Perturbative motion of a small object as defined in the Gralla-Wald (left) and self-consistent (right) approximations. In the Gralla-Wald case, the buffer region is centered on a zeroth-order worldline $\gamma_0\subset\man_0$, which in a given coordinate system $x^\mu:\man_0\to\mathbb{R}^4$ has coordinates $z^\mu_0( s)$. In a generic gauge, the self-force drives the object (shown as a sequence of red circles) away from $\gamma_0$, and this deviation is represented by $z_1^\mu( s)$, a vector defined on $\gamma_0$. In the self-consistent case, the buffer region is centered on an accelerated worldline $\gamma_\e\subset\man_0$, which in the same coordinates $x^\mu$ has coordinates $z^\mu( s,\epsilon)$. The accelerated worldline faithfully tracks the object's motion, such that the object is always at the ``center'' of the region enclosed by the buffer. On timescales much smaller than the dephasing time, we have that the two approximations are related by $z^\mu( s,\epsilon) = z^\mu_0( s)+\e z^\mu_1( s)+\O(\e^2)$.}
\end{center}
\end{figure}

Now consider the Gralla-Wald case. Here $\gamma_0$ is the worldline around which the inner expansion is performed. In a generic gauge (in particular, in the Lorenz gauge I work in), the small object's center of mass deviates from this worldline, as assumed in the expansion~\eqref{z_expansion}. Gauges in which the object does not deviate from $\gamma_0$ (on short timescales) are discussed in Sec.~\ref{gauge and motion}, but in a generic gauge, clearly we do not have $M^i=0$ in coordinates centered on $\gamma_0$. Instead, we calculate $M^i$ in those coordinates and then define the first-order correction to the motion to be 
\beq
z^i_1\equiv M^i/m.
\eeq
Figure \ref{Fig:GW-vs-SC} illustrates the difference between this setup and the self-consistent one.

			\section{Algorithm for an $n$th-order self-consistent approximation: general solution in the buffer region}\label{algorithm}


We are now positioned to actually obtain an outer expansion. In this section, I present an algorithm for finding the outer expansion in the buffer region. In Sec.~\ref{puncture}, I describe an algorithm for obtaining a global solution using as input the results from the buffer region. In both cases, I specialize to the self-consistent case.

My method of finding the general solution in the buffer region is modeled on the post-Minkowskian methods of Blanchet and Damour~\cite{Blanchet-Damour:86}. Like them, I write the general solution in terms of a set of \emph{algorithmic multipole moments}, which are defined simply from the algorithm of solving the relaxed Einstein equation, but a subset of which can be concretely identified with the physical multipole moments of $\gin_{\mu\nu}$. All the moments are completely unconstrained so long as we solve only the relaxed Einstein equations~\eqref{SC_EFEs}. Once we impose the gauge condition, as in Eqs.~\eqref{gauge_conditions_SC}, the moments become constrained, and in particular, evolution equations arise for the mass monopole, mass dipole, and spin dipole moments. The equation for the mass dipole moment will be used to identify the center-of-mass worldline $\gamma$ for which $M^i=0$.

The solution thus obtained will be a general solution in the sense that it is made to satisfy the matching condition (BC2) but not any particular global boundary conditions (BC1). (Refer back to the opening portion of Sec.~\ref{SC_GW_osculating}.) Because the explicit calculations, as well as their results, are exceedingly lengthy, I merely sketch the algorithm and the form of the results. I refer the reader to Refs.~\cite{Pound:10a,Poisson-Pound-Vega:11,Pound:12b,Pound-Miller:14} for more detailed expositions.

\subsection{Setup}
For concreteness, I work in Fermi-Walker coordinates $(t,x^a)$ centered on $\gamma$, in which $t$ is proper time on $\gamma$, 
$r=\sqrt{\delta_{ij}x^ix^j}$ is the proper distance from $\gamma$ along a spatial geodesic $\beta$ that intersects $\gamma$ perpendicularly, $n^i$ is a unit radial vector that labels the direction along which $\beta$ is sent out, and the spatial coordinate is $x^i=rn^i$. Ref.~\cite{Poisson-Pound-Vega:11} contains a pedagogical introduction.

The background metric in these coordinates is given by
\begin{subequations}\label{g-Fermi}%
\begin{align}
g_{tt} &= -(1+a_ix^i)^2-R_{0i0j}x^ix^j+\O(r^3),\\
g_{ta} &= -\tfrac{2}{3}R_{0iaj}x^ix^j+\O(r^3),\\
g_{ab} &= \delta_{ab}-\tfrac{1}{3}R_{aibj}x^ix^j+\O(r^3),
\end{align}
\end{subequations}
where the Riemann terms are evaluated on the worldline and contracted with members of a tetrad $(u^\alpha,e^\alpha_a)$ on $\gamma$ that satisfies $\frac{\partial y^\mu}{\partial x^i}=e^\mu_i$ for any coordinates $y^\mu$. 
For example, $R_{0iaj}(t)\equiv R_{\alpha\mu\beta\nu}(z(t)) u^\alpha e^\mu_ie^\beta_ae^\nu_j$. An overdot will indicate a covariant derivative along the worldline, such as $\dot R_{0iaj}\equiv R_{\alpha\mu\beta\nu;\rho}\big|_\gamma u^\alpha e^\mu_ie^\beta_ae^\nu_ju^\rho$. Relating back to the discussion below Eqs.~\eqref{TH-pdot}--\eqref{TH-Sdot}, the Riemann tensor on $\gamma$ can be written in terms of two tidal moments:
\beq\label{tides}
\E_{ab} \equiv R_{0a0b}, \qquad \B_{ab} \equiv -\frac{1}{2}\epsilon^{pq}{}_{(a}R_{b)0pq},
\eeq
and $R_{abcd}=\E_{ac}\delta_{bd}+\E_{bd}\delta_{ac}+\E_{ad}\delta_{bc}+\E_{bc}\delta_{ad}$. 

Reference~\cite{Pound:12b} displays the metric~\eqref{g-Fermi} to higher order in $r$. At all orders in $r$, all terms are made from the acceleration, Riemann tensor, and derivatives of the Riemann tensor. Because the coordinates are tethered to an $\e$-dependent worldline, the background picks up an $\e$ dependence, seen most obviously in the acceleration terms $a^i$. However, this is purely a coordinate effect. In any global coordinates covering $\D_{\gamma,\varsigma(\e)}$, $g_{\mu\nu}$ is $\e$-independent by definition. In a Gralla-Wald expansion, the $\e$-dependent coordinate transformation to Fermi-Walker coordinates would be expanded in powers of $\e$, splitting it into an $\e$-independent coordinate transformation (to Fermi coordinates centered on $\gamma_0$) plus a gauge transformation. Because I wish to at no point expand $z^\mu$, I do not expand the transformation in this way. 


In these coordinates, I expand $h_{\mu\nu}^{n}$ as in Eq.~\eqref{most_singular_part}. I further assume that the coefficients in that expansion are smooth, and I make logarithms explicit; these logarithms appear generically in solutions to inhomogeneous hyperbolic equations.\footnote{Intuitively, the logarithms are caused by the object perturbing the spacetime's light cones. One can expect the solution to the exact Einstein equation to propagate on (and within) null cones of the exact spacetime, and given that the mass of the body induces a logarithmic correction to the retarded time, logarithmic corrections then naturally appear in $h_{\mu\nu}^n$. This effect is well known from solutions to the Einstein equation in harmonic coordinates (see, e.g., Refs.~\cite{Blanchet-Damour:86,Blanchet:87}). For generality, I allow logarithms at any value of $n$, but I assume that for each finite $n$, $p$, and $\ell$, the highest power of $\ln r$ is a finite number $\qmax(n,p,\ell)$. For simplicity, to make sure that term-by-term differentiation is valid without worrying about issues of convergence, I also assume for a given, finite $n$ and $p$, $\ell$ has a maximum $\lmax(n,p)$.} We then have an expansion of the form
\begin{equation}
h^n_{\mu\nu}(t,x^a;z)=\sum_{p\geq-n}\sum_{\ell=0}^\lmax \sum_{q=0}^{\qmax}r^p(\ln r)^qh_{\mu\nu L}^{npq\ell}(t;z)\nhat^L, \label{hn_form}
\end{equation}
where $L=i_1\cdots i_\ell$ is a multi-index, $h_{\mu\nu L}^{npq\ell}$ is a smooth function of $t$, and $\nhat^L\equiv n^{\langle i_1}\cdots n^{i_\ell\rangle}$ is an STF combination of unit vectors. The decomposition in terms of $\nhat^L$ is equivalent to an expansion in spherical harmonics, and like a spherical harmonic, $\nhat^L$ is an eigenfunction of the flat-space Laplacian, satisfying $\partial^i\partial_i \nhat^L=-\frac{\ell(\ell+1)}{r^2}\nhat^L$. References~\cite{Blanchet-Damour:86,Damour-Iyer:91} contain excellent introductions to this type of decomposition, along with many useful identities. Favoring $n^a$ over angles $(\theta,\phi)$, and $\nhat^L$ over spherical harmonics, is useful because $g_{\mu\nu}$ is naturally written, as above, in terms of $x^a=r n^a$.

\subsection{Seed solutions}\label{seeds}

By definition, $h^n_{\mu\nu}$ is the solution to the $n$th-order relaxed Einstein equation, which has the form $E_{\mu\nu}[h^n]=S^n_{\mu\nu}[h^1,\ldots,h^{n-1}]$. Before considering the general solution to this equation, let us first consider solutions to the corresponding homogeneous equation $E_{\mu
\nu}[h^n]=0$. We shall find that every solution of the form~\eqref{hn_form} is a member of a set $\{h^\seed_{\mu\nu}(x;z,I^n_\ell):0\leq\ell\leq n\}$ or of a set $\{h^\free_{\mu\nu}(x;z,k^n_\ell):\ell\geq0\}$, or it is a superposition of such members. For a given background metric, the \emph{seed solutions} $h^\seed_{\mu\nu}(x;z,I^n_\ell)$ and the \emph{free solutions} $h^\free_{\mu\nu}(x;z,k^n_\ell)$ are completely determined by the worldline $z^\mu$ and functions of time $I^n_{\mu\nu L}(t)$ or $k^n_{\mu\nu L}(t)$ along that worldline. The quantities $I^n_{\mu\nu L}$ will be identified as \emph{algorithmic multipole moments} that describe the object's multipole structure, while each of the quantities $k^n_{\mu\nu L}$ will identify an effectively external, freely propagating perturbation.

We arrive at these conclusions simply by substituting the expansion~\eqref{hn_form} into $E_{\mu\nu}[h^n]=0$. First note that spatial derivatives reduce the order in $r$ by one power, while temporal derivatives do not affect the order in $r$. This means we can write $E_{\mu\nu}[h^n]=\partial^i\partial_i h^n_{\mu\nu}+W_{\mu\nu}[h^n]$, where $W_{\mu\nu}[h^n]\sim h^n_{\mu\nu}/r$. Solving for $h^{npq\ell}_{\mu\nu}$ order by order in $r$ is hence reduced to solving a sequence of Poisson equations $\partial^i\partial_i h^n_{\mu\nu} = -W_{\mu\nu}[h^n]$. Substituting the expansion~\eqref{hn_form} into this, we find that the coefficient of $r^{p-2}(\ln r)^q \nhat^L$ in the equation is
\begin{align}
[p(p+1)&-\ell(\ell+1)]h_{\mu\nu L}^{npq\ell} +(q+1)(2p+1)h_{\mu\nu L}^{n,p,q+1,\ell}+(q+1)(q+2)h_{\mu\nu L}^{n,p,q+2,\ell}\nonumber\\
 &= \left\{\sum_{p'=-n}^{p-1}\sum_{\ell'=0}^{\lmax}\sum_{q'=0}^{\qmax}W_{\mu\nu}\left[r^{p'}(\ln r)^{q'}\nhat^{L'} h_{\mu\nu L'}^{np'q'\ell'}\right]\right\}^{p-2,q,\ell} \label{nth_Poisson_expanded},
\end{align}
where $\{\cdot\}^{pq\ell}$ means ``pick off the coefficient of $r^{p-2}(\ln r)^q \nhat^L$''. The important thing to note is that the right-hand side depends only on coefficients with $p'<p$. So start with the term with the lowest power of $r$, $1/r^{n+2}$, such that the right-hand side of  Eq.~\eqref{nth_Poisson_expanded} vanishes. Further, start with $q=\qmax$, and suppose $\qmax>0$. By assumption, $h_{\mu\nu L}^{n,-n,\qmax+1,\ell}$ and $h_{\mu\nu L}^{n,-n,\qmax+1,\ell}$ vanish, and so Eq.~\eqref{nth_Poisson_expanded} becomes $[n(n-1)-\ell(\ell+1)]h_{\mu\nu L}^{n,-n,\qmax,\ell}=0$. This has a nontrivial solution only if $\ell={n-1}$. Now look at the equation for the same $\ell$ and $p$ but for $q=\qmax-1$. It reads  $[n(n-1)-\ell(\ell+1)]h_{\mu\nu L}^{n,-n,\qmax-1,\ell}+\qmax(1-2n)h_{\mu\nu L}^{n,-n,\qmax,\ell}=0$. But the first term vanishes because $\ell=n-1$. Hence,  $h_{\mu\nu N-1}^{n,-n,\qmax,n-1}=0$ if $\qmax>0$, and we conclude that $h_{\mu\nu N-1}^{n,-n,0,n-1}\nhat^{N-1}/r^n$ is the only nontrivial solution to the $1/r^{n+2}$ term in the homogeneous equaion $E_{\mu\nu}[h^n] = 0$. I define the algorithmic moment
\beq
I^n_{\mu\nu L} \equiv h^{n,-n,0,n-1}_{\mu\nu L}.
\eeq
Now proceed to sequentially higher orders in $r$, and at every order, set to zero all free functions that arise. The result is a solution to $E_{\mu\nu}[h^n] = 0$ of the form
\beq
h^\seed_{\mu\nu}(x;z,I^n_{n-1}) = \frac{I^n_{\mu\nu N-1}(t)\nhat^{N-1}}{r^n}+\O(1/r^{n-1}) 
\eeq
in which every single term is linear in $I^n_{\mu\nu N-1}(t)$ (and $t$-derivatives of it), specifically consisting of $I^n_{\mu\nu N-1}(t)$ (and its derivatives) contracted with the other available Cartesian tensors $n^i$, $a^i$, $\delta^{ij}$, $\epsilon^{ijk}$, pieces of the background Riemann tensor, and pieces of derivatives of the background Riemann tensor.

Now set $I^n_{\mu\nu N-1}$ to zero and solve the $1/r^{n+1}$ and higher-order terms in Eq.~\eqref{nth_Poisson_expanded}, following exactly the same steps as above. Again the result is a homogeneous solution that is completely determined by a single function, in this case $I^n_{\mu\nu N-2}\equiv h^{n,-n+1,0,n}_{\mu\nu N-2}$. Now keep doing the same by beginning at each following order up to (and including) $1/r^3$, always setting every free function to zero except the first one encountered. This procedure completely populates the set $\{h^\seed_{\mu\nu}(x;z,I^n_\ell):0\leq\ell\leq n\}$ with solutions of the form
\beq
h^\seed_{\mu\nu}(x;z,I^n_\ell) = \frac{I^n_{\mu\nu L}(t)\nhat^{L}}{r^{\ell+1}}+\O(1/r^{\ell}), 
\eeq
where every term in the solution is proportional to $I^n_{\mu\nu L}(t)\equiv h^{n,-\ell-1,0,\ell}_{\mu\nu L}(t)$ or its derivatives. The algorithmic multipole moments $I^n_{\mu\nu L}$ are symmetric in their first two indices and STF in their last $\ell$ indices.

Continuing the same procedure to higher orders in $r$, we completely populate the other set, $\{h^\free_{\mu\nu}(x;z,I^n_\ell):\ell\geq 0\}$, with solutions
\beq
h^\free_{\mu\nu}(x;z,k^n_\ell) = r^\ell k^n_{\mu\nu L}(t)\nhat^{L} + \O(r^{\ell+1}), 
\eeq 
where every term in the solution is proportional to $k^n_{\mu\nu L}(t)\equiv h^{n,\ell,0,\ell}_{\mu\nu L}(t) $ or its derivatives. The change in behavior from $1/r^{\ell+1}$ to $r^\ell$ arises because the factor $[p(p+1)-\ell(\ell+1)]$ in Eq.~\eqref{nth_Poisson_expanded} vanishes for $p=-\ell-1$ if $p<0$ and for $p=\ell$ if $p\geq0$. As with $I^n_{\mu\nu L}$, the quantities $k^n_{\mu\nu L}$ are symmetric in their first two indices and STF in their last $\ell$ indices.

It is easy to see from the above steps that the set of seed solutions $h^\seed_{\mu\nu}(x;z,I^n_\ell)$ and free solutions $h^\free_{\mu\nu}(x;z,k^n_\ell)$ form a complete basis of solutions to $E_{\mu\nu}[h^n]=0$ of the form~\eqref{hn_form}. If extended down to $r=0$, each seed solution diverges at $r=0$, while each free solution is smooth there.

For later purposes, it will be convenient to split each algorithmic multipole moment into a mass and current moment,
\begin{equation}
I_{\mu\nu L}^n = M_{\mu\nu L}^n+S_{\mu\nu L}^n,
\end{equation}
where the mass moment $M_{\mu\nu L}^n$ is the even-parity part of $I_{\mu\nu L}^n$, satisfying $M_{\mu j L}^n=M_{\mu(j i_1)i_2\cdots i_\ell}^n$, and the current moment $S_{\mu\nu L}^n$ is the odd-parity part, satisfying $S_{\mu jL}^n=S_{\mu[ji_1]i_2\cdots i_\ell}^n$. This splits each seed solution into two, $h^\seed_{\mu\nu}(x;z,M^n_\ell)$ and $h^\seed_{\mu\nu}(x;z,S^n_\ell)$.

\subsection{General solution}
Now turn to the inhomogeneous equation $E_{\mu\nu}[h^n]=S^n_{\mu\nu}[h^1,\ldots,h^{n-1}]$. Start at the $n=1$ equation, \eqref{SC_EFE1}. The source vanishes, and the general solution is made up of a single seed solution,
\beq
h^\seed_{\mu\nu}(x;z,I^1_0)=\frac{I^1_{\mu\nu}(t)}{r}+\O(r^0),
\eeq
plus the sum of all free solutions $h^\free_{\mu\nu}(x;z,k^1_\ell)$. That is,
\beq
h^1_{\mu\nu} = h^\seed_{\mu\nu}(x;z,I^1_0)+\sum_{\ell\geq0}h^\free_{\mu\nu}(x;z,k^1_\ell).
\eeq

Next move to the $n=2$ equation,~\eqref{SC_EFE2}. The $1/r^{p+2}$ term in the equation looks essentially the same as Eq.~\eqref{nth_Poisson_expanded}, except that the right-hand side contains a term coming from $S^2_{\mu\nu}[h^1,h^1]$. One can straightforwardly solve for the functions $h^{2pq\ell}_{\mu\nu L}$ to find that each one of them is either the starting point for a new seed or free solution ($h^{2,-2,0,1}_{\mu\nu i}$, $h^{2,-1,0,0}_{\mu\nu }$, or $h^{2,\ell,0,\ell}_{\mu\nu L}$ for $\ell\geq0$) or it is directly proportional to a $pq\ell$ mode of the source. Since the source is constructed from quadratic combinations of $h^1_{\mu\nu}$, it follows that the general solution is made up of the two seed solutions
\begin{align}
h^\seed_{\mu\nu}(x;z,I^2_1) &= \frac{I^2_{\mu\nu i}(t)n^i}{r^2}+\O(1/r),\\
h^\seed_{\mu\nu}(x;z,I^2_0) &= \frac{I^2_{\mu\nu}(t)}{r}+\O(r^0),
\end{align}
plus the sum of all free solutions $h^\free_{\mu\nu}(x;z,k^2_\ell)$, plus a particular inhomogeneous solution made up entirely of terms that are quadratic combinations of members of the set $\{I^1_{\mu\nu},k^1_{\mu\nu L}:\ell\geq0\}$ (in other words, a particular solution in which all seed solutions and free solutions are set to zero). That is,
\begin{align}
h^2_{\mu\nu} &= h^\seed_{\mu\nu}(x;z,I^2_1)+h^\seed_{\mu\nu}(x;z,I^2_0)+\sum_{\ell\geq0}h^\free_{\mu\nu}(x;z,k^2_\ell) \nonumber\\
&\quad+ h^{2IH}_{\mu\nu}(x;z,I^1_0,\{k^1_\ell\}_{\ell\geq0}).\label{h2-wM}
\end{align}
The promised logarithms of $r$ first appear in $h^{2IH}_{\mu\nu}(x;z,I^1_0,\{k^1_\ell\}_{\ell\geq0})$~\cite{Pound:10a,Pound:12b}.

Carrying this procedure to $n$th order, we find
\begin{align}
h^n_{\mu\nu} &= \sum_{\ell=0}^{n-1} h^\seed_{\mu\nu}(x;z,I^n_\ell)+\sum_{\ell\geq0}h^\free_{\mu\nu}(x;z,k^n_\ell) \nonumber\\
&\quad+ h^{nIH}_{\mu\nu}(x;z,\{I^m_\ell\}_{m<n,\ell<n},\{k^m_\ell\}_{m<n,\ell\geq0}),\label{nth_general}
\end{align}
where all functions $h^{npq\ell}_{\mu\nu L}$ appearing in $h^{nIH}_{\mu\nu}$ are made up of explicit products of members of the set $\{I^m_{\mu\nu L}:m<n,\ell<n-1\}\cup\{k^m_{\mu\nu L}:m<n,\ell\geq0\}$ (or their $t$ derivatives) contracted with combinations of $n^i$, $a^i$, $\delta^{ij}$, $\epsilon^{ijk}$, pieces of the background Riemann tensor, and pieces of derivatives of the background Riemann tensor. 

\emph{In any particular solution of the form~\eqref{hn_form} to the relaxed Einstein equation outside a small object, every term, to all orders in $r$ and $\e$, is a linear or nonlinear combination of the functions $I^n_{\mu\nu L}(t)=h^{n,-\ell-1,0,\ell}_{\mu\nu L}(t)$ and $k^n_{\mu\nu L}(t)=h^{n,\ell,0,\ell}_{\mu\nu L}(t)$.} Specifying all the functions $I^n_{\mu\nu L}(t)$ and $k^n_{\mu\nu L}(t)$ corresponds to making a specific choice of small object and of global boundary conditions.

\subsection{Imposing matching and gauge conditions}\label{imposing_gauge}
To satisfy the unrelaxed Einstein equation, the general solution described above must also satisfy the gauge conditions~\eqref{gauge_conditions_SC}, which we can write generically as 
\beq
L_\mu[h^n,F_0]=P_\mu[\{h^m,F_m\}_{m=1}^{n-1}],
\eeq
for some $P_\mu$ that is linear in the $h$'s but not usually in the $F$'s. Recasting this as a constraint on the seed and free solutions in terms of lower-order fields, we have
\begin{align}
\sum_{\ell=0}^{n-1} L_\mu[h^\seed&[I^n_\ell],F_0]+\sum_{\ell\geq0} L_\mu[h^\free[k^n_\ell] ,F_0] \nonumber\\
&= P_\mu[\{h^m,F_m\}_{m=1}^{n-1}]-L_\mu[h^{nIH}].\label{nth_gauge}
\end{align}
Like the relaxed Einstein equation, Eq.~\eqref{nth_gauge} can be solved order by order in $r$, thereby constraining the functions $I^n_{\mu\nu L}$ and $k^n_{\mu\nu L}$, which in the general solution to the relaxed Einstein equation were entirely arbitrary. Working through sequential orders in $r$ first constrains each $I^n_{\mu\nu L}$ in sequence, from $\ell=n-1$ to $\ell=0$, and then each $k^n_{\mu\nu L}$ in sequence from $\ell=0$ to $\ell=\infty$. The most important outcome is evolution equations for the moments $I^n_{\mu\nu L}(t)$.


At $n=1$, the term with the lowest power of $r$ in Eq.~\eqref{SC_gauge1}, $1/r^2$, is $\partial^b(h^\seed_{\alpha b}[I^1_0]-\tfrac{1}{2}\eta_{\alpha b}\eta^{\mu\nu}h_{\mu\nu}^\seed[I^1_0])=\O(1/r)$, which determines $I^1_{\mu\nu}=2m(t)\delta_{\mu\nu}$. The overall factor is written as $2m(t)$ for good reason: from Eq.~\eqref{gin_buffer}, the inner background metric has an expansion $\gin_{\mu\nu}=\frac{I^1_{\mu\nu}}{r}+\O(1/r^2)$, and substituting that expansion into the formula for the Arnowitt-Deser-Misner (ADM) mass, one finds that $m(t)$ is precisely the ADM mass of  $\gin_{\mu\nu}$. We call this the leading-order mass of the object. Proceeding to the next order in $r$, $1/r$, time derivatives and acceleration terms appear, and we find our sought evolution equations
\beq
\frac{dm}{dt} =0, \qquad F_0^\mu =0,
\eeq
which tell us that the object behaves approximately as a test body in $g_{\mu\nu}$. The remainder of the content of the $n=1$ gauge condition, at all orders in $r$, is to enforce various uninteresting relationships between the functions $k^n_{\mu\nu L}$.

At $n=2$, the lowest order in $r$ in the gauge condition~\eqref{SC_gauge2}, $1/r^3$, similarly determines (after a very slight gauge refinement that remains within the Lorenz gauge~\cite{Pound:10a}) that $I^2_{\mu\nu i}$ divides into mass and current moments of the form
\beq\label{dipole-moments}
M^2_{\mu\nu i} = 2M_i(t)\delta_{\mu\nu}, \qquad S^2_{\mu\nu i} = 4u_{(\mu}\epsilon_{\nu)ji}S^j(t)
\eeq
where $\epsilon_{tij}\equiv 0$. $M_i$ and $S^i$ are, respectively, the mass dipole moment and ADM angular momentum of $\gin_{\mu\nu}$; we call these the object's leading-order mass dipole moment relative to $\gamma$ and leading-order spin about $\gamma$. Defining $\delta m_{\mu\nu}(t)\equiv I^2_{\mu\nu}(t)$, we also find (from the order-$1/r^2$ and $1/r$ terms in the gauge condition) that
\begin{align}
\delta m_{\alpha\beta} &= \tfrac{1}{3}m\left(2k^1_{\alpha\beta}+g_{\alpha\beta}g^{\mu\nu}k_{\mu\nu}^1\right) +m(g_{\alpha\beta}+2u_{\alpha} u_{\beta})u^\mu u^\nu k^1_{\mu\nu}\nonumber\\
&\quad +4u_{(\alpha}(mk^1_{\beta)\mu}u^\mu+2\dot M_{\beta)}),\label{dm_free}
\end{align}
where $\dot M^{\beta}\equiv \frac{DM^\beta}{d\tau}$. $\delta m_{\alpha\beta}$ can be thought of loosely as a correction to the object's monopole moment, but it is not invariant; a smooth gauge perturbation $2\xi_{(\alpha;\beta)}$ that is nonvanishing on $\gamma$ will add directly to $k^1_{\alpha\beta}$. This is why the moments are called algorithmic: they arise in the algorithm of solving the field equations, but they are not invariant descriptions of the perturbed object. Furthermore, $\delta m_{\alpha\beta}$ is determined only up to a constant $c\delta_{\alpha\beta}$, but I choose to incorporate that term into $m\delta_{\alpha\beta}$.

Finally, the $1/r^2$ term in the $n=2$ gauge condition determines the evolution equation $\frac{dS^{i}}{dt}=0$, which tells us that the leading-order spin is parallelly propagated along $\gamma$, and the $1/r$ term determines 
\beq
\frac{d^2M_a}{dt^2} = - R_{a0b0}M^b - mF^1_a - m\frac{dk^1_{ta}}{dt}+\tfrac{1}{2}mk^1_{tta} +\tfrac{1}{2}R_{0aij}S^{ij},\label{master_equation}
\eeq
where I have defined $S^{ij}=\epsilon^{ijk}S_k$. I call this the \emph{master equation of motion}. It describes how the small object moves relative to any choice of nearby worldline $\gamma$. That motion is influenced by (i) geodesic deviation (evinced by the first term on the right), (ii) the acceleration of $\gamma$ (manifesting as $F^1_a$), (iii) the ambient free fields in the neighbourhood, and (iv) the Mathisson-Papapetrou force created by coupling of the spin to the curvature of the external background metric. In the self-consistent expansion, we choose $\gamma$ such that $M^i=0$. Therefore,
\beq
F^1_a = -\frac{dk^1_{ta}}{dt}+\tfrac{1}{2}k^1_{tta} +\tfrac{1}{2m}R_{0aij}S^{ij}.\label{F1_free}
\eeq
This is the MiSaTaQuWa force (albeit in an atypical form) plus the Mathisson-Papapetrou force. We see more clearly now that the practical consequence of the self-consistent approach, by utilizing a worldline relative to which $M^i$ vanishes, is to prevent the term $~M_in^i/r^2$ from appearing in the second-order field~\eqref{h2-wM}. According to Eq.~\eqref{master_equation}, this term would otherwise grow quadratically with time; this is the same undesirable growth discussed in Sec.~\ref{long_term_dynamics}.

As at $n=1$, at $n=2$ the gauge condition imposes relationships between the functions $k^2_{\mu\nu L}$, these relationships do not involve any of the algorithmic moments, and they leave each $k^n_{\mu\nu L}$ written in terms of arbitrary functions of time. 

Extrapolating to arbitrary order $n$, we draw several conclusions: (i) The moment $I^n_{\mu\nu N-1}$ appearing in the leading order term $h^{n,-n}_{\mu\nu}/r^n$ is directly identified with a moment of $\gin_{\mu\nu}$. Explicitly, as the mass appears at order $\e$ and the spin at order $\e^2$, so the quadrupole moment will appear at order $\e^3$, the octupole at order $\e^4$, etc.\footnote{To relate back to the Thorne and Hartle results~\eqref{TH-pdot}--\eqref{TH-Sdot}, canonical mass and current quadrupole moments $Q_{ij}$ and $\mathcal{Q}_{ij}$ can be defined from $M^3_{\mu\nu ij}$ and $S^3_{\mu\nu ij}$ according to
\beq
M^3_{\mu\nu ij} = 3Q_{ij}\delta_{\mu\nu}, \qquad S^3_{\mu\nu ij} = 8u_{(\mu}\epsilon_{\nu)ki}\mathcal{Q}^k{}_j,
\eeq
similarly to Eq.~\eqref{dipole-moments}. In choosing the normalization factors of these moments, I follow Ref.~\cite{Damour-Iyer:91}.} We interpret these moments $I^n_{\mu\nu N-1}$ as the leading-order moments of the object. (ii) The others, $I^n_{\mu\nu \ell}$ with $\ell<n-1$, we interpret as perturbations to the object's moments, and the gauge condition determines that each of these corrections involve some linear or nonlinear factors of moments of $\gin_{\mu\nu}$. (iii) Even after the gauge and matching conditions are imposed, every coefficient $h^{npq\ell}_{\mu\nu L}(t)$ appearing in the free fields is made up entirely of arbitrary functions of time that are determined only once global boundary conditions (BC1) are also imposed.\footnote{A proof of statement (iii) will be presented elsewhere. Here it can be taken as a conjecture, although it is known to be true at all orders in $\e$ and $r$ that have been explicitly considered. Intuitively, it can be inferred from the fact that we can choose boundary conditions for which all $k^n_{\mu\nu L}$ vanish at a given $n$, and with that choice we must still be able to satisfy the gauge condition; hence, the constraints on the $I^n_{\mu\nu L}$'s cannot involve the $k^n_{\mu\nu L}$'s of the same $n$.} 

When going to higher order, it is not obvious what condition should be imposed to make the worldline a good representation of the object's bulk motion. A natural candidate would be to set the corrections $M^{n>2}_{\mu\nu i}$ to the object's mass dipole moment to zero. However, one would have to investigate the third- and higher-order gauge conditions to see whether (i) doing so is possible and (ii) doing so prevents unwanted secular growth that can be associated with a growing displacement of the object away from the chosen representative worldline. Section~\ref{gauge and motion} offers a different method of choosing a good worldline, at least at second order.



\subsection{Split into self-field and effective field}\label{singular-regular}
I now define a convenient split of $h^n_{\mu\nu}$ into a self-field $h^{\S n}_{\mu\nu}$ and an effectively external field $h^{\R n}_{\mu\nu}$. Specifically, I define $h^{\R n}_{\mu\nu}$ to be the piece of Eq.~\eqref{nth_general} that contains no linear or nonlinear combinations of the algorithmic moments $I^n_{\mu\nu L}$:
\begin{align}
h^{\R n}_{\mu\nu} &= \sum_{\ell\geq0}h^\free_{\mu\nu}(x;z,k^n_\ell) + h^{\R nIH}_{\mu\nu}(x;z,\{k^m_\ell\}_{m<n,\ell\geq0}),\label{nth_R}
\end{align}
where $h^{\R nIH}_{\mu\nu}$ is the part of $h^{nIH}_{\mu\nu}$ containing no factors of $I^n_{\mu\nu L}$.  This implicitly defines the self-field $h^{\S n}_{\mu\nu}=h^n_{\mu\nu}-h^{\R n}_{\mu\nu}$ to be 
\begin{align}
h^{\S n}_{\mu\nu} &= \sum_{\ell=0}^{n-1} h^\seed_{\mu\nu}(x;z,I^n_\ell) + h^{\S nIH}_{\mu\nu}(x;z,\{I^m_\ell\}_{m<n,\ell<n},\{k^m_\ell\}_{m<n,\ell\geq0}),\label{nth_S}
\end{align}
where every term in $h^{\S nIH}_{\mu\nu}$ contains at least one factor of at least one algorithmic moment. I also define $h^\R_{\mu\nu}=\sum_n \e^n h^{\R n}_{\mu\nu}$ and $h^\S_{\mu\nu}=\sum_n \e^n h^{\S n}_{\mu\nu}$. Conclusion (iii) at the end of the last section implies that each of these two fields separately satisfies the Lorenz gauge condition.\footnote{If the statements were not true, one could always slightly alter the singular-regular split to make the two fields independently satisfy the gauge condition. Doing so would involve appropriately moving part of the free fields into $h^\S_{\mu\nu}$~\cite{Pound:12b}.}

More explicitly, at first order, $h^1_{\mu\nu}=h^{\S1}_{\mu\nu}+h^{\R1}_{\mu\nu}$, with
\begin{align}
h^{\S1}_{\mu\nu} &= h^\seed_{\mu\nu}(x;z,m) =\frac{2m\delta_{\mu\nu}}{r}+\O(r^0),\label{hS1}\\
h^{\R1}_{\mu\nu} &= \sum_{\ell\geq0}h^\free_{\mu\nu}(x;z,k^1_\ell)=k^1_{\mu\nu}(t)+rk^1_{\mu\nu i}(t) n^i+\O(r^2).\label{hR1}
\end{align}
Recall that for any particular solution of the form~\eqref{hn_form}, the functions in the free fields are $k^n_{\mu\nu L}=h^{n,\ell,0,\ell}_{\mu\nu L}$.

At second order, $h^2_{\mu\nu}=h^{\S2}_{\mu\nu}+h^{\R2}_{\mu\nu}$, with
\begin{align}
h^{\S2}_{\mu\nu} &= h^\seed_{\mu\nu}(x;z,S)+h^\seed_{\mu\nu}(x;z,\delta m) + h^{\S 2IH}_{\mu\nu}(x;z,m,\{k^1_\ell\}_{\ell\geq0}),\label{hS2}\\
h^{\R2}_{\mu\nu} &= \sum_{\ell\geq0}h^\free_{\mu\nu}(x;z,k^2_\ell) + h^{\R 2IH}_{\mu\nu}(x;z,\{k^1_\ell\}_{\ell\geq0}),\label{hR2}
\end{align}
where
\begin{align}
h^\seed_{\mu\nu}(x;z,S) &= \frac{4 S^jn^i}{r^2}u_{(\mu}\epsilon_{\nu)ji}+\O(1/r),\\
h^\seed_{\mu\nu}(x;z,\delta m) &= \frac{\delta m_{\mu\nu}}{r}+\O(r^0),\\
h^{\R2}_{\mu\nu}& = k^2_{\mu\nu}(t)+rk^2_{\mu\nu i}(t) n^i+\O(r^2),
\end{align}
$h^{\S 2IH}_{\mu\nu}(x;z,m,\{k^1_\ell\}_{\ell\geq0})$ is made up exclusively of terms quadratic in $m$ or products of $m$ and $h^{\R1}_{\mu\nu}$ (and derivatives of $h^{\R1}_{\mu\nu}$), and $h^{\R 2IH}_{\mu\nu}(x;z,\{k^1_\ell\}_{\ell\geq0})$ is made up of terms quadratic in $h^{\R1}_{\mu\nu}$ (and derivatives of $h^{\R1}_{\mu\nu}$). 

I remind the reader that in Sec.~\ref{Detweiler-Whiting} I identified three ``nice'' properties I wish an effectively external metric to satisfy: 
\begin{enumerate}
\item it should be a vacuum solution
\item it should be causal on the worldline
\item the object should move as a test body in it.
\end{enumerate}
There is no guarantee a priori that these conditions can simultaneously be satisfied at all orders in $\e$, but with the definitions given above, the first nice property is met: the effective metric $\tilde g_{\mu\nu}=g_{\mu\nu}+h^\R_{\mu\nu}$ is a smooth solution to the vacuum Einstein equation $R_{\mu\nu}[\tilde g]=0$ \emph{even at $r=0$}. This can be seen from the facts that (i) by construction, for $r>0$ we still have a solution even if we set all moments $I^n_{\mu\nu L}$ to zero (i.e., if there is no object in $\B_\gamma$), and (ii) for the free fields the construction applies even at $r=0$. Note that here I am taking a particular series~\eqref{nth_R} obtained outside the object and simply analytically extending it down to $r=0$. In this way, I define an effective spacetime $(\tilde g_{\mu\nu},\tilde\man)$ in $\B_\gamma$, where the object lies in the physical spacetime.

More explicitly, the first- and second-order regular fields satisfy the vacuum equations
\begin{align}
E_{\mu\nu}[h^{\R1}]&=0,\qquad\quad  L_\mu[h^{\R1},F_0]=0,\\
E_{\mu\nu}[h^{\R2}]&=2\delta^2R_{\mu\nu}[h^{\R1},h^{\R1}],\qquad L_{\mu}[h^{\R2},F_0]=-\delta L_\mu[h^{\R1},F_1],
\end{align}
for all $r\geq0$. The self-fields are left to satisfy 
\begin{align}
E_{\mu\nu}[h^{\S1}]&=0, \qquad L_\mu[h^{\S1},F_0]=0,\\
E_{\mu\nu}[h^{\S2}]&=2\delta^2R_{\mu\nu}[h^{\S1},h^{\S1}]+4\delta^2R_{\mu\nu}[h^{\S1},h^{\R1}],\\
L_{\mu}[h^{\S2},F_0] &=-\delta L_\mu[h^{\S1},F_1].
\end{align}

The third ``nice'' condition is also met, at least through order $\e$: combining Eqs.~\eqref{hR1} and \eqref{F1_free}, we get, in covariant form, 
\beq
\frac{D^2z^\mu}{d\tau^2} = -\frac{1}{2}P^{\mu\nu}(2h^{{\rm R1}}_{\rho\nu;\sigma}-h^{{\rm R1}}_{\rho\sigma;\nu})u^\rho u^\sigma 
											+ \frac{1}{2m}R^\mu{}_{\nu\rho\sigma}u^\nu S^{\rho\sigma}+\O(\e^2),\label{SC_eq_mot_wSpin}
\eeq
where I have defined the antisymmetric 4-tensor $S^{\mu\nu}=e^\mu_i e^\nu_j S^{ij}$.\footnote{The monopole correction~\eqref{dm_free} can also be trivially rewritten in terms of $h^{\R1}_{\mu\nu}$ as
\begin{align}
\delta m_{\alpha\beta} &= \tfrac{1}{3}m\left(2h^{\R1}_{\alpha\beta}+g_{\alpha\beta}g^{\mu\nu}h^{\R1}_{\mu\nu}\right) +m(g_{\alpha\beta}+2u_{\alpha} u_{\beta})u^\mu u^\nu h^{\R1}_{\mu\nu} +4u_{(\alpha}mh^{\R1}_{\beta)\mu}u^\mu,\label{dm_hR}
\end{align}
where I have set $M^i=0$, and all fields are evaluated on $\gamma.$}  Together with $\frac{dS^i}{dt}=0$, Eq.~\eqref{SC_eq_mot_wSpin} is the equation of motion of a test body in $\tilde g_{\mu\nu}$. It can be put in the form of the Thorne-Hartle equations~\eqref{TH-pdot}--\eqref{TH-Sdot} by using the results of Appendix~\ref{geodesic_expansion_in_h} to absorb the $h^{\R1}_{\mu\nu}$ terms into the covariant derivative and proper time in $\tilde g_{\mu\nu}$, using $p^\mu=m\tilde u^\mu+\O(\e^3)$ and the fact that $m$ is constant, and using Eq.~\eqref{tides} to write the Riemann tensor in terms of the tidal field $\B_{ij}$ (noting $\mathfrak{B}_{ij}=\B_{ij}+\O(\e)$). Doing so allows us to identify $\tilde g_{\mu\nu}$ with Thorne and Hartle's ``external'' metric $\varext_{\mu\nu}$, at least in the weak sense that $\tilde g_{\mu\nu}=\varext_{\mu\nu}+\O(\e^2,\e r^2)$; since only first derivatives of $h^{\R1}_{\mu\nu}$ appear in Eq.~\eqref{SC_eq_mot_wSpin}, we can make no claims on agreement at order $r^2$ or higher.

I show in the next section that my definition of the regular field also satisfies the remaining nice property: it is causal on the worldline. I also note that at least through order $r^2$, my definition of $h^{\R1}_{\mu\nu}$ agrees with the Detweiler-Whiting definition~\cite{Pound-Miller:14}, despite the fact that the Detweiler-Whiting field is defined in a wholly different manner based on Green's functions~\cite{Detweiler-Whiting:03,Poisson-Pound-Vega:11}. However, despite all these reassuring facts, I stress that in general, \emph{even if the effective metric I define is found to satisfy each of the three ``nice'' properties at a given order $\e^n$, it is not the unique field satisfying those properties}. One can simply shift any given free field $h^\free_{\mu\nu}(x;z,k^{n'}_{\ell})$ (with $n'\leq n$ and $\ell$ sufficiently large for $k^{n'}_{\mu\nu L}$ to not appear in the $n$th equation of motion), and its nonlinear combinations, from the regular field into the self-field. The strongest of my ``nice conditions'' appears to be that the object behaves as a test \emph{body} in the effective metric: since higher multipole moments couple to higher derivatives of the effective metric, this strongly constrains which part of the full metric should go into the effective one. At any finite perturbative order, however, one can always alter the effective metric's higher derivatives without spoiling any of its nice properties. Nonetheless, I deem my ``nice'' choice most natural as a part of the process of solving the relaxed Einstein equations using the local expansion~\eqref{hn_form}: before making reference to global boundary conditions, I simply put all the terms that involve the object's multipole moments into the self-field, and I put all the terms made up entirely of unknown functions into the effective field.


\subsection{Summary at first and second order}
To extract the key pragmatic information of the preceding sections, I restate the conclusions at first and second order: The first-order field is given by Eqs.~\eqref{hS1}--\eqref{hR1}, where the self-field $h^{\S1}_{\mu\nu}$ is locally determined by the (constant) mass $m$, and the effective field $h^{\R1}_{\mu\nu}$ is to be determined by global boundary conditions. The second-order field is given by Eqs.~\eqref{hS2}--\eqref{hR2}, where the self-field $h^{\S2}_{\mu\nu}$ is locally determined by (i) the mass $m$, (ii) the first-order effective field $h^{\R1}_{\mu\nu}$, (iii) the (parallely propagated) spin $S^\mu$, and (iv) the monopole correction $\delta m_{\mu\nu}$ given in Eq.~\eqref{dm_hR}; the effective field $h^{\R2}_{\mu\nu}$ is to be determined by global boundary conditions and by $h^{\R1}_{\mu\nu}$. I encourage readers to examine the more explicit expressions for these fields given in, e.g., Ref.~\cite{Pound-Miller:14}.

Finally, the object moves on a worldline governed by Eq.~\eqref{SC_eq_mot_wSpin}.

		\section{Algorithm for an $n$th-order self-consistent approximation: point particles, punctures, and global solutions}\label{puncture}


With the general solution in the buffer region ready at hand, I now describe how to use it to generate a global solution. In short, this relies on a rigorous procedure of replacing the physical field in $\B_\gamma$ with a fictitious field \emph{without altering the field elsewhere}. What are the fictitious fields? Simply the analytical expansion~\eqref{nth_general} from the buffer region, extended to apply to all $r>0$. This was done already for $h^\R_{\mu\nu}$ in the previous section. Continuing the expression for $h^{\rm S}_{\mu\nu}$ into $\B_\gamma$ makes it into a field that diverges at $r=0$: the true self-field in the interior of the body, whatever it may be, is replaced with this divergent field, and the self-field becomes the \emph{singular field}. Since this extension into $\B_\gamma$ does not affect the field in the buffer region, it also does not affect the field values outside the buffer, out in the external universe. 

Section \ref{skeleton} describes how this analytical extension can be used to ascribe a certain pointlike stress-energy distribution to the object, thereby recovering the point particle picture at linear order (though this stress-energy I construct cannot be taken as a physical source in the Einstein equations beyond linear order). Section~\ref{puncture_scheme} then describes how the analytical extension can be used to obtain a global solution at any order; this will be the first point at which a global boundary condition (BC1) is finally imposed.

\subsection{Skeletal stress-energy}\label{skeleton}

Consider the solution~\eqref{nth_general} extended down to $r>0$. For reasons described in Sec.~\ref{extended_bodies}, there is no known distributional source for this solution on a domain that includes $r=0$. However, we can devise the following setup: each of the seed solutions can be thought of as being sourced by a pointlike stress-energy supported on $\gamma$. Everything else in the general solution then grows from these seeds, either being generated by nonlinearities or in the case of the free fields, being determined by global boundary conditions.

I refer to the stress-energy for the seed solutions as the skeletal stress energy, taking after a similar phrase in Ref.~\cite{Mathisson:37}; the idea is that in some sense, the object (or body) can be replaced by a skeleton. That skeleton is made up of multipole moments living on the object's worldline.


For each seed solution $h^\seed_{\mu\nu}(x;z,I^n_\ell)$, I roughly follow the approach taken by Gralla and Wald at first order~\cite{Gralla-Wald:08}, defining the (trace-reversed) distributional stress-energy tensor 
\begin{equation}
\bar T_{\mu\nu}^{n\ell} \equiv -\frac{1}{16\pi}E_{\mu\nu}[h^\seed[I^n_\ell]].\label{Tp}
\end{equation}
The right-hand side can be written as   
\begin{align}\label{split}
E_{\mu\nu}[h^\seed[I^n_\ell]] = \partial^i\partial_i \frac{I^n_{\mu\nu L}\nhat^L}{r^{\ell+1}} +N_{\mu\nu}[h^\seed[I^n_\ell]],
\end{align}
where $N_{\mu\nu}[h^\seed[I^n_\ell]]=\partial^i\partial_i\left(h^\seed_{\mu\nu}[I^n_\ell]-I^n_{\mu\nu L}\nhat^L/r^{\ell+1}\right) +W_{\mu\nu}[h^\seed[I^n_\ell]]$.

First examine the most singular term. Using the identities $\partial_L r^{-1}=(-1)^\ell(2\ell-1)!!\frac{\nhat_L}{r^{\ell+1}}$ and $\partial^i\partial_i r^{-1}=-4\pi\delta^3(\vec x)$, we have 
\begin{align}
\partial^i\partial_i\frac{I_{\mu\nu L}^n\nhat^L}{r^{\ell+1}} &= \frac{4\pi(-1)^{\ell+1}I_{\mu\nu L}^n}{(2\ell-1)!!}\partial_L\delta^3(\vec x).\label{ddI}
\end{align}
Integrating the right-hand side against a test function $\psi^{\mu\nu}$, we find
\begin{align}
\int \psi^{\mu\nu}I_{\mu\nu L}^n\partial_L\delta^3(\vec x)\sqrt{-g}d^3xdt 
				&= (-1)^\ell\int_\gamma\partial_L\left(\sqrt{-g}\psi_{\mu\nu} I^{\mu\nu L}_n\right)dt\\
				&= (-1)^\ell\int_\gamma\left(\psi_{\mu\nu} I^{\mu\nu\alpha_1\cdots\alpha_\ell}_n\right)_{;\alpha_1\cdots\alpha_\ell}dt,\label{delta}
\end{align}
where in going from the second line to the third I have utilized the identity $\partial_i\sqrt{-g}=\Gamma^\beta_{\beta i}\sqrt{-g}$. Here I have defined $I^{\mu\nu \alpha_1\cdots\alpha_\ell}_n$ to be the tensor that agrees with $I^{\mu\nu L}_n$ if all $\alpha_i$ are spatial indices and zero otherwise, meaning that $I^{\mu\nu \alpha_1\cdots\alpha_\ell}_n$ is STF with respect to $g_{\mu\nu}$ and that $I^{\mu\nu \alpha_1\cdots\alpha_i\cdots\alpha_\ell}_n u_{\alpha_i}=0$. Equation~\eqref{delta} shows that
\begin{equation}
I^{\mu\nu L}_{(n)}\partial_L\delta^3(\vec x) = \int_\gamma I^{\mu\nu\alpha_1\cdots\alpha_\ell}_{(n)}\delta(x,z)_{;\alpha_1\cdots\alpha_\ell}d\tau,
\end{equation}

Equation~\eqref{split} now reads
\begin{align}
E_{\mu\nu}[h^\seed[I^n_\ell]]	&= 
		\frac{4\pi(-1)^{\ell+1}}{(2\ell-1)!!}\int_\gamma I^{\mu\nu \alpha_1\cdots\alpha_\ell}_{(n)}\delta(x,z)_{;\alpha_1\cdots\alpha_\ell}d\tau+N_{\mu\nu}.
\end{align}
Now note that, by construction, $N_{\mu\nu}[h^\seed[I^n_\ell]]$ vanishes pointwise for $r>0$. If it is nonvanishing as a distribution,\footnote{That it must be well defined as a distribution follows from it being the result of linear operations on $h_{\mu\nu}^\seed[I^n_\ell]$, and $h_{\mu\nu}^\seed[I^n_\ell]$ itself being a sum of terms constructed from linear operations on an integrable function. The latter fact follows from the first term in the sum being expressible as the linear operation $\partial_L$ on an integrable function proportional to $r^{-1}$ [as in the text above Eq.~\eqref{ddI}], and all higher-order terms in the sum being constructed from linear operations on lower order terms in the sum (as described in Sec.~\ref{seeds}).} it must have support only on $\gamma$, in which case it must be proportional to $\delta^3(x)$ or a derivative thereof; but from the calculation just performed, that would lead to a different algorithmic moment $I^n_{\mu\nu L'}$ (with $\ell'\neq\ell$) appearing in the solution, which would contradict the definition of the seed field $h_{\mu\nu}^\seed[I^n_\ell]$. So $N_{\mu\nu}[h^\seed[I^n_\ell]]$ vanishes as a distribution.

Therefore, Eq.~\eqref{Tp} becomes 
\begin{equation}\label{Tnl}
\bar T^{\mu\nu}_{n}(x;z,I^n_\ell)=\frac{(-1)^\ell}{4(2\ell-1)!!}\int_\gamma I^{\mu\nu \alpha_1\cdots\alpha_\ell}_n\delta(x,z)_{;\alpha_1\cdots\alpha_\ell}d\tau.
\end{equation}
If we add up all the multipole moments, we arrive at a skeletal stress-energy tensor
\begin{align}
\bar T^{\mu\nu} &= \sum_{n,\ell} \bar T^{\mu\nu}_{n}(x;z,I^n_\ell) = \sum_\ell\int_\gamma I^{\mu\nu\alpha_1\cdots\alpha_\ell}\delta(x,z)_{;\alpha_1\cdots\alpha_\ell}d\tau,\label{total_T}
\end{align}
where I have defined the corrected and normalized moments
$I^{\mu\nu \alpha_1\cdots\alpha_\ell} \equiv \sum_n\frac{(-1)^\ell}{4(2\ell-1)!!}I^{\mu\nu \alpha_1\cdots\alpha_\ell}_n$.
Notably, this skeletal stress-energy agrees in form with that of the traditional multipolar expansion of a material body's stress-energy tensor~\cite{Mathisson:37, Tulczyjew:59, Steinhoff-Puetzfeld:10}.


At first order in $\e$, there is only one seed field, $h^\seed_{\mu\nu}(x;z,m)$, and Eq.~\eqref{Tnl} gives
\beq
T^{\mu\nu}_1(x;z,m) = \int_\gamma m u^\mu u^\nu\delta(x,z)d\tau
\eeq
---precisely the point particle stress-energy used in Sec.~\ref{point_particle_picture}. Because $h^1_{\mu\nu}=h^{\R1}_{\mu\nu}+h^\seed_{\mu\nu}(x;z,m)$, and $E_{\mu\nu}[h^{\R1}]=0$, we can conclude that $E_{\mu\nu}[h^1]=T^{\mu\nu}_1(x;z)$. In other words, \emph{the first-order field is identical to one sourced by a point particle}. Of course, this is derived from the analytically extended field, but it also applies to the physical field at distances $r\gg\e$ from the worldline.

At second order, there are seed fields, $h^\seed_{\mu\nu}(x;z,\delta m)$, $h^\seed_{\mu\nu}(x;z,S)$, and $h^\seed_{\mu\nu}(x;z,M)$; although the last of these three we set to zero with our choice of $z^\mu$, it is worth displaying the skeletal stress-energy that would source it if we chose a different $z^\mu$. Equation~\eqref{Tnl} gives
\begin{align}
T^{\mu\nu}_{2}(x;z,\delta m) &= \frac{1}{4}\int_\gamma \overline{\delta m}^{\mu\nu}\delta(x,z)d\tau\label{Tdm},\\
T^{\mu\nu}_{2}(x;z,S) &= - \int_\gamma u^{(\mu}S^{\nu)\alpha}\nabla_{\alpha}\delta(x,z)d\tau,\\
T^{\mu\nu}_{2}(x;z,M) &= - \int_\gamma u^{\mu}u^\nu M^{\alpha}\nabla_{\alpha}\delta(x,z)d\tau,\label{TM}
\end{align}
where the overline indicates trace reversal. Appendix~\ref{T1_expansion} shows that $T^{\mu\nu}_{2}(x;z_0,M)$ [plus a piece of $T^{\mu\nu}_{2}(x;z_0,\delta m)$] is precisely equal to the linear term in the expansion of $T^{\mu\nu}_{1}(x;z,m)$ given $z^\mu=z_0^\mu+\e M^\mu/m+\O(\e)$.

\subsection{Puncture schemes}\label{puncture_scheme}
I now describe how one can use the results from the buffer region to obtain a global solution. The method is called a \emph{puncture scheme}. It has become standard in the linearized problem~\cite{Barack-Golbourn:07,Vega-Detweiler:07,Dolan-Barack:11,Vega-Wardell-Diener:11}, and it is the only known practical way to obtain numerical results at second order and higher~\cite{Detweiler:12,Pound:12a,Gralla:12,Pound:12b,Pound-Miller:14}.


A puncture scheme begins with the construction of a \emph{puncture} $h^\P_{\mu\nu}$, defined by truncating the local expansion of the singular field from the buffer region at some specified order. One then defines the residual field 
\begin{equation}
h^\res_{\mu\nu} \equiv h_{\mu\nu}-h^\P_{\mu\nu}
\end{equation}
and in a region covering the object, writes a field equation for $h^\res_{\mu\nu}$, rather than one for (the analytically continued) physical field $h_{\mu\nu}$. Since $h^\P_{\mu\nu}\approx h^\S_{\mu\nu}$, so too $h^\res_{\mu\nu}\approx h^\R_{\mu\nu}$. The better $h^\P_{\mu\nu}$ represents $h^\S_{\mu\nu}$, the better $h^\res_{\mu\nu}$ represents $h^R_{\mu\nu}$. For example, if $\lim_{x\to\gamma}[h^\P_{\mu\nu}(x)-h^\S_{\mu\nu}(x)]=0$, then $\lim_{x\to\gamma}h^\res_{\mu\nu}(x)=\lim_{x\to\gamma}h^\R_{\mu\nu}(x)$; that is, the residual field agrees with the regular field on the worldline. If $h^\P_{\mu\nu}$ agrees with $h^\S_{\mu\nu}$ to one order higher, meaning $h^\P_{\mu\nu}-h^\S_{\mu\nu}=o(r)$, then $\lim_{x\to\gamma}\nabla_{\!\rho} h^\res_{\mu\nu}=\lim_{x\to\gamma}\nabla_{\!\rho} h^\R_{\mu\nu}$; since the self-force is constructed from first derivatives of $h^\R_{\mu\nu}$, this condition guarantees that the force can be calculated from $h^\res_{\mu\nu}$, as in Eq.~\eqref{motion_SC} below.

There are several schemes that can be developed from the starting point of the puncture. Here I describe a worldtube scheme in the tradition of Refs.~\cite{Barack-Golbourn:07,Dolan-Barack:11}. In this type of scheme one uses the field variables $h^{\res n}_{\mu\nu}$ inside a worldtube $\Gamma$ surrounding $\gamma$, the field variables $h^n_{\mu\nu}$ outside that worldtube, and the change of variables $h^n_{\mu\nu}=h^{\res n}_{\mu\nu}+h^{\P n}_{\mu\nu}$ when moving between the two regions.\footnote{One does not solve the problem in each domain separately, since the separate problems would be ill-posed. Instead, when calculating $h^n_{\mu\nu}$ at a point just outside $\Gamma$ that depends on points on past time slices inside $\Gamma$, one makes use of the values of $h^{\res n}_{\mu\nu}$ already calculated at those earlier points, and vice versa; see Sec.~VB of Ref.~\cite{Barack-Golbourn:07}.} Concretely, a second-order puncture scheme is then summarized by the coupled system of equations\footnote{The effective sources $S^{{\rm eff}n}_{\mu\nu}$ are usually written to include the skeletal stress-energy terms, which are canceled by distributional content in $E_{\mu\nu}[h^{\P n}]$. Here I have instead followed Ref.~\cite{Gralla:12} by writing the source pointwise, off $\gamma$; if the puncture agrees with the singular field sufficiently well, the source at points on $\gamma$ can be defined as the limit from off $\gamma$.}
\begin{subequations}\label{h1_SC}%
\begin{align}
E_{\mu\nu}[h^{\res1}] &= -E_{\mu\nu}[h^{\P1}]\equiv S^{{\rm eff}1}_{\mu\nu} & \text{inside }\Gamma,\\
E_{\mu\nu}[h^{1}] &= 0 & \text{outside }\Gamma,
\end{align}
\vspace{-1.5\baselineskip}
\end{subequations}
\begin{subequations}\label{h2_SC}%
\begin{align}
E_{\mu\nu}[h^{\res2}] &= 2\delta^2R_{\mu\nu}[h^1,h^1] - E_{\mu\nu}[h^{\P2}]\equiv S^{{\rm eff}2}_{\mu\nu} & \text{inside }\Gamma,\\
E_{\mu\nu}[h^2] &= 2\delta^2R_{\mu\nu}[h^1,h^1] & \text{outside }\Gamma,
\end{align}
\end{subequations}
\vspace{-1.5\baselineskip}
\begin{align}
\frac{D^2 z^\mu}{d\tau^2} &= -\frac{1}{2}P^{\mu\nu}\left(g_\nu{}^\gamma-h^\res_\nu{}^\gamma\right)
		\left(2h^\res_{\gamma\alpha;\beta}-h^\res_{\alpha\beta;\gamma}\right)u^\alpha u^\beta,\label{motion_SC}
\end{align}
where the puncture diverges on the worldline $z^\mu$ determined by Eq.~\eqref{motion_SC}. Here I have looked ahead by using the second-order equation of motion (for an object whose spin and quadrupole moments vanish); the first-order equation~\eqref{Detweiler-Whiting-form1} could be used instead, though as discussed in Sec.~\ref{accuracy}, the results would be accurate in a smaller region of spacetime. 

In this scheme, Eqs.~\eqref{h1_SC}--\eqref{motion_SC} must be solved together, as a coupled system for the variables $z^\mu$, $h^{\res1}_{\mu\nu}\slash h^1_{\mu\nu}$ (inside/outside $\Gamma$), and $h^{\res2}_{\mu\nu}\slash h^2_{\mu\nu}$. The residual fields govern the position of the puncture, and the position of the puncture effectively sources the residual fields.  This system of equations is to be solved subject to some global boundary conditions. For simplicity, we can consider using specified initial data on a Cauchy surface.\footnote{Because the approximation is accurate only within a finite region of size $1/\e$, one might better solve the equations in the future domain of dependence of a partial Cauchy surface.} 

I make two techical asides: $h^{\P 2}_{\mu\nu}$, according to Eq.~\ref{hS2}, contains terms proportional to the first-order regular field $h^{\R1}_{\mu\nu}$. So at each timestep in a numerical evolution, one must first calculate the first-order residual field from the first-order puncture, then use that residual field to calculate the second-order effective source, and so forth. Also, the divergence of the individual sources is quite strong at the worldline, with the terms $2\delta^2R_{\mu\nu}[h^1,h^1]$ and $E_{\mu\nu}[h^{\P2}]$ in Eq.~\eqref{h2_SC} each going as $1/r^4$; by construction, these divergences analytically cancel each other, but the cancellation is numerically delicate.


In principle, there is no obstacle to using a puncture scheme at any order in $\e$. Outside $\Gamma$, one may solve the physical problem $E_{\mu\nu}[h^n] = S^n_{\mu\nu}[h^1,\ldots, h^{n-1}]$, inside $\Gamma$ one may solve the effective problem $E_{\mu\nu}[h^{\res n}] = S^n_{\mu\nu}[h^1,\ldots, h^{n-1}]-E_{\mu\nu}[h^{\P n}]\equiv S^{{\rm eff}n}_{\mu\nu}$, and when crossing $\Gamma$, one may change variables from the residual field to the full field via $h^n_{\mu\nu}=h^{\res n}_{\mu\nu}+h^{\P n}_{\mu\nu}$. Of course, this requires sufficiently high-order expressions for the puncture and for the equation of motion governing how the puncture moves. But we know how to obtain both expressions from a local analysis in the buffer region.

\begin{figure}[t]
\begin{center}
\includegraphics[width=\textwidth]{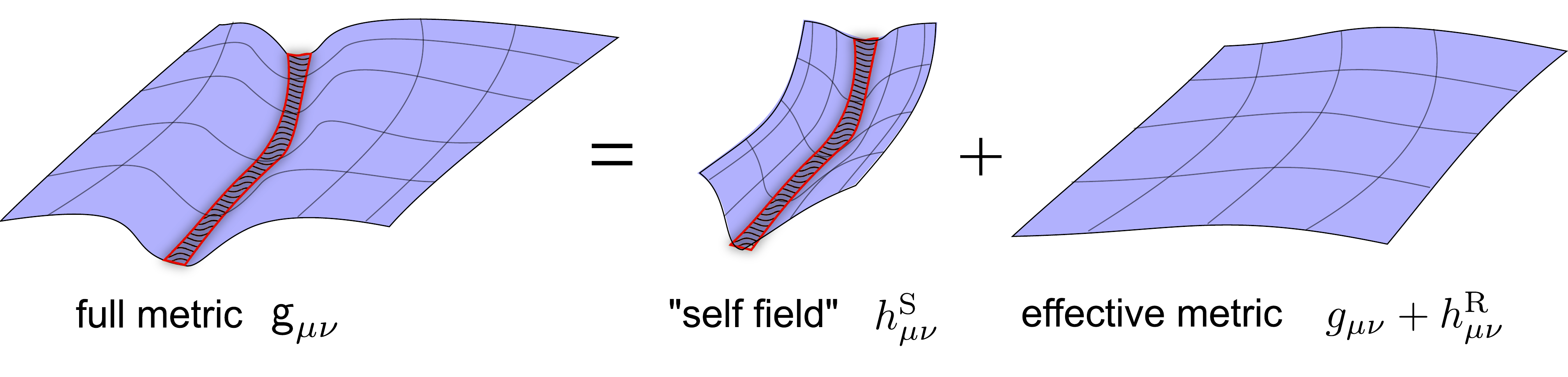}
\includegraphics[width=\textwidth]{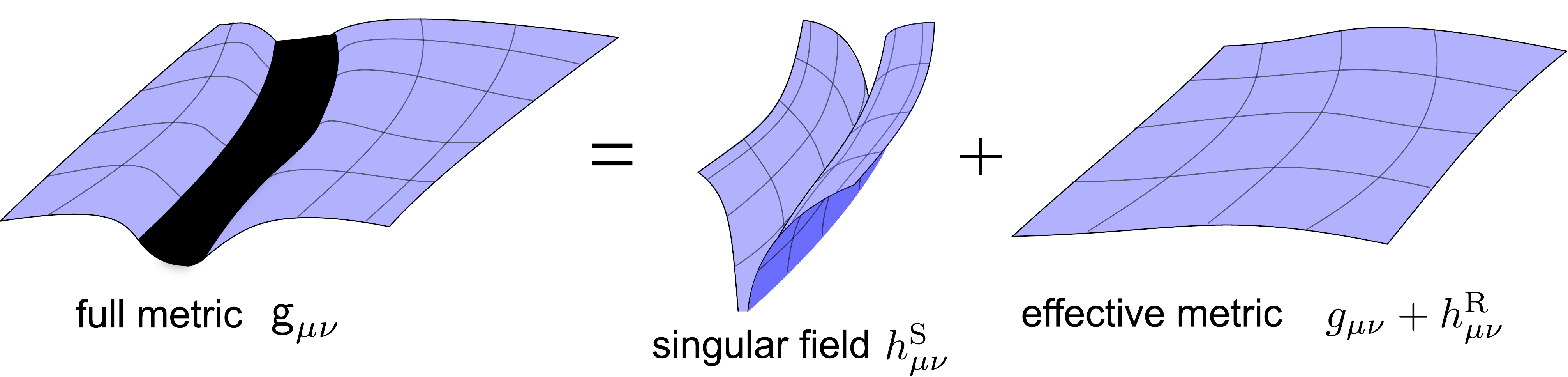}
\caption{\label{Fig:puncture}Replacement of the physical metric with an effective metric plus a puncture. In the top row, the physical metric ${\sf g}_{\mu\nu}$ is split into a self-field $h^\S_{\mu\nu}$ plus an ``effectively external" metric $\tilde g_{\mu\nu}=g_{\mu\nu}+h^\R_{\mu\nu}$. In the bottom row, we ``black out'' the region in and very near the object, and we replace the physical metric with a singular field  $h^\S_{\mu\nu}$ (any local approximation to which is called a puncture) plus the ``effectively external" metric. The replacement of the self-field with the singular field is made only very near the object: the self-field and singular field are identical to one another  in the buffer region and beyond; the effective metric $\tilde g_{\mu\nu}$ is completely unaltered everywhere it is defined.}
\end{center}
\end{figure}

The basic picture of a puncture scheme is illustrated in Fig.~\ref{Fig:puncture}. Physically, the scheme replaces the actual problem in the region covering the body, with all its matter fields, singularities (in the case of a black hole), or other oddities (in the case of exotic matter), with an \emph{effective} problem. But it yields the correct physical field outside the object. 
Hence, while we must abandon the point particle model at nonlinear orders, we can replace it with the more general concept of a puncture, a local singularity that encodes all the necessary information about the object in its multipole moments.


\subsection{The causality of the regular field}

I now return to the question of how ``nice'' the regular field is, by which I specifically mean the three enumerated properties in Sec.~\ref{singular-regular}. I have already shown that at all orders, the regular field I defined is a vacuum solution. I have also shown that at first order, the object moves as a test body in the effective geometry it induces, and (for spherical, nonspinning objects) this result will be extended to second order in Sec.~\ref{gauge and motion}. Now the final property follows immediately, at all orders in $\e$, from the design of the puncture scheme: the residual field and its derivatives manifestly depend only on the causal past, and by construction, the residual field and any number of its derivatives agree with those of the regular field on the worldline. Taking the limit of an infinite number of derivatives, we get the desired result.

\subsection{A note on ``regularization''}

In the gravitational self-force literature, one often speaks of ``regularizing'' the field or the self-force. This can mistakenly give the impression that one has introduced infinities into the problem, and that one must regularize them to recover the physical result. But in the formalism I have described here, one only ever deals with finite quantities. The force on the object, for example, is derived from the field equations outside the object, and it is written in terms of manifestly finite fields in that region. Those fields outside the object are then written in terms of quantities on the worldline by identifying $S^i(t)$, a field off the worldline, with $e^i_\mu S^\mu(z(t))$, for example; or similarly, by defining the regular field such that $h^{\R1}_{\mu\nu}(z(t))=k^1_{\mu\nu}(t)$ and $h^{\R1}_{\mu\nu,i}(z(t))=k^1_{\mu\nu i}(t)$. 

From this perspective, the various ``regularization'' methods that have been used in the self-force problem to remove the `singular part' of the field~\cite{Barack:09,Poisson-Pound-Vega:11} arise only as a practical necessity: we cannot easily determine the physical metric \emph{inside} the object, nor are we interested in doing so, which prompts us to replace it with the fiction of a singular field solely as a means of calculating the physical metric \emph{outside} the object. Computational techniques such as puncture schemes and mode-sum ``regularization''~\cite{Barack:09,Poisson-Pound-Vega:11} are not methods of removing singularities; they are simply methods of calculating the particular finite quantities in question. In mode-sum regularization, for example, one rigorously writes a spherical-harmonic mode decomposition of $h^{\R1}_{\mu\nu}(z)$ by decomposing $h^{\R1}_{\mu\nu}(z)=\lim_{x\to z}(h^1_{\mu\nu}-h^{\S1}_{\mu\nu})$, with $h^1_{\mu\nu}$ being the field of a point mass. Every quantity in the calculation is finite every step of the way.

																					\section{Gralla-Wald and osculating-geodesics approximations}\label{worldline}

Given results in the self-consistent approximation, one can always obtain analogous results in a Gralla-Wald or osculating-geodesics approximation by performing an expansion of the worldline, as described in Secs.~\ref{GW-approx} and \ref{osc-approx}. In this section I present that procedure and those results.

\subsection{Gralla-Wald approximation}
There are two ways to obtain the Gralla-Wald approximation from prior results: by expanding the worldline in the self-consistent results; or more simply, by returning to the results in the buffer region and simply choosing the worldline at $r=0$ to be $\gamma_0$, the zeroth-order worldline. 

For the moment, let us take the second route. All the local results are unchanged, except that $F_{n>0}^\mu$ is set to zero and the mass dipole moment $M^i$ is \emph{not} set to zero. This means that the metric perturbations $\check{h}^n_{\mu\nu}=\check{h}^{\S n}_{\mu\nu}+\check{h}^{\R n}_{\mu\nu}$ are modified from Eqs.~\eqref{hS1}--\eqref{hR2} to be
\begin{align}
\check{h}^{\S1}_{\mu\nu} &= h^\seed_{\mu\nu}(x;z_0,m) =\frac{2m\delta_{\mu\nu}}{r}+\O(r^0),\label{GW_hS1}\\
\check{h}^{\R1}_{\mu\nu} &= \sum_{\ell\geq0}h^\free_{\mu\nu}(x;z_0,k^1_\ell),\label{GW_hR1}
\end{align}
and
\begin{align}
\check{h}^{\S2}_{\mu\nu} &= h^\seed_{\mu\nu}(x;z_0,M) + h^\seed_{\mu\nu}(x;z_0,S)+h^\seed_{\mu\nu}(x;z_0,\check{\delta m}) \nonumber\\
&\quad+ h^{\S 2IH}_{\mu\nu}(x;z_0,m,\{k^1_\ell\}_{\ell\geq0}),\label{GW_hS2}\\
\check{h}^{\R2}_{\mu\nu} &= \sum_{\ell\geq0}h^\free_{\mu\nu}(x;z_0,k^2_\ell) + h^{\R 2IH}_{\mu\nu}(x;z_0,\{k^1_\ell\}_{\ell\geq0}),\label{GW_hR2}
\end{align}
where 
\beq\label{M_seed}
h^\seed_{\mu\nu}(x;z_0,M) = \frac{2 M_in^i}{r^2}\delta_{\mu\nu}+\O(1/r),\\
\eeq
and $\check{\delta m}_{\mu\nu}$ differs from Eq.~\eqref{dm_hR} by the inclusion of the $M_i$ term in Eq.~\eqref{dm_free}, becoming
\begin{align}
\check{\delta m}_{\alpha\beta} &= \tfrac{1}{3}m\left(2\check{h}^{\R1}_{\alpha\beta}+g_{\alpha\beta}g^{\mu\nu}\check{h}^{\R1}_{\mu\nu}\right) +m(g_{\alpha\beta}+2u_{\alpha} u_{\beta})u^\mu u^\nu \check{h}^{\R1}_{\mu\nu}\nonumber\\
&\quad +4u_{(\alpha}(m\check{h}^{\R1}_{\beta)\mu}u^\mu+2\dot M_{\beta)}).\label{GW_dm_hR}
\end{align}
Otherwise, all functionals remain completely unchanged, except that they are evaluated as functionals of $z_0^\mu$ rather than of $z^\mu$. The regular field naturally remains a vacuum solution even on $\gamma_0$, its first two orders satisfying $E_{\mu\nu}[\check{h}^{\R1}]=0$, $E_{\mu\nu}[\check{h}^{\R2}_{\mu\nu}]=2\delta^2 R_{\mu\nu}[\check{h}^{\R1},\check{h}^{\R1}]$, and $L_\mu[\check{h}^{\R n}]=0$.

Finally, if we define $z_{1\perp}^\mu\equiv M^\mu/m$ to be the deviation perpendicular to $\gamma_0$ (defining $M^t=0$), then from the master equation~\eqref{master_equation}, we get the Gralla-Wald equation
\begin{align}
\frac{D^2z^\mu_{1\perp}}{d\tau_0^2} &= - R^\mu{}_{\alpha\nu\beta}u^\alpha_{0}z_{1\perp}^\nu u^\beta_0 
											-\frac{1}{2}P_0^{\mu\nu}(2\check{h}^{{\rm R1}}_{\rho\nu;\sigma}-\check{h}^{{\rm R1}}_{\rho\sigma;\nu})u_0^\rho u_0^\sigma \nonumber\\
							&\quad		+ \frac{1}{2m}R^\mu{}_{\nu\rho\sigma}u_0^\nu S^{\rho\sigma}.\label{GW_wSpin}
\end{align}

Just as in the self-consistent case, these local results can be used in the design of a puncture scheme to obtain global results. However, before describing that scheme, I return to the other method of obtaining the Gralla-Wald approximation: by substituting an expansion of the worldline into the self-consistent results.

\subsubsection{Expansion of the worldline}
We would like to express the expansion covariantly, in terms of vectors that live on $\gamma_0$. Suppose we begin with an expansion $z^\mu(s,\e)=z^\mu_0(s)+\e z^\mu_1(s)+\e^2 z_2^\mu(s)+\ldots$, where $z^\mu_n(s)=\frac{1}{n!}\frac{\partial^n z^\mu}{\partial\e^n}\big|_{\e=0}$. The linear term, $z^\mu_1(s)$, is automatically a vector on $\gamma_0$, since it is a first derivative along the curve of constant $s$ and increasing $\e$. However, beyond linear order, the terms are no longer tensorial; each of them is simply a collections of four scalars dependent upon the particular chart $x^\mu$ in which $z^\mu(s,\e)=x^\mu(\gamma(s,\e))$.

So let us approach the problem more geometrically. To facilitate the expansion, I introduce a Lie derivative $\varLie$ that acts on a functional's dependence on $z^\mu$:
\beq
\varLie_\xi A_{\Lambda}(x;z) \equiv \frac{d}{d\lambda}A_{\Lambda}(x;z+\lambda\xi)\big|_{\lambda=0},
\eeq
where $\Lambda$ is a multi-index of any covariant and contravariant rank. This is closely related to a Lie derivative acting at $x^\mu$. $\Lie$ moves the field point relative to the worldline; $\varLie$ moves the worldline relative to the field point. The two operations are not identical, since the tensorial character of the functional is different at the two points. For example, in Eq.~\eqref{h1_z} we see that $h^1_{\mu\nu}(x;z)$ is a rank-two tensor at $x^\mu$ but a scalar at $z^\mu$.

An expansion of a functional of the worldline $z^\mu(s,\e)$ in the limit $\e\to0$ is really an expansion along the flow of increasing $\e$. We can write this as 
\beq\label{A_expansion}
A_{\Lambda}(x;z_\e) = A_{\Lambda}(x;z_0) + \e\delta A_{\Lambda}(x;z_0)+\e^2\delta^2 A_{\Lambda}(x;z_0)+\ldots, 
\eeq
where
\beq\label{dnA}
\delta^n A_{\Lambda}(x;z_0) = \frac{1}{n!}\varLie^n_v A_{\Lambda}(x;z_0);
\eeq
here $v^\mu\equiv \frac{\partial z^\mu}{\partial\lambda}$ is the generator of the flow. 

Now, let $z^\mu_{2F}=\frac{1}{2}v^\mu\nabla_\mu v^\nu\big|_{\e=0}$. This is a vector on $\gamma_0$. It is equal to $z^\mu_2$ if the expansion of the worldline is performed in Fermi normal coordinates. One can obtain a covariant evolution equation for it from a second-order self-consistent equation of motion, as described in Appendix~\ref{geodesic_expansion_in_h_and_dz}. Beginning from Eq.~\eqref{2nd-geo}, this procedure leads to Eq.~\eqref{z2_generic}, for example. More importantly for the present analysis, we can also write the expansion~\eqref{dnA} in terms of this quantity. As an instance of that, the second-order term is 
\beq
\frac{1}{2}\varLie_v^2 A_{\Lambda}(x;z_0) = z_{2F}^{\mu'}\nabla_{\mu'}A_{\Lambda}(x;z_0) + \frac{1}{2}z^{\mu'}_1 z^{\nu'}_1\nabla_{\mu'}\nabla_{\nu'}A_{\Lambda}(x;z_0), 
\eeq
where primed indices refer to the point $z^\mu_0$.
One can do the same at any order: from the equation of motion for $z^\mu$, obtain evolution equations for vectors on $z_0^\mu$, and write the expansion~\eqref{A_expansion} in terms of those vectors.

Applying this expansion to the perturbations from the self-consistent approximation, we have 
\begin{align}
\e^n h^n_{\mu\nu}(x;z) = \e^n h^n_{\mu\nu}(x;z_0)+\e^{n+1} \varLie_{z_1}h^n_{\mu\nu}(x;z_0)+\O(\e^{n+2}).
\end{align}
Therefore, the Gralla-Wald expansion~\eqref{GW_h_expansion} reads
\begin{equation}\label{h_GW_expansion}
h_{\mu\nu} = \e \check{h}^1_{\mu\nu}(x;z_0) + \e^2 \check{h}^2_{\mu\nu}(x;z_0,z_1) + \O(\epsilon^3),
\end{equation}
with $\check{h}^1_{\mu\nu}(x;z_0)$ being the same functional as $h^1_{\mu\nu}(x;z)$, but with $z^\mu_0$ having replaced $z^\mu$ in its argument, and with the second-order perturbation given by the new functional
\beq
\check{h}^2_{\mu\nu}(x;z_0,z_1) = h^2_{\mu\nu}(x;z_0) + \delta h^1_{\mu\nu}(x;z_0,z_1).
\eeq
where $\delta h^n_{\mu\nu}(x;z_0,z_1)=\varLie_{z_1} h_{\mu\nu}(x;z_0)$.

The same expansions are applied in precise analogy for the singular and regular fields, yielding $\check{h}^n_{\mu\nu}=\check{h}^{\S n}_{\mu\nu}+\check{h}^{\R n}_{\mu\nu}$ with
\begin{align}
\check{h}^{\S/\R1}_{\mu\nu} &= h^{\S/\R1}_{\mu\nu}(x;z_0),\\
\check{h}^{\S/\R2}_{\mu\nu} &= h^{\S/\R2}_{\mu\nu}(x;z_0) + \delta h^{\S/\R1}_{\mu\nu}(x;z_0,z_1).
\end{align}

Appendix~\ref{pp_quantities_expansion} describes these expansions in more detail. But in a few words, the end result is that the procedure exactly recovers Eqs.~\eqref{GW_hS1}--\eqref{GW_wSpin}. All of the meat of the result is in $\delta h^1_{\mu\nu}(x;z_0,z_1)$: the action of $\varLie$ on $h^1_{\mu\nu}$ precisely generates the mass dipole moment seed field $h^\seed_{\mu\nu}(x;z_0,M)$ and the contribution of the mass dipole moment to $\check{\delta m}_{\mu\nu}$. Appendix~\ref{pp_quantities_expansion} also shows that (i) the generation and modification of these two seed fields precisely corresponds to the linear term in the expansion of the skeletal stress-energy $T^{\mu\nu}_1(x;z)$ around $z^\mu=z^\mu_0$, and (ii) the term $\delta h^1_{\mu\nu}(x;z_0,z_1)$ [and analogously, $\delta h^{\S/\R1}_{\mu\nu}(x;z_0,z_1)$] can be written as a functional $\delta h^1_{\mu\nu}(x;z_0,z_{1\perp})$ that depends only on the perpendicular piece of $z_1^\mu$. Result (ii) is in agreement with the fact that the seed fields depend only on $M^\mu=mz^\mu_{1\perp}$. Physically, this results from the field equations' indifference to the piece of the deviation that is tangential to the worldline, which can always be set to zero by reparametrizing the family $z^\mu(s,\e)$ with a parameter $s'=s'(s,\e)$ that ensures curves of fixed $s'$ strike $z_0^\mu$ orthogonally at $\e=0$.

\subsubsection{Global metric}


A global solution can be obtained from the local results just as in the self-consistent case. I first note that the local singularity structure of the singular field $\check{h}^{\S}_{\mu\nu}=\e\check{h}^{\S1}_{\mu\nu}+\e^2\check{h}^{\S2}_{\mu\nu}$ is identical to the self-consistent singular field, but for two important alterations:
\begin{itemize}
\item The divergent terms diverge on $\gamma_0$, not on $\gamma$.
\item The second-order singular field depends on the correction $z^\mu_1$ to the position.
\end{itemize}

Because the point at which the puncture diverges is independent of the perturbations $\check{h}^n_{\mu\nu}$ in this expansion, the puncture scheme becomes a sequence of equations, rather than a coupled system: first, the zeroth-order worldline is prescribed as a solution to the background geodesic equation,
\begin{equation}\label{z0_GW}
\frac{D^2z^\mu_0}{d\tau_0^2} = 0,
\end{equation}
then the first order field is found from
\begin{subequations}\label{h1_GW}
\begin{align}
E_{\mu\nu}[\check{h}^{\res1}] &= -E_{\mu\nu}[\check{h}^{\P1}_{\alpha\beta}] & \text{inside }\Gamma_0,\\
E_{\mu\nu}[\check{h}^{1}] &= 0 & \text{outside } \Gamma_0,
\end{align}
\end{subequations}
then that field is used to find the first-order correction to the position by solving the Gralla-Wald equation~\eqref{GW_wSpin},
and finally the second-order field is found from
\begin{subequations}\label{h2_GW}
\begin{align}
E_{\mu\nu}[\check{h}^{\res2}] &=2\delta^2R_{\mu\nu}[\check{h}^1,\check{h}^1]  - E_{\mu\nu}[\check{h}^{\P2}] & \text{inside }\Gamma_0,\\
E_{\mu\nu}[\check{h}^2] &= 2\delta^2R_{\mu\nu}[\check{h}^1,\check{h}^1] & \text{outside }\Gamma_0.
\end{align}
\end{subequations}
Here $\Gamma_0$ is a tube around $\gamma_0$; unlike in the self-consistent case, neither $\gamma_0$ nor $\Gamma_0$ need be updated over the course of the numerical simulation. Like the self-consistent puncture scheme, these equations can be solved given data on a Cauchy surface.\footnote{Since the approximation is held to be valid in a region of size $\varsigma(\e)\ll1/\sqrt{\e}$, a reasonable approach would be to solve the equations in the causal future of a partial Cauchy surface of that size.}

Again, there is no obstacle to carrying a puncture scheme like this to arbitrary order, given the local solution in the buffer region.


\subsection{Osculating-geodesics approximation}
I refer the reader back to Sec.~\ref{osc-approx} for a reminder of how the osculating-geodesics approximation works. In brief, it obtains a self-consistent approximation by first using a sequence of Gralla-Wald approximations to find the the self-consistent worldline $z^\mu$, then solving the relaxed field equations with $z^\mu$ already determined.  Concretely, through second order one seeks a solution to the self-consistent system~\eqref{h1_SC}--\eqref{motion_SC} by applying the Gralla-Wald approximation~\eqref{h_GW_expansion} to the right-hand side of Eq.~\eqref{motion_SC} at each instant $\tau$, with the expansion $z^\mu(\tau',\e) = z_{0(\tau)}^\mu(\tau')+\e z_{1(\tau)}^\mu+\O(\e)$ being around the geodesic $z_{0(\tau)}^\mu$ that is instantaneously tangential to $z^\mu$ at time $\tau$. The terms $\check{h}^{\res n}_{\mu\nu}(z(\tau);z_0,\ldots,z_{n-1})$ that then appear in Eq.~\eqref{motion_SC} are found by solving Eqs.~\eqref{z0_GW}--\eqref{h2_GW}. By using the resulting sequence of forces $\e F^\mu_1(\tau;z_{0(\tau)})+\e^2 F^\mu_2(\tau;z_{0(\tau)},z_{1(\tau)})$, one can solve Eq.~\eqref{motion_SC}. Finally, one can solve Eqs.~\eqref{h1_SC}--\eqref{h2_SC} using the $z^\mu$ one has already found.

What is the advantage of this over simply solving Eqs.~\eqref{h1_SC}--\eqref{motion_SC} directly? At linear order, the advantage is that one can solve Eq.~\eqref{h1_GW} very easily in the frequency domain, while the self-consistent equations do not seem to admit a frequency decomposition. One can then compute a table of values of $F^\mu_1(\tau;z_{0(\tau)})$ for different geodesics and easily solve $\frac{D^2z^\mu}{d\tau^2}=\e F^\mu_1(\tau;z_{0(\tau)})$ to obtain an approximation to $z^\mu$~\cite{Warburton-etal:12}.  But at second-order, it is not obvious whether the scheme is simpler than a direct solution of the self-consistent equations. The term $\delta h^1_{\mu\nu}(x;z_{0(\tau)},z_{1(\tau)})$ that appears in the second-order field (and a piece of which appears in the puncture $\check{h}^{2\P}_{\mu\nu}$) grows with time away from the osculation point $x^\mu=z^\mu(\tau)$; can it be handled in an efficient way? Furthermore, one must ensure that appropriate boundary conditions are used in each Gralla-Wald expansion to reproduce desired global boundary conditions (such as no incoming radiation) on the self-consistent field; since the approximations only apply in a region of size $\ll 1/\sqrt{\e}$, is there any simple way to find these boundary conditions? I leave these questions open.


															\section{Gauge transformations in perturbative descriptions of motion}\label{gauge}


Thus far, all the explicit results I have presented have been confined to a single choice of gauge. But as I described in Sec.~\ref{long_term_dynamics}, gauge and motion are intimately related in perturbation theory: any worldline $z^\mu$ one finds can always be shifted by an amount $\e\xi^\mu$. Precisely how this impacts one's approximation scheme depends strongly on how one represents the worldline.

The effect of a gauge transformation on the worldline (or equivalently, on the self-force) was first explored by Barack and Ori~\cite{Barack-Ori:01}. Their results were extended to discontinous gauge transformations by Gralla and Wald~\cite{Gralla-Wald:08,Gralla:11}\footnote{The extension of Gralla and Wald's result to the self-consistent case~\cite{Pound:10b} unfortunately contained a significant error, leading to a result that held only in gauges continuously related to Lorenz, as in the Barack-Ori analysis.} and to even more singular transformations in Ref.~\cite{Pound-Merlin-Barack:14}. In this section I focus, for simplicity, on smooth transformations, and I aim mostly at (i) clarifying the gauge freedom in each of the different representations of motion and their corresponding approximation schemes, and (ii) defining appropriate rules for the gauge transformations of the singular and regular fields. The presentation is adapted from Ref.~\cite{Pound:15b}, which includes additional results and proves some key statements that are here asserted without proof.

\subsection{Review of gauge freedom in perturbation theory}\label{gauge_review}

Before considering the question of gauge in self-force theory, I briefly remind the reader of the basics, following Refs.~\cite{Geroch:69, Stewart-Walker:74,Bruni-etal:96}. It will be convenient in this section to adopt index-free notation for tensors. In perturbation theory, we consider a family of metrics ${\sf g}_{\mu\nu}(x,\e)$, or simply ${\sf g}$ in the absence of a chart. This family lives on a family of manifolds $\man_\e$, and a given choice of gauge refers to an identification map $\phi^X_\e:\man_0\to\man_\e$. The identification map induces a flow through the family down to the base manifold $\man_0$ where the background metric $g\equiv{\sf g}_0$ lives.  
 Call the generator of this flow $X\equiv\frac{d\phi^X_\e}{d\e}$. We wish to approximate a tensor ${\sf A}$ at a point $\phi^X_\e(p)\in\man_\e$ as an expansion around its value at $\e=0$. This expansion is given by 
\beq\label{X-expansion}
(\phi^X_\e{}^*{\sf A})(p) = (e^{\e\Lie_X}{\sf A})(p) = \sum_{n\geq0}\frac{\e^n}{n!} (\Lie^n_X{\sf A})(p),
\eeq
where $\phi^X_\e{}^*$ denotes the pullback of $\phi^X_\e$, $\Lie$ is the Lie derivative, and $p\in\man_0$. We define the $n$th-order perturbation $A^X_n$ in this gauge to be
\beq
A^X_n(p) \equiv \frac{1}{n!} (\Lie^n_X{\sf A})(p).
\eeq

Now say we work in a different gauge. This corresponds to a different choice of identification map $\phi^Y_\e:\man_0\to\man_\e$ and flow generator $Y\equiv\frac{d\phi^Y_\e}{d\e}$. The approximation of the tensor ${\sf A}$ in terms of tensors at the point $p\in\man_0$ is now given by 
\beq\label{Y-expansion}
(\phi^Y_\e{}^*{\sf A})(p) = (e^{\e \Lie_Y}{\sf A})(p),
\eeq
and the $n$th-order perturbation is
\beq
A^Y_n(p) \equiv \frac{1}{n!} (\Lie^n_Y{\sf A})(p).
\eeq

The gauge transformation of ${\sf A}$ is the difference between the two expansions when evaluated at a point in $\man_0$: in more common notation, we say $A_n \to A_n'=A_n+\Delta A_n$, where the primed tensor refers to the $Y$ gauge, the unprimed to the $X$ gauge, and their difference is
\beq
\Delta A_n(p) = \frac{1}{n!} (\Lie^n_Y{\sf A})(p)-\frac{1}{n!} (\Lie^n_X{\sf A})(p).
\eeq
The first- and second-order terms are easily expressed in the familiar form 
\begin{subequations}\label{DeltaA}
\begin{align}
\Delta A_1 &= \Lie_{\xi_1} A_0,\label{DeltaA1}\\
\Delta A_2 &= \Lie_{\xi_2} A_0 + \frac{1}{2}\Lie^2_{\xi_1}A_0 + \Lie_{\xi_1} A_1,\label{DeltaA2}
\end{align}
\end{subequations}
where $\xi_1\equiv Y-X$ and $\xi_2\equiv \frac{1}{2}[X,Y]$ are the usual gauge vectors. 
Higher-order terms are straightforwardly written down, but since concrete self-force results are not available beyond second order, I stop here.

In a chart, one can show that a gauge transformation can equivalently be thought of as a near-identity coordinate transformation rather than a change in identification map. First lay a chart $x^\mu$ on each $\man_\e$ using some identification, say $x^\mu(\phi^X_\e(p))=x^\mu(p)$ for each $p\in\man_0$. The two identification maps $\phi^X_\e$ and $\phi^Y_\e$ identify the point $p$ in $\man_0$ with two different points $q=\phi^X_\e(p)$ and $q'=\phi^Y_\e(p)$ in $\man_\e$, which are related by the active diffeomorphism $q\mapsto \phi^Y_\e((\phi^X_\e)^{-1}(q))$ and which have slightly different coordinate values. Since the chart $x^\mu$ consists of four ordinary scalar fields, we can apply the general expansion~\eqref{Y-expansion} to write the coordinates at $q'$ as
\beq\label{qtoq'}
x^\mu(q') = x^\mu(q) + \e \xi^\mu_1(x(q)) + \e^2\left[\xi^\mu_2(x(q))+\frac{1}{2}\xi^\nu_1(x(q))\partial_\nu\xi^\mu_2(x(q))\right]+\O(\e^3),
\eeq
where I have used $\Lie_X x^\mu=0$ to express $Y$ derivatives as $\xi$ derivatives, and I have used the fact that $x^\mu(p)=x^\mu(q)$ to express the components on the right-hand side as functions of $x^\mu(q)$. Now, rather than an active diffeomorphism on $\man_\e$, let us consider this as a passive change in the chart, $x^\mu(q)\mapsto x'^\mu(x(q))$. Define the coordinate transformation such that $x'^\mu(q')=x^\mu(q)$. Rewriting Eq.~\eqref{qtoq'} as an equation for $x^\mu(q)$ as a function of $x^\mu(q')$, we get
\begin{align}
x'^\mu(q') &= x^\mu(q')-\epsilon \xi^\mu_1(x(q'))-\epsilon^2\!\! \left[\xi^\mu_2(x(q'))-\frac{1}{2}\xi^\nu_1(x(q'))\partial_\nu\xi_1^\mu(x(q'))\right]\nonumber\\
 &\quad +O(\epsilon^3).\label{coord_transformation}
\end{align}

Gauge transformation laws for components of tensors can be derived directly from this coordinate transformation. For example, by rewriting Eq.~\eqref{coord_transformation} as an equation for $x^\mu(x'(q'))$ and substituting it into the ordinary transformation law for the components of the metric, one finds
\begin{subequations}\label{component_transformation}%
\begin{align}
{\sf g}'_{\mu\nu}(x',\e) &= \frac{\partial x^\alpha}{\partial x'^\mu}\frac{\partial x^\beta}{\partial x'^\nu}{\sf g}_{\alpha\beta}(x(x'),\e)\\
								&= {\sf g}_{\mu\nu}(x',\e) +\Lie_\xi{\sf g}_{\mu\nu}(x',\e)+\frac{1}{2}\Lie^2_\xi{\sf g}_{\mu\nu}(x',\e) +O(\e^3),  
\end{align}
\end{subequations}
where $\xi^\mu=\e\xi_1^\mu+\e^2\xi^\mu_2+\O(\e^3)$; the analogous transformation law for a tensor of arbitrary rank is also easily found. To relate this to the language used above, note that we are now using a single identification map $\phi_\e$, and given that identification, components of ${\sf g}$ have identical coordinate values as $(\phi_\e{}^*{\sf g})$. Equation~\eqref{component_transformation} applies even if ${\sf g}_{\mu\nu}(x',\e)$ is not expanded for small $\e$. If it is so expanded, then Eq.~\eqref{component_transformation} returns Eq.~\eqref{DeltaA}.


Of course, we are ultimately interested in solving the Einstein equation, and so in addition to transformations of the metric, we must consider transformations of curvature tensors. Useful identities for the transformations of curvature tensors are derived in Appendix~\ref{gauge-identities}. One of their consequences is that when examining perturbations of a curvature tensor, one can derive transformation laws in two equally natural ways: directly from Eq.~\eqref{DeltaA} or from the transformations of the metric perturbations. For example, in a vacuum background, Eq.~\eqref{DeltaA} directly implies
\begin{align}
\Delta \delta R_{\mu\nu}[h^1] &= \Lie_{\xi_1}R_{\mu\nu}[g]=0,\label{DEFE1}\\
\Delta (\delta R_{\mu\nu}[h^2]+\delta^2 R_{\mu\nu}[h^1,h^1]) &= \Lie_\xi \delta R_{\mu\nu}[h^1];\label{DEFE2}
\end{align}
or the same equations can be found by instead using Eq.~\eqref{DeltaA} for the metric itself, writing
\begin{align}\label{DEFE2-steps}
\Delta (\delta &R_{\mu\nu}[h^2]+\delta^2 R_{\mu\nu}[h^1,h^1])\nonumber\\
		&= \delta R_{\mu\nu}[h'^2]+\delta^2 R_{\mu\nu}[h'^1,h'^1] - (\delta R_{\mu\nu}[h^2]+\delta^2 R_{\mu\nu}[h^1,h^1]) \nonumber\\
		&=  \delta R_{\mu\nu}[\Lie_{\xi_2}g]+\tfrac{1}{2}\delta R_{\mu\nu}[\Lie^2_{\xi_1}g]+\delta R_{\mu\nu}[\Lie_{\xi_1}h^1]\nonumber\\
		&\quad	+2 \delta^2 R_{\mu\nu}[h^1,\Lie_{\xi_1}g]+\delta^2 R_{\mu\nu}[\Lie_{\xi_1}g,\Lie_{\xi_1}g]
\end{align}
and then applying Eqs.~\eqref{Lie R}--\eqref{Lie dR}.

\subsection{Gauge in the Gralla-Wald approximation}\label{gauge_GW}
I reverse my usual ordering by first considering transformations in the Gralla-Wald approximation, which allows the most straightforward treatment. Since the Gralla-Wald expansion is an ordinary one, with coefficients independent of $\e$, all the ordinary rules apply.

\subsubsection{Transformation of the metric and the worldline}\label{gauge_GW-general}

First, let us examine the transformation of the deviation terms in the expansion of the worldline. According to Eq.~\eqref{coord_transformation}, under a gauge transformation the coordinates $z^\mu(s,\e)=x^\mu(\gamma_\e(s))$ on the worldline become
\begin{align}
z'^{\mu}(s,\e) &= z^\mu(s,\e)-\epsilon \xi^\mu_1(z)-\epsilon^2 \left[\xi^\mu_2(z)-\frac{1}{2}\xi^\nu_1(z)\partial_\nu\xi_1^\mu(z)\right]+O(\epsilon^3),\label{z_transformation}
\end{align}
where functions of $z^\mu$ are evaluated at $z^\mu(s)$. By expanding this in powers of $\e$, we immediately find
\begin{subequations}\label{DzGW}
\begin{align}
z_0'^\mu(s) &=z_0^\mu(s),\label{z0_transformation} \\
z_1'^{\mu}(s) &= z_1^\mu(s) - \xi_1^\mu(z_0), \label{z1_transformation}\\
z_2'^{\mu}(s) &= z_2^\mu(s) - \xi_2^\mu(z_0)+\frac{1}{2}\xi^\nu_1(z_0)\partial_\nu\xi_1^\mu(z_0)-z_1^\nu(s)\partial_\nu\xi^\mu_1(z_0),\label{z2_transformation}
\end{align}
\end{subequations}
where functions of $z_0^\mu$ are evaluated at $z^\mu_0(s)$. Note that the zeroth-order worldline is unchanged; the effect of the transformation is to alter the deviations relative to that worldline. 

Now let us examine the transformation of the metric perturbations. Applying the transformation laws~\eqref{DeltaA} to the metric, we find $\check{h}^n_{\mu\nu}\to \check{h}'^n_{\mu\nu}+\Delta \check{h}^n_{\mu\nu}$ with
\begin{subequations}\label{DhGW}
\begin{align}
\Delta \check{h}^1_{\mu\nu}(x;z_0) &=\Lie_{\xi_1} g_{\mu\nu},\label{Deltah1GW}\\
\Delta \check{h}^2_{\mu\nu}(x;z_0) &=\Lie_{\xi_2} g_{\mu\nu} + \frac{1}{2}\Lie^2_{\xi_1}g_{\mu\nu} + \Lie_{\xi_1} \check{h}^1_{\mu\nu}(x;z_0). \label{Deltah2GW}
\end{align}
\end{subequations}
Corresponding to the treatment of the worldline, all $\check{h}'^n_{\mu\nu}$, like all $\check{h}^n_{\mu\nu}$, diverge at $z^\mu_0(s)$. Instead of altering the curve on which the fields diverge, the gauge transformation alters the singularity on that curve, by altering the functions $z_{n>0}^\mu$ that appear in $\check{h}^{n>1}_{\mu\nu}$. 

In fact, one need not even appeal to Eq.~\eqref{z_transformation} to determine how the deviation vectors are transformed: the transformation laws for $z_{n>0}^\mu$, as given in Eq.~\eqref{DzGW}, can be derived directly from those for $\check{h}^{n>1}_{\mu\nu}$. For example, $z_1^\mu$ appears in the metric as the mass dipole moment $M^\mu/m$, and the transformation of $M^\mu$ can be determined from the transformation law for $\check{h}^2_{\mu\nu}$. By applying Eq.~\eqref{Deltah2GW} to the results \eqref{GW_hS1}--\eqref{GW_hR2} in Fermi normal coordinates\footnote{At this stage the transformation is applied only to the sums $\check{h}^n_{\mu\nu}=\check{h}^{\R n}_{\mu\nu}+\check{h}^{\S n}_{\mu\nu}$; one does not yet need transformation laws for the individual pieces $\check{h}^{\R n}_{\mu\nu}$ and $\check{h}^{\S n}_{\mu\nu}$.} and comparing to Eq.~\eqref{M_seed}, one quickly finds that $\Lie_{\xi_1} \check{h}^1_{\mu\nu}$ is the only part of $\Delta\check{h}^2_{\mu\nu}$ that contributes a mass dipole term. More specifically, $\Lie_{\xi_1} \frac{2m\delta_{\mu\nu}}{r}$ contributes the relevant term, given by $\frac{-2m\xi^i_1n_i}{r^2}\delta_{\mu\nu}$. This modifies $M^i$ by an amount $\Delta M^i=-m\xi^i_1$, from which we read off the same result as in Eq.~\eqref{z1_transformation}.

\subsubsection{Transformation of the singular and regular fields}

Though deriving the transformation laws for $\check{h}^n_{\mu\nu}$ was trivial, we must put some thought into how to do the same for the singular and regular fields $\check{h}^{\S n}_{\mu\nu}$ and $\check{h}^{\R n}_{\mu\nu}$. We are free to split $\check{h}'^n_{\mu\nu}$ into a singular piece and a regular piece however we like, just as we were with  $\check{h}^n_{\mu\nu}$. However, ideally, we can do so in such a way that we preserve the nice properties of the split. Allow me to specialize to a nonspinning, spherical object, such that there is no force from coupling of moments to external curvature. This means the properties I wish to preserve are that the transformed worldline $z'^\mu$ [in its expanded form~\eqref{DzGW}] should satisfy the geodesic equation in the transformed effective metric $\tilde {g}'_{\mu\nu}=g_{\mu\nu}+h'^{\R}_{\mu\nu}$, and that effective metric should satisfy the vacuum Einstein equation.

Appropriate transformation laws can be found by noting that for a smooth metric and smooth worldline, both the geodesic equation and the vacuum Einstein equation are manifestly invariant under a generic smooth coordinate transformation. It follows that if the worldline and the metric in those two equations are expanded in a power series, and the metric and the expanded geodesic transform according to the standard laws of gauge transformations, then the perturbative expansions of the worldline and metric continue to satisfy the perturbative geodesic equation and the perturbative Einstein equation. In our case, we already have the worldline transforming according to the standard transformation law. Accordingly, if we define the regular field to transform like an ordinary smooth perturbation, as $\check{h}'^{\R n}_{\mu\nu}=\check{h}^{\R n}_{\mu\nu}+\Delta \check{h}^{\R n}_{\mu\nu}$, where
\begin{subequations}\label{DhRGW}%
\begin{align}
\Delta \check{h}^{\R1}_{\mu\nu}(x;z_0) &=\mathcal{L}_{\xi_1} g_{\mu\nu},\\
\Delta \check{h}^{\R2}_{\mu\nu}(x;z_0) &=\mathcal{L}_{\xi_2} g_{\mu\nu} + \frac{1}{2}\mathcal{L}^2_{\xi_1} g_{\mu\nu}
				+\mathcal{L}_{\xi_1} \check{h}^{\R1}_{\mu\nu}(x;z_0),
\end{align}
\end{subequations}
then all the nice properties of the regular field are maintained under a gauge transformation. 
However, at the same time, the total field $\check{h}^n_{\mu\nu}=\check{h}^{\S n}_{\mu\nu}+\check{h}^{\R n}_{\mu\nu}$ must satisfy Eq.~\eqref{DhGW}. This means the singular field must satisfy 
\beq
\check{h}'^{\S n}_{\mu\nu}=\check{h}^{\S n}_{\mu\nu}+(\Delta\check{h}^n_{\mu\nu}-\Delta\check{h}^{\R n}_{\mu\nu})
\eeq
---or more explicitly,
\begin{subequations}\label{DhSGW}%
\begin{align}
\Delta \check{h}^{\S1}_{\mu\nu}(x;z_0) &=0,\\
\Delta \check{h}^{\S2}_{\mu\nu}(x;z_0) &=\mathcal{L}_{\xi_1} \check{h}^{\S1}_{\mu\nu}(x;z_0).\label{DhS2}
\end{align}
\end{subequations}

\subsubsection{Governing equations in alternative gauges}

By design, the transformation laws~\eqref{DzGW}--\eqref{DhSGW} ensure that \emph{the governing equations of the Gralla-Wald representation are invariant under a gauge transformation}: in any gauge, the regular field satisfies the vacuum equations $\delta R_{\mu\nu}[\check{h}^{\R1}]=0$, $\delta R_{\mu\nu}[\check{h}^{\R2}_{\mu\nu}]=-\delta^2 R_{\mu\nu}[\check{h}^{\R1},\check{h}^{\R1}]$; $z^\mu_0$ is unaltered and the deviations from it satisfy~\eqref{GW_ddotz1}--\eqref{GW_ddotz2};\footnote{Rather than saying the equation of motion~\eqref{GW_ddotz1} is invariant, self-force literature usually talks about a transformation of the self-force. At first order, the equation in the new gauge is $\frac{D^2z'^\mu_{1\perp}}{d\tau_0^2}+R^\mu_{\alpha\nu\beta}u_0^\alpha z'^\nu_{1\perp}u_0^\beta=\check{F}'^\mu_1$. The force is given by
\begin{align}
\check{F}'^\mu_1 &= -\frac{1}{2}P^{\mu\nu}_0(2\check{h}'^{\R1}_{\nu\alpha;\beta}-\check{h}'^{\R1}_{\alpha\beta;\nu})u_0^\alpha u_0^\beta
= \check{F}_1^\mu -\left(\frac{D^2\xi_{1\perp}^\mu}{d\tau_0^2}+R^\mu{}_{\alpha\nu\beta}u_0^\alpha \xi^\nu_{1\perp}u_0^\beta\right),\label{DF}
\end{align}
where the second equality follows from $\check{h}'^{\R1}_{\mu\nu}=\check{h}^{\R1}_{\mu\nu}+\Lie_{\xi_1}g_{\mu\nu}$. With $z'^\mu_{1\perp}=z_{1\perp}^\mu-\e\xi_{1\perp}^\mu$ in the left-hand side of the equation of motion, the equation's invariance is transparent.}
 and assuming the gauge choice admits a well-posed initial value problem, the full field off the worldline can be calculated using the puncture scheme encapsulated in Eqs.~\eqref{z0_GW}--\eqref{h2_GW}. These facts follow from the invariance of the geodesic equation and Eqs.~\eqref{DEFE1}--\eqref{DEFE2-steps}. However, because of Eq.~\eqref{DhS2}, the singularity structure of the second-order field, and hence the puncture in the puncture scheme, is altered by a gauge transformation. We can assess the impact of this change by examining how it manifests in the second-order field equation. Using Eq.~\eqref{Lie dR}, we have
\begin{subequations}
\begin{align}
\delta R_{\mu\nu}[\Lie_{\xi_1} \check{h}^{\S 1}] &= \Lie_{\xi_1}\delta R_{\mu\nu}[\check{h}^{\S1}] - 2\delta^2 R_{\mu\nu}[h^{\S1},\Lie_\xi g]\\
			&= 8\pi\Lie_{\xi_1}\bar T^1_{\mu\nu} - 2\delta^2 R_{\mu\nu}[\check{h}^{\S1},\Delta \check{h}^{\R1}].\label{dRLiehS1}
\end{align}
\end{subequations}
The second term on the right is precisely the term required to make $h'^{\S2}_{\mu\nu}$ a solution to the correct equation, $\delta R_{\mu\nu}[\check{h}'^{\S 2}]=-\delta^2 R_{\mu\nu}[\check{h}'^{\S1},\check{h}'^{\S1}]-2\delta^2 R_{\mu\nu}[\check{h}'^{\S1},\check{h}'^{\R1}]$ at points off $\gamma_0$. The first term on the right-hand side of Eq.~\eqref{dRLiehS1} indicates a change in the skeletal stress-energy: $\bar T^2_{\mu\nu}\to \bar T^2_{\mu\nu}+\Lie_{\xi_1}\bar T^1_{\mu\nu}$. This change in skeleton corresponds to a change in the seed solutions $h^\seed_{\mu\nu}(x;z_0,\check{\delta m})$ and $h^\seed_{\mu\nu}(x;z_0,M)$. An explicit expression for $\Lie_{\xi_1}\bar T^1_{\mu\nu}$ can be derived from one for $\Lie_{\xi_1}T_1^{\mu\nu}$, given in Eq.~\eqref{LieT}. From that latter equation, we see that $\Lie_{\xi_1}T_1^{\mu\nu}$ both alters $\delta m_{\mu\nu}$ and more notably, shifts the mass dipole moment by an amount $\Delta M^\mu=-m\xi^\mu_{1\perp}$, in agreement with the discussion above.

Of course, to make use of the fact that the governing equations are the same in all smoothly related gauges, one must first find an effective metric satisfying them in a particular gauge. And given that multiple effective metrics can satisfy the same governing equations, one must realize that under a gauge transformation, one is referring to the transformation of one's particular choice of effective metric. Furthermore, there is the potentially more dangerous caveat that the entire analysis of this section assumes smooth transformations. Under a transformation that is singular on the worldline, the singularity structure of the field can be significantly altered, the transformation laws for the deviation vectors are no longer given by Eq.~\eqref{DzGW}, and the relation between the ``nice" effective metrics in the two gauges will not be given by Eq.~\eqref{DhRGW}. 

\subsection{Gauge in the self-consistent approximation}\label{gauge_SC}

I now consider the freedom in the self-consistent expansion, which is slightly thornier than the Gralla-Wald case. 

\subsubsection{Transformation of the metric and the worldline}

First, I note that the self-consistently determined worldline $z^\mu$ transforms according to Eq.~\eqref{z_transformation}. Unlike in the Gralla-Wald case, neither $z^\mu$ nor $z'^\mu$ are expanded around $z_0^\mu$.

Now to the metric. Let us adopt the passive view. Suppose we performed the self-consistent expansion in two slightly different coordinate systems $x^\mu$ and $x'^\mu$. In the unprimed coordinates, the expansion is performed by writing ${\sf g}_{\mu\nu}$ as ${\sf g}_{\mu\nu}(x,\e;z)$ and expanding for small $\e$ while holding $x^\mu$ and $z^\mu$ fixed. In the primed coordinates, we write ${\sf g}'_{\mu\nu}$ as ${\sf g}'_{\mu\nu}(x',\e;z')$ and expand while holding $x'^\mu$ and $z'^\mu$ fixed. In terms of the perturbations $h_{\mu\nu}\equiv {\sf g}_{\mu\nu}-g_{\mu\nu}$ and $h'_{\mu\nu}\equiv {\sf g}'_{\mu\nu}-g_{\mu\nu}$, we have expansions
\begin{align}
h_{\mu\nu}(x,\e;z) &= \sum \e^n {h}^{n}_{\mu\nu}(x;z),\label{h-exp}\\
h'_{\mu\nu}(x',\e;z') &= \sum \e^n {h'}^{n}_{\mu\nu}(x';z').\label{h'-exp} 
\end{align}
The metric in the two coordinate systems are related, as in Eq.~\eqref{component_transformation}, according to
\begin{subequations}
\begin{align}
{\sf g}'_{\mu\nu}(x',\e;z') &= \frac{\partial x^\alpha}{\partial x'^\mu}\frac{\partial x^\beta}{\partial x'^\nu}{\sf g}_{\alpha\beta}(x(x'),\e;z(z')),\\
			&= {\sf g}_{\mu\nu}(x',\e;z(z')) + \Lie_\xi {\sf g}_{\mu\nu}(x',\e;z(z')) \nonumber\\&\quad+ \frac{1}{2}\Lie^2_\xi g_{\mu\nu}(x',\e;z(z')) + \O(\e^3).
\end{align}
\end{subequations}
Expanding ${\sf g}_{\mu\nu}(x',\e;z(z'))$ returns 
\begin{align}
h'_{\mu\nu}(x',\e) &=   \e\big[h^1_{\mu\nu}(x';z) + \Lie_{\xi_1} g_{\mu\nu}(x')\big] + \e^2 \big[ h^2_{\mu\nu}(x';z)+ \Lie_{\xi_2} g_{\mu\nu}(x')  \nonumber\\
			&\quad + \Lie_{\xi_1} h^1_{\mu\nu}(x';z) + \tfrac{1}{2}\Lie^2_{\xi_1} g_{\mu\nu}(x') \big] +\O(\e^3).\label{h'z}
\end{align}

In Eq.~\eqref{h'z} I have simply left $z^\mu(z')$ as $z^\mu$. Just as in the Gralla-Wald case, the term $\Lie_{\xi_1} h^1_{\mu\nu}(x';z)$ introduces a mass dipole moment relative to the reference worldline---in this case, relative to $z^\mu$. But if we now expand $z^\mu(z')$ around $z^\mu=z'^\mu$, the perturbation reads 
\begin{align}
h'_{\mu\nu}(x',\e) &= \e\big[h^1_{\mu\nu}(x';z') + \Lie_{\xi_1} g_{\mu\nu}(x')\big] + \e^2\big[h^2_{\mu\nu}(x';z') + \Lie_{\xi_2} g_{\mu\nu}(x') \nonumber\\
			&\quad + (\Lie_{\xi_1}+\varLie_{\xi_1}) h^1_{\mu\nu}(x';z')+ \tfrac{1}{2}\Lie^2_{\xi_1} g_{\mu\nu}(x') \big] +\O(\e^3).\label{h'z'}
\end{align}
where $\varLie_{\xi_1}h^{1}_{\mu\nu}(x';z')=\delta h^{1}_{\mu\nu}(x';z',\xi)$ uses the Lie derivative introduced in Sec.~\ref{worldline}. The two Lie derivatives $(\Lie_{\xi_1}+\varLie_{\xi_1}) h^1_{\mu\nu}(x';z')$ do not precisely cancel one another, but $\varLie_{\xi_1}$ \emph{does} precisely remove the mass dipole moment induced by $\Lie_{\xi_1}$; simply put, $\Lie_{\xi_1}$ moves everything relative to the worldline, and  $\varLie_{\xi_1}$ moves the worldline by the same amount, such that there is no net change relative to the worldline. In other words, as we expect, $h'_{\mu\nu}$ contains no mass dipole term when written as a functional of the transformed worldline $z'^\mu$. The fact that the dipole moment cancels in this way can be seen more explicitly by acting with $(\Lie_{\xi_1}+\varLie_{\xi_1})$ on the leading term in $h^1_{\mu\nu}$, $\frac{2m\delta_{\mu\nu}}{r}\sim \frac{2m\delta_{\mu\nu}}{|x^i-z^i|}$; it can also be inferred from the result~\eqref{LieT+varLieT}.

Equation~\eqref{h'z'} is an expansion at fixed $x'^\mu$ and $z'^\mu$. Hence, we can read off the coefficients of $\e^n$ and say that the individual terms in the two expansions~\eqref{h-exp} and \eqref{h'-exp} are related as
\begin{align}
{h'}^1_{\mu\nu}(x';z') &=  h^1_{\mu\nu}(x';z') + \Lie_{\xi_1} g_{\mu\nu}(x') ,\label{h1'}\\
{h'}^2_{\mu\nu}(x';z') &=  h^2_{\mu\nu}(x';z') +  (\Lie_{\xi_1}+ \varLie_{\xi_1})h^{1}_{\mu\nu}(x';z') \nonumber\\
			&\quad+ \Lie_{\xi_2} g_{\mu\nu}(x') + \frac{1}{2}\Lie^2_{\xi_1} g_{\mu\nu}(x').\label{h2'}
\end{align}

We can see from this analysis that a gauge transformation acts quite differently here than in the Gralla-Wald case. In the Gralla-Wald approximation, the zeroth-order worldline, on which the singular field diverges, is invariant; in the self-consistent approximation, a gauge transformation actually shifts the curve on which the singular field diverges.


\subsubsection{Transformation of the singular and regular fields}


As in the Gralla-Wald case, I define the split into singular and regular fields in the primed gauge in the simplest way that preserves the properties of the split. In other words, we must have the effective metric $\tilde g_{\mu\nu}$ transform as any ordinary smooth tensor field would: $\tilde g'_{\mu\nu}(x')=\frac{\partial x^\alpha}{\partial x'^\mu}\frac{\partial x^\beta}{\partial x'^\nu} \tilde g_{\alpha\beta}(x(x'))$, which implies $\tilde g'_{\mu\nu}(x',\e;z')=\frac{\partial x^\alpha}{\partial x'^\mu}\frac{\partial x^\beta}{\partial x'^\nu} \tilde g_{\alpha\beta}(x(x'),\e;z(z'))$, and from there,
\begin{align}
{h'}^{\R1}_{\mu\nu}(x';z') &=  h^{\R1}_{\mu\nu}(x';z') + \Lie_{\xi_1} g_{\mu\nu}(x') ,\label{DhR1SC}\\
{h'}^{\R2}_{\mu\nu}(x';z') &=  h^{\R2}_{\mu\nu}(x';z') +  (\Lie_{\xi_1}+\varLie_{\xi_1}) h^{\R1}_{\mu\nu}(x';z') \nonumber\\
			&\quad+ \Lie_{\xi_2} g_{\mu\nu}(x') + \frac{1}{2}\Lie^2_{\xi_1} g_{\mu\nu}(x'). \label{DhR2SC}
\end{align}
At the same time we must satisfy Eqs.~\eqref{h1'}--\eqref{h2'}, which leaves the singular field to transform as
\begin{align}
{h'}^{\S1}_{\mu\nu}(x';z') &=  h^{\S1}_{\mu\nu}(x';z') ,\\
{h'}^{\S2}_{\mu\nu}(x';z') &=  h^{\S2}_{\mu\nu}(x';z') +  (\Lie_{\xi_1}+\varLie_{\xi_1}) h^{\S1}_{\mu\nu}(x';z').\label{DhS2SC}
\end{align}

With these choices, the worldline $z'^\mu$ is a geodesic of $\tilde g'_{\mu\nu}$, and $\tilde g'_{\mu\nu}$ is a smooth solution to the vacuum Einstein equation.

\subsubsection{Governing equations in alternative gauges}
Again as in the Gralla-Wald case, by design, the transformation laws~\eqref{DhR1SC}--\eqref{DhS2SC} ensure that \emph{the governing equations of the self-consistent approximation are invariant under a gauge transformation}: in any gauge $X$, the regular field satisfies $\delta R^X_{\mu\nu}[h^{\R1}]=0$, $\delta R^X_{\mu\nu}[h^{\R2}]=-\delta^2 R^X_{\mu\nu}[h^{\R1},h^{\R1}]$, where $\delta R^X_{\mu\nu}$ is the linearized Ricci tensor in the $X$ gauge; the center-of-mass worldline on which the field diverges satisfies~\eqref{SC_eq_mot_2nd}; and assuming well-posedness, the full field off the worldline can be computed using the puncture scheme~\eqref{h1_SC}--\eqref{motion_SC}, with the replacement $E_{\mu\nu}\to-2\delta R^X_{\mu\nu}$. The only change is to the form of the second-order puncture. The nature of this change can be inferred from the analogue of Eq.~\eqref{dRLiehS1}, which reads
\begin{align}
\delta R_{\mu\nu}[(\Lie_{\xi_1}+\varLie_{\xi_1}) h^{\S 1}] &= 8\pi(\Lie_{\xi_1}+\varLie_{\xi_1})\bar T^1_{\mu\nu} - 2\delta^2 R_{\mu\nu}[h^{\S1},\Delta h^{\R1}].
\end{align}
This is just as in the Gralla-Wald case but for the presence of the $\varLie$ term. The two Lie derivatives induce a change in the skeletal stress-energy, and hence in the seed solutions contained in the puncture, but unlike in the Gralla-Wald case, this change does not alter the mass dipole moment; the presence of $\varLie_{\xi_1}$ ensures the mass dipole moment remains zero, as discussed above. So $h^\seed_{\mu\nu}(x;z,M)$ vanishes in all smoothly related gauges, and only $h^\seed_{\mu\nu}(x;z,\delta m)$ is altered.

\subsection{Gauge in the osculating-geodesics approximation}\label{gauge_osculating}

Gauge freedom in the osculating-geodesics approximation is more complicated than in either of the previous two cases. Here, there is freedom at two levels: in the calculations in the Gralla-Wald expansion at each osculation instant, and in the calculation of the self-consistent perturbation $h_{\mu\nu}(x;z)$ that stitches together the Gralla-Wald expansions.

If we always use the same gauge at the two levels, then there is no real complication: a gauge transformation at the self-consistent level can be expanded out to find the induced gauge transformation at the Gralla-Wald level in a straightforward way. However, in principle, calculations at the two levels can be performed in different gauges. In that case, one must have good control over the relationships between the gauges, since the final self-consistent approximation $h_{\mu\nu}(x;z)=\sum_{n=1}^{N}\e^n h^n_{\mu\nu}(x;z)$ satisfies the Einstein equation through order $\e^N$ only if $z^\mu$ is in the same gauge as $h_{\mu\nu}$. 

Suppose we wish to calculate $h^n_{\mu\nu}(x;z)$ in a gauge $X$ (denoted by unprimed symbols), but at an osculation instant $\tau$ we wish to calculate the Gralla-Wald perturbations $\check{h}'^n_{\mu\nu}(x;z_{0(\tau)})$ in a gauge $Y$ (denoted by primed symbols). To find the curve $z^\mu$ in gauge $X$, we need to solve
\beq\label{SC-osc}
\frac{D^2z^\mu}{d\tau^2} = F^\mu(\tau;z)= \e\check{F}_1^\mu(\tau;z_{0(\tau)})+\e^2\check{F}_1^\mu(\tau;z_{0(\tau)})
\eeq
with the forces in gauge $X$. But as input we have only the $Y$-gauge perturbations $\check{h}'^n_{\mu\nu}(x;z_{0(\tau)})$. For simplicity, consider only the first-order force in gauge $X$. It is related to $Y$-gauge quantities as [see Eq.~\eqref{DF}]
\beq\label{DF-osc}
\check{F}^\mu_{1}(\tau) =\check{F}'^\mu_1(\tau) +\left(\frac{D^2\xi_{1\perp}^\mu}{d\tau_0^2}+R^\mu{}_{\alpha\nu\beta}u_{0(\tau)}^\alpha \xi^\nu_{1\perp}u_{0(\tau)}^\beta\right),
\eeq
where $F'^\mu_1(\tau)=-\frac{1}{2}P^{\mu\nu}_0[2\check{h}'^{\R1}_{\nu\alpha;\beta}(x;z_{0(\tau)})-\check{h}'^{\R1}_{\alpha\beta;\nu}(x;z_{0(\tau)})]u_0^\alpha u_0^\beta$. To know the correct force for our self-consistent evolution, we need to know the gauge vector $\xi^\mu_{1}$. However, this gauge vector is constrained by the osculation condition: we must have $z^\mu(\tau)=z^\mu_{0(\tau)}(\tau)$ and $u^\mu(\tau)=u^\mu_{0(\tau)}(\tau)$. Hence, we should impose $\xi^\mu_1(z(\tau))=0$, reducing Eq.~\eqref{DF-osc} to  
\beq
\check{F}^\mu_{1} = \check{F}'^\mu_1 +\frac{D^2\xi_{1\perp}^\mu}{d\tau_0^2}.
\eeq
Using this relation, one can calculate the force that appears in Eq.~\eqref{SC-osc} using the force calculated from the Gralla-Wald expansion in gauge $Y$---but one must know $\frac{D^2\xi_{1\perp}^\mu}{d\tau_0^2}$ at the osculation instant $\tau$. The same analysis can be straightforwardly generalized to the second-order force $\check{F}_2^\mu(\tau;z_{0(\tau)})$.

												\section{Equations of motion from a rest gauge}\label{gauge and motion}


In the algorithm of Sec.~\ref{algorithm}, the $n$th-order equation of motion is obtained by solving the $(n+1)$th-order Einstein equation in the buffer region. However, at least in certain cases, we can instead obtain an equation of motion using only the $n$th-order solution. This is accomplished by finding the relationship between one's desired gauge---call it the \emph{practical gauge}---and what I call a \emph{rest gauge}, which is, essentially, any of the gauges that have been used to study tidally perturbed objects in the context of matched expansions (see Ref.~\cite{Poisson:14} and the many references therein for examples of these studies). Methods along these lines were the first ones used to derive the MiSaTaQuWa equation~\cite{Mino-Sasaki-Tanaka:97} and currently, they are the only ones that have been used to derive second-order equations of motion~\cite{Rosenthal:06b,Pound:12a,Pound:14a,Gralla:12}, although different notations and descriptions have obscured the shared underlying features of these derivations. 

Throughout this section, I specialize to the case of an approximately spherical, nonspinning object whose leading-order dipole and quadrupole moments vanish: $I^2_{\mu\nu i}=I^3_{\mu\nu ij}=0$. I only sketch the calculations in this section refer the reader to Refs.~\cite{Pound:14a,Gralla:12} for the details of the calculations in this section.

\subsection{Metric in a rest gauge}\label{rest gauge}

By a rest gauge, I mean a gauge in which the object is manifestly at rest relative to a worldline $z^\mu$ and manifestly centered on that worldline. To motivate this idea, I return to the themes of the introduction. In Sec.~\ref{Detweiler-Whiting}, I argued that any equation of motion can be written as the geodesic equation in some smooth piece of the metric. In Sec.~\ref{matched_expansions_intro}, I recalled the results of Thorne and Hartle, who showed something slightly stronger: that the equations of motion and precession for any compact object can be written as those for a test body immersed in some effectively external metric. 
 In Eq.~\eqref{SC_eq_mot_wSpin}, the self-force program has recovered their result for the center-of-mass motion, and it has shown that the ``external'' metric in their formalism is in fact (at least through order $\e r$)  the effective metric $\tilde g_{\mu\nu}=g_{\mu\nu}+h^\R_{\mu\nu}$.

At second order in $\e$ and beyond, the algorithm of Sec.~\ref{algorithm} has not yet revealed how $\tilde g_{\mu\nu}$ relates to Thorne and Hartle's external metric. But if we can establish the relationship, we can establish the equation of motion. Notably, they derived their equations by working in an inertial frame of their external metric; if the object possesses no spin or quadrupole moments, then there is no force on the object in this frame, and it is what I call the rest gauge. Why I call it a gauge rather than a frame will become clear as we move forward.

The metric in this gauge is most easily constructed through the inner expansion~\eqref{inner_expansion} rather than the outer. To do so, I first note that just as the matching condition fixes a lowest allowed power of  $r$ in the outer expansion, so it determines a highest power of $r$ in the inner expansion: expanding the $n$th-order inner perturbation $\e^n H^n_{\mu\nu}(t,\bar x^i)$ in the buffer region, we find $\e^n \sum_p(\e/r)^p H^{np}_{\mu\nu}(t,n^i)$; the condition that no negative powers of $\e$ occur in the outer expansion then determines
\beq\label{H-generic}
H^n_{\mu\nu}(t,\bar x^i)= \sum_{p\leq n}\bar r^p H^{np}_{\mu\nu}(t,n^i).
\eeq
Here and in what follows, I assume some locally Cartesian coordinates centered on the center-of-mass worldline $z^\mu$. Given an appropriate choice of rest gauge, the coordinates will later be identified with Fermi-Walker coordinates centered on $z^\mu$.

For our approximately spherical, approximately nonspinning object, the inner background metric, when expanded in the buffer region according to Eq.~\eqref{gin_buffer}, is equal to the Schwarzschild metric through order $1/\bar r^4$. It has the schematic form
\beq\label{gin_expansion-rest}
\gin\sim \eta+ \frac{m}{\bar r}+\frac{m^2}{\bar r^2}+\frac{m^3}{\bar r^3}+\O(1/\bar r^4);
\eeq
here I do not specify any particular coordinate system, though I insist that the coordinates are mass-centered, such that no mass dipole moment appears in the expansion of $\gin$; every term in Eq.~\eqref{gin_expansion-rest}, at the orders displayed, is fully determined by $m$. Using our previous results, we know $m$ is constant. Hence, we can find the perturbations $H^n_{\mu\nu}$ in the buffer region using the well-developed formalism of perturbations of Schwarzschild. Furthermore, given the assumption that the perturbations are quasistationary (because, as explained in Sec.~\ref{matched expansions}, the inner expansion scales space but not time), time derivatives appear as higher-order terms in the Einstein equations. That is, one first solves the time-independent linearized Einstein equation 
\beq\label{inner-1}
\delta R^0_{\mu\nu}[\gin, H^1]=0 
\eeq
for $H^1_{\mu\nu}$, with $t$ fixed; here $\delta R_{\mu\nu}[\gin,H^1]=0$ means the Ricci tensor linearized off the background $\gin_{\mu\nu}$, and the superscript ``0'' means ``terms in $\delta R_{\mu\nu}$ containing no $t$ derivatives''. Next, one solves 
\beq\label{inner-2}
\delta R^0_{\mu\nu}[\gin, H^2]=-\delta^2 R^0_{\mu\nu}[\gin, H^1,H^1]-\delta R^1_{\mu\nu}[\gin, H^1]
\eeq
for $H^2_{\mu\nu}$ at fixed $t$ and for the time derivative of $H^1_{\mu\nu}$, where the superscript ``1'' in $\delta R^1_{\mu\nu}$ means ``terms in $\delta R_{\mu\nu}$ containing one $t$ derivative''.  
At this point, the pattern  should be obvious.

Just as when constructing the outer expansion in the buffer region, we here solve a sequence of spatial differential equations. Furthermore, it is well known that in an appropriate gauge, stationary vacuum perturbations of Schwarzschild behave as $1/{\bar r}^{\ell+1}$ or $\bar r^\ell$ at large $\bar r$, where $\ell$ is the spherical harmonic index, just as in the outer expansion. Let us label the $1/r^{\ell+1}$ solutions as $H^{n,\ell,-}_{\mu\nu}(t,\bar x^i)$ and the $r^\ell$ solutions as $H^{n,\ell,+}_{\mu\nu}(t,\bar x^i)$. Combining this with Eq.~\eqref{H-generic}, we find that at first order, where there is no source, the solution behaves as
\beq
H^1\sim \bar r H^{1,1,+}+ \bar r^0 H^{1,0,+}  + \frac{1}{\bar r}H^{1,0,-} + \frac{1}{\bar r^2}H^{1,1,-}+\O(1/\bar r^3).
\eeq
It is well known that the monopole and dipole solutions $r^0H^{n,0,+}_{\mu\nu}$ and $rH^{n,0,+}_{\mu\nu}$ are pure gauge. Hence, I set them to zero. The gauge-invariant content in $\frac{\e^2}{r}H^{1,0,-}_{\mu\nu}$ is a correction to $m$, and in $\frac{\e^3}{r^2}H^{1,1,-}_{\mu\nu}$ it is a perturbation toward the Kerr metric---that is, a small spin. With a bit of work, one can show from Eq.~\eqref{inner-2} that these perturbations are independent of $t$, and we can freely set them to zero, absorbing the mass correction into $m$ and specifying that the object remains unspinning at this order. Physically, this time-independence corresponds to the fact that physical changes in mass and angular momentum are caused by tidal heating and torquing, which are caused by nonlinear effects at much higher order in $\e$. So in the end, we have
\beq\label{H1}
H^1_{\mu\nu}=0.
\eeq

Since $H^1_{\mu\nu}$ vanishes, Eq.~\eqref{inner-2} is again the linearized vacuum equation: $\delta R^0_{\mu\nu}[\gin, H^2]=0$. The solution behaves as
\beq
H^2\sim \bar r^2 H^{2,2,+} + \bar r H^{2,1,+}+ \bar r^0 H^{2,0,+} + \frac{1}{\bar r}H^{2,0,-}+\O(1/\bar r^2).
\eeq
Again, $H^{2,1,+}_{\mu\nu}$ and $H^{2,0,+}_{\mu\nu}$ can be gauged away. The invariant part of $H^{2,0,-}_{\mu\nu}$ is another correction to $m$, which again can be absorbed into a redefinition of $m$. Writing $H^{2,2,+}_{\mu\nu}=H^{2,2,+}_{\mu\nu ij}\nhat^{ij}$, we then have
\beq
H^2_{\mu\nu}\sim \bar r^2 H^{2,2,+}_{\mu\nu ij}(t,\bar r)\nhat^{ij} + \O(1/\bar r^2).
\eeq
$\lim_{\bar r\to\infty}H^{2,2,+}_{\mu\nu ij}(t,\bar r)$ can be written in terms of two STF functions $\E_{ij}(t)$ and $\B_{ij}(t)$, and in the expansion of $H^{2,2,+}_{\mu\nu ij}(t,\bar r)$ for large $r$, every term is directly proportional to one of those two functions (i.e., constructed from contractions of one of them with $\delta_{ij}$ or $\epsilon_{ijk}$). These two quantities can be interpreted as quadrupole tidal moments in the neighbourhood of the object, and from the matching condition they will be identified with those of the external background metric, defined in Eq.~\eqref{tides}.

The third-order Einstein equation now reads
\beq
\delta R^0_{\mu\nu}[\gin, H^3] = -\delta R^1_{\mu\nu}[\gin, H^2].
\eeq
Following the same steps as at the previous orders, we find
\beq
H^3_{\mu\nu}\sim \bar r^3 H^{3,3,+}_{\mu\nu ijk}(t,\bar r)\nhat^{ijk}+\bar r^3\dot H^{2,2,+}_{\mu\nu ij}(t,\bar r)\nhat^{ij}+\bar r^2 H^{3,2,+}_{\mu\nu ij}(t,\bar r)\nhat^{ij} + \O(1/\bar r).\label{H3}
\eeq
$H^{3,3,+}_{\mu\nu ijk}(t,\bar r)$ can be written in terms of two STF functions $\E_{ijk}(t)$ and $\B_{ijk}(t)$, such that every term in its large-$\bar r$ expansion is directly proportional to one or the other. These two quantities can be interpreted as octupole tidal moments in the neighbourhood of the object, and they will be identified with those of the external background metric, which are constructed from covariant derivatives of the Riemann tensor of $g_{\mu\nu}$. Equation~\eqref{H3} additionally contains corrections to the quadrupole fields, both in the form of the time derivatives $\dot\E_{ij}$  and $\dot\B_{ij}$ appearing in $\dot H^{2,2,+}_{\mu\nu ij}$, and in the field $H^{3,2,+}_{\mu\nu ij}$, which like $H^{2,2,+}_{\mu\nu ij}$ can be written in terms of two STF functions, call them $\delta\E_{ij}$ and $\delta\B_{ij}$.

Now let us put the results together. After we rewrite it in terms of unscaled coordinates $r=\e\bar r$ and re-expand in $\e$, the metric ${\sf g}_{\mu\nu}=\gin_{\mu\nu}(\bar r)+\e^2H^2_{\mu\nu}(\bar r)+\e^3 H^3_{\mu\nu}(\bar r)+\O(\e^4)$ reads ${\sf g}_{\mu\nu}=g_{\mu\nu}(r)+\e h'^1_{\mu\nu}(r)+\e^2 h'^2_{\mu\nu}(r)+\e^3 h'^3_{\mu\nu}(r)+\O(\e^4)$, with
\beq\label{metric-rest}\begin{array}{rclclclc}
g &=& \eta & +& r^2\E_{ij} &+& r^3(\E_{ijk}+ \dot \E_{ij})&+\O(r^4)\\
h'^1 &\sim& \frac{m}{r} & +& mr\E_{ij} +r^2 \delta\E_{ij} &+& mr^2(\E_{ijk}+ \dot \E_{ij})&+\O(r^3),\\
h'^2 &\sim& \frac{m^2}{r^2} & +& m^2\E_{ij} + mr \delta\E_{ij} &+& m^2r(\E_{ijk}+ \dot \E_{ij})&+ \O(r^2),\\
h'^3 &\sim& \frac{m^3}{r^3} & +& \frac{m^3}{r}\E_{ij} +m^2 \delta\E_{ij}&+&m^3(\E_{ijk}+ \dot \E_{ij})&+ \O(r),
\end{array}
\eeq
where primes refer to quantities in the rest gauge, and $\eta={\rm diag}(-1,1,1,1)$. `+' signs here mean ``plus terms directly proportional to''. For compactness, I have omitted the magnetic-type moments $\B_{ij}$, $\dot \B_{ij}$, $\delta \B_{ij}$, and $\B_{ijk}$; they appear in exact analogy with the electric type moments $\E_{ij}$, $\dot \E_{ij}$, $\delta\E_{ij}$, and $\E_{ijk}$. All terms in the tableau, at the displayed orders, are directly proportional to one of these moments and/or to a power of $m$. Every term in the first column is $\ell=0$, every term in the second is $\ell=2$, and every term in the third is either $\ell=2$ (the $\dot \E_{ij}$ and $\dot \B_{ij}$ terms) or $\ell=3$ (the $\E_{ijk}$ and $\B_{ijk}$ terms). In this gauge, the object is manifestly at rest on $z^\mu$: there are no mass dipole moment terms, which have the even-parity dipolar behavior $n^i/r^2$, nor any acceleration terms, which have the even-parity dipolar behavior $r n^i$.

By choosing the gauge of the inner expansion appropriately, we can put the external background metric $g_{\mu\nu}$, given by the top row of Eq.~\eqref{metric-rest}, in precisely the form~\eqref{g-Fermi} that it takes in Fermi-Walker coordinates---except that no acceleration terms appear. Since we know that the center-of-mass worldline is accelerated in $g_{\mu\nu}$, this tells us that Eq.~\eqref{metric-rest} is actually incomplete. \emph{The construction we have followed here is valid for any worldline $z^\mu$, regardless of whether it is accelerating in $g_{\mu\nu}$ or not.} The fact that we see no explicit acceleration terms is equivalent to the empty statement that any worldline $z^\mu$ can be written as a geodesic in \emph{some} smooth piece of the full metric. Ergo, at this stage we know nothing at all about $z^\mu$, and to make $g_{\mu\nu}$ the external background spacetime, we must insert into it by hand the acceleration terms appearing in Eq.~\eqref{g-Fermi}. At the same time, in our choice of gauge, no acceleration terms appear in the full metric ${\sf g}_{\mu\nu}$. Therefore, we must insert $F_n^\mu$ terms into the perturbations $h'^n_{\mu\nu}$ such that the sum  $g_{\mu\nu}(r)+\e h'^1_{\mu\nu}(r)+\e^2 h'^2_{\mu\nu}(r)+\e^3 h'^3_{\mu\nu}(r)+\O(\e^4)$ contains no acceleration terms when one applies the expansion~\eqref{SC_F_expansion}.

The form of ${\sf g}_{\mu\nu}$ in the rest gauge invites us to make a new split into an effectively external metric $\varext_{\mu\nu}= g_{\mu\nu}+\varh^{\R}_{\mu\nu}$ that contains the $m$-independent tidal terms in Eq.~\eqref{metric-rest}, plus a self-field $\varh^\S_{\mu\nu}={\sf g}_{\mu\nu}-\varext_{\mu\nu}$, every term of which is directly proportional to a power of $m$. With an appropriate choice of rest gauge, $\varext_{\mu\nu}$ looks exactly like Eq.~\eqref{g-Fermi} with none of the acceleration terms; that is, our gauge makes the Fermi coordinates refer to proper distances, times, Riemann curvature, and parallel transport defined with respect to $\varext_{\mu\nu}$. We have
\begin{subequations}\label{ext-Fermi}%
\begin{align}
\varext_{tt} &= -1-\mathfrak{R}_{0i0j}x^ix^j+\O(r^3),\\
\varext_{ta} &= -\tfrac{2}{3}\mathfrak{R}_{0iaj}x^ix^j+\O(r^3),\\
\varext_{ab} &= \delta_{ab}-\tfrac{1}{3}\mathfrak{R}_{aibj}x^ix^j+\O(r^3),
\end{align}
\end{subequations}
where $\mathfrak{R}_{\mu\alpha\nu\beta}=R_{\mu\alpha\nu\beta}+\e\delta R_{\mu\alpha\nu\beta}[\varh^{\R1}]+\O(\e^2)$ is the Riemann tensor of $\varext_{\mu\nu}$. As per the discussion of acceleration terms in the preceding paragraph, the new ``regular field'' reads
\begin{subequations}\label{varhR1-rest}%
\begin{align}
\varh^{\R1}_{tt} &= 2F^i_1n_i-\delta R_{0i0j}[\varh^\R]x^ix^j+\O(r^3),\\
\varh^{\R1}_{t a} &= -\tfrac{2}{3}\delta R_{0iaj}[\varh^\R]x^ix^j+\O(r^3),\\
\varh^{\R1}_{ab} &= -\tfrac{1}{3}\delta R_{aibj}[\varh^\R]x^ix^j+\O(r^3),
\end{align}\vspace{-15pt}
\end{subequations} 
\beq
\varh^{\R2}_{tt} = 2F^i_2n_i+\O(r^2),\qquad \varh^{\R2}_{t a} = \varh^{\R2}_{ab} =\O(r^2),\label{varhR2-rest}
\eeq
and $\varh^{\R3}_{\mu\nu}=\O(r)$. For visual clarity, I have truncated these equations at lower orders in $r$ than Eq.~\eqref{metric-rest}.

In summary, in a rest gauge the metric perturbations take the form
\beq\label{h-rest}
h'^n_{\mu\nu} = \varh^{\S n}_{\mu\nu}+\varh^{\R n}_{\mu\nu}, 
\eeq
where $\varh^{\S n}_{\mu\nu}$ is made up entirely of terms containing explicit factors of $m^n$, no mass monopole corrections, no spin moment, no quadrupole moment, no mass dipole moment, and no acceleration terms, and $\varh^{\R n}_{\mu\nu}$ is given by Eq.~\eqref{varhR1-rest}--\eqref{varhR2-rest}.
Equation~\eqref{h-rest} is the \emph{general solution in the buffer region outside an approximately spherical and nonspinning object}, just written in a particular gauge. In this gauge, $z^\mu$ is manifestly a geodesic of an effectively external metric, and $\varext_{\mu\nu}$ is being \emph{defined} as that effectively external metric in which $z^\mu$ is a geodesic.\footnote{Note that without some additional input beyond that definition, the split of ${\sf g}_{\mu\nu}$ into $g_{\mu\nu}$, $\varh^\R_{\mu\nu}$, and $\varh^\S_{\mu\nu}$ is quite ambiguous. Suppose there were a zeroth-order force acting on the object. One could still write the equation of motion as a geodesic in some smooth piece of the metric, but to do so, one would have to shift part of $g_{\mu\nu}$ into $\varh^\S_{\mu\nu}$; one would not simply be splitting the perturbations $h'^n_{\mu\nu}$. With the present setup, the ambiguity is lifted by assuming the expansion~\eqref{SC_F_expansion} and utilizing the independently determined fact that $F_0^\mu=0$. } A priori, \emph{$\varext_{\mu\nu}$ need not be related to $\tilde g_{\mu\nu}$}; the latter was defined in a very different manner and is not yet known to be the metric in which the motion is geodesic (except at zeroth and first order).

\subsection{Self-consistent approximation: worldline-preserving gauge transformations}
Metrics in local rest gauges are common in the literature, usually derived in the context of tidally perturbed compact objects. 
 The goal of the self-force game is not simply to construct such a metric, but to find the equation of motion in whichever ``practical gauge'' one wants to use to compute a global solution.  

Let us take the Lorenz gauge as the practical gauge. I wish to find a unique gauge transformation between the rest-gauge perturbations \eqref{h-rest} and the Lorenz-gauge perturbations~\eqref{hS1}--\eqref{hR2}. 
That is, I seek gauge vectors $\xi_1^\mu$ and $\xi_2^\mu$ satisfying
\begin{align}
\Lie_{\xi_1}g_{\mu\nu} & = h'^1_{\mu\nu} -h^1_{\mu\nu},\label{to-rest1}\\
\Lie_{\xi_2}g_{\mu\nu} & = h'^2_{\mu\nu} -h^2_{\mu\nu}-\tfrac{1}{2}\Lie^2_{\xi_1}g_{\mu\nu}-\Lie_{\xi_1}h^1_{\mu\nu}.\label{to-rest2}
\end{align}
Note that here I do not utilize the third-order perturbation, though its form $h'^3_{\mu\nu}$ was found in the rest-frame gauge. What is its relevance? Only that it contains no mass dipole moment term; in other words, it shows that with this choice of gauge, the correction to the mass dipole moment can unambiguously be set to zero to define a center-of-mass-worldline.

These equations are to be solved subject to one crucial condition: \emph{they must preserve the location of the worldline.} The rest-gauge solution shows that the object is at rest on some worldline, the Lorenz-gauge solution is written in coordinates centered on some worldline, and in order to claim that the worldline of the Lorenz-gauge solution is the desired center-of-mass worldline, we must enforce that it be identical to the worldline of the rest-gauge solution. In the simplest scenario, 
Eqs.~\eqref{to-rest1}--\eqref{to-rest2} can be solved with a smooth gauge transformation of the form
\beq
\xi_n^\mu = \sum_{p\geq0}\sum_{\ell\leq p}r^p\xi^{\mu L}_{np\ell}(t)\nhat_L.
\eeq
Given Eq.~\eqref{z_transformation}, the condition that the worldline not be altered is that
\beq
\xi^{\mu}_{n00}(t)=\xi^\mu_n\big|_{\gamma} = 0.
\eeq
I call a transformation satisfying this condition \emph{a worldline-preserving transformation}. This condition was first used in Ref.~\cite{Mino-Sasaki-Tanaka:97}. A more general condition is used in Refs.~\cite{Pound:12a,Pound:14a}. 

By working through sequential orders of $r$ in Eqs.~\eqref{to-rest1}--\eqref{to-rest2}, one uniquely determines $\xi^{\mu L}_{np\ell}(t)$ (up to residual gauge freedom in the rest gauge and Lorenz gauge). I do not present details of that process here, but the explicit solution to the first-order equation~\eqref{to-rest1}, with particular choices of rest gauge, can be found in various references; see, e.g., Ref.~\cite{Pound:10b}. The explicit solution at second order is presented in Ref.~\cite{Pound:14a}. What is the essential result in these solutions? At first order, they uniquely determine $F_1^\mu$, which appears via Eq.~\eqref{varhR1-rest}, to be the MiSaTaQuWa force~\eqref{F1_free} (with $S^i=0$), as we already know. At second-order, they uniquely determine the second-order force $F^\mu_2$, which appears via Eq.~\eqref{varhR2-rest}, to be
\beq
F_i^2 = \frac{1}{2}\partial_i h^{{\rm R}2}_{tt} - \partial_t h^{{\rm R}2}_{ti} - \frac{11}{3}m\dot F^1_i 
	+  F^1_i h^{{\rm R}1}_{tt} - \frac{1}{2}h^{{\rm R}1}_{ti}\partial_th^{{\rm R}1}_{tt}.
\eeq

Following Appendix~\ref{geodesic_expansion_in_h}, the results for $F^\mu_1$ and $F^\mu_2$ can be combined to write the equation of motion $\frac{D^2 z^\mu}{d\tau^2}=\e F^\mu_1+\e^2 F^\mu_2+\O(\e^3)$ as the geodesic equation $\frac{\tilde D^2 z^\mu}{d\tilde \tau^2}=\O(\e^3)$ in $\tilde g_{\mu\nu}=g_{\mu\nu}+h^\R_{\mu\nu}$. This is the result promised way back in Sec.~\ref{SC-approx}.

What is the meaning of this result? In the rest gauge, we knew that $z^\mu$ was a geodesic in $\varext_{\mu\nu}$. Now we know it is a geodesic in $\tilde g_{\mu\nu}$. Either of $\varext_{\mu\nu}$ or $\tilde g_{\mu\nu}$ can be thought of as a ``nice'' effectively external metric. Are they the same metric? Not necessarily. The two geodesic equations only restrict the metrics $\varext_{\mu\nu}$ and $\tilde g_{\mu\nu}$ on the worldline and their first derivatives on the worldline; they place no constraint on higher derivatives. But one of the results of the process of finding $\xi_1^\mu$ and $\xi_2^\mu$ is that the two metrics are related by 
\begin{align}
\varh^{\R1}_{\mu\nu}&=h^{\R1}_{\mu\nu}+\Lie_{\xi_1}g_{\mu\nu}+\O(r^2)\\
\varh^{\R2}_{\mu\nu}&=h^{\R2}_{\mu\nu}+\Lie_{\xi_2}g_{\mu\nu} +\tfrac{1}{2}\Lie^2_{\xi_1}g_{\mu\nu}+\Lie_{\xi_1}h^{\R1}_{\mu\nu}+\O(r^2).
\end{align}
In other words, they are equivalent up to possible $\O(r^2)$ differences. Are they the same at order $r^2$ and higher? The answer depends on precisely how the rest gauge, and the tidal moments $\delta\E_{ij}$, $\delta\B_{ij}$ (and at higher order in $r$, octupolar and higher order moments) appearing in $\varh^{\R1}_{\mu\nu}$, is constructed; this construction is, unfortunately, not unique. Reference~\cite{Pound:14a} demonstrates explicitly that the two effective fields can differ at order $r^2$. 

Is there a way to determine which effective field is the ``best one''? As described in Sec.~\ref{singular-regular}, the most promising route is by solving the field equations outside an object with spin and higher moments, since those moments will couple to \emph{some} tidal moments, and we can determine \emph{which} metric the tidal moments belong to, and thence the metric in which the object moves as a test body. But just with the results at hand, we can confidently say that $\tilde g_{\mu\nu}$ has all the nice properties one might desire of it for a spherical, nonspinning object: it is a vacuum solution, causal on the worldline, and the object moves on a geodesic of its geometry.



\subsection{Gralla-Wald approximation}
Starting from the self-consistent results, one can readily derive the second-order equation of motion~\eqref{GW_ddotz2} following the expansion procedure of Appendix~\ref{geodesic_expansion_in_h_and_dz}.

Alternatively, one can derive a second-order equation of motion directly, as was done by Gralla~\cite{Gralla:12}. In the Gralla-Wald case, the rest gauge is constructed in coordinates centered not on $z^\mu$ but on $z_0^\mu$. The results are the same as in Sec.~\ref{rest gauge}, but all acceleration terms are set to zero in $g_{\mu\nu}$ (and likewise for the $F_n^\mu$ terms in $h'^n_{\mu\nu}$ that cancel them). 
When transforming to a ``practical gauge'', one uses
\begin{align}
\check{h}^1_{\mu\nu} &= \check{h}'^1_{\mu\nu} + \Lie_{\xi_1}g_{\mu\nu},\\
\check{h}^2_{\mu\nu} &= \check{h}'^2_{\mu\nu} + \Lie_{\xi_2}g_{\mu\nu} +\tfrac{1}{2}\Lie^2_{\xi_1}g_{\mu\nu}+\Lie_{\xi_1}\check{h}^1_{\mu\nu},
\end{align}
and \emph{one does not impose the worldline-preserving condition}. The gauge transformation is allowed to move the object relative to $z_0^\mu$. Since $z'^\mu_1=z'^\mu_2=0$ in the rest gauge, we have from Eq.~\eqref{DzGW} that in the practical gauge
\begin{align}
z_1^{\mu}(s) &= - \xi_1^\mu(z_0),\\
z_2^{\mu}(s) &= - \xi_2^\mu(z_0)+\frac{1}{2}\xi^\nu_1(z_0)\partial_\nu\xi_1^\mu(z_0),
\end{align}
and the covariant second deviation is $z_{2F}^{\mu}(s) = - \xi_2^\mu(z_0)+\frac{1}{2}\xi^\nu_1(z_0)\nabla_\nu\xi_1^\mu(z_0)$.

Following Gralla, one can use this procedure to obtain results in any gauge smoothly related to a particular rest gauge. Defining regular fields
\begin{align}
\check{h}^{\R1}_{\mu\nu} &= \check{\varh}^{\R1}_{\mu\nu} + \Lie_{\xi_1}g_{\mu\nu},\\
\check{h}^{\R2}_{\mu\nu} &= \check{\varh}^{\R2}_{\mu\nu} + \Lie_{\xi_2}g_{\mu\nu} +\tfrac{1}{2}\Lie^2_{\xi_1}g_{\mu\nu}+\Lie_{\xi_1}\check{\varh}^{\R1}_{\mu\nu},
\end{align}
one finds evolution equations for $z^\mu_n$ in terms of $\check{h}^{\R n}_{\mu\nu}$ by taking appropriate derivatives of these relations; any desired gauge condition is imposed on $\check{h}^{\R n}_{\mu\nu}$, $\check{h}^{\R n}_{\mu\nu}$ is computed via a puncture scheme, and $z^\mu_n$ is computed from $\check{h}^{\R n}_{\mu\nu}$. Because of the particular method of construction, in which the rest gauge is defined with respect to $z_0^\mu$ rather than $z^\mu$, the evolution equations for $z^\mu_n$ do not correspond to an expansion of the geodesic equation in the metric $\varext_{\mu\nu}=g_{\mu\nu}+\varh^{\R}_{\mu\nu}$ that Gralla defines; this lack of geodesic motion in $g_{\mu\nu}+\varh^{\R}_{\mu\nu}$ should amount to a slightly different definition of $\varh^{\R}_{\mu\nu}$ than the one used in the self-consistent expansion, although no one has yet carried out a detailed comparison of Gralla's second-order results to those of the self-consistent method.



																					\section{Conclusion}\label{conclusion}


\subsection{Summary}

In this review, we began with the idea of a point particle interacting with its own backscattered field; we ended with the idea of using laws of gauge transformation to obtain laws of motion for extended objects. And yet, reassuringly, from the latter description we have recovered the former. 

Along the circuitous way, I have emphasized several core ideas. First, the laws of motion of an extended object, to all orders in perturbation theory, are determinable from the metric \emph{outside} the object; all information about the object's shape and internal composition is encoded in a discrete set of multipole moments, themselves defined from the form of the metric in the buffer region outside the object. To obtain an equation of motion of order $\e^n$ in perturbation theory, moments of multipole order $\ell=n$ are required. Second, the laws of motion found in this way show that at least through second order in perturbation theory, a small, sufficiently spherical, sufficiently slowly spinning compact object moves on a geodesic of an effectively external metric that (i) satisfies the vacuum Einstein equations and (ii) behaves causally on the object's representative worldline. Furthermore, the analysis of the metric outside the object provides a clean way of separating the metric into two constituents: a suitable effectively external metric satisfying the above properties, and a self-field that, loosely speaking, locally characterizes the object. 

This treatment weds several traditions that run through the history of the problem of motion in general relativity, from the characterization of motion of a material body in terms of suitably defined multipole moments in the tradition of Mathisson~\cite{Mathisson:37} and Dixon~\cite{Dixon:74}, to the method of algorithmically characterizing the metric in terms of multipole moments in the multipolar post-Minkowskian theory of Blanchet and Damour~\cite{Blanchet-Damour:86,Blanchet:14}, to the derivation of laws of motion of asymptotically small objects from the Einstein equation outside the objects in the tradition of Einstein~\cite{Einstein-Infeld:49}, D'Eath~\cite{DEath:75}, and Thorne and Hartle~\cite{Thorne-Hartle:85}. The perturbative treatment here also complements the non-perturbative treatment of Harte~\cite{Harte:12}. Just as Harte brought Dixon's project to fruition by finding a utile separation of the metric into a self-field and an effectively external metric, so the same can be said about the method here bringing to fruition the project of Thorne and Hartle.

An important aspect of any perturbative treatment of motion is the relationship between that motion and gauge freedom. It is well known that on short timescales, the self-force is pure gauge; it can be freely set to zero by a suitable choice of gauge. Equivalently, one can say that on short timescales, the perturbed worldline of the object can be transformed to a geodesic of the background spacetime.

What physical role, then, does the self-force play? On short timescales, one can say the following: the deviation from geodesic motion appears explicitly in the second-order field. Hence, in order to get gauge-invariant information about the second-order field from the solution to the Einstein equation in any given gauge, one must include the effect of the self-force. (Note that the argument doesn't run in reverse. Since the first-order deviation is determined entirely by the first-order gauge choice, one does not need the entire second order perturbation to obtain invariant information involving the first-order deviation~\cite{Barack-Sago:11}.) However, on these short timescales, the perturbative correction to the motion need not be accounted for to obtain physical information from the first-order metric.

As I have argued, this situation changes when one considers long scales. If one seeks a solution valid on a large domain, such as over the course of a binary inspiral, then one \emph{must} account for the correction to the motion in order to obtain a well-defined perturbation theory. If one were to attempt to put the perturbative shift in the motion into the second-order field, then on this timescale the second-order field would grow larger than the first-order field. Hence, one must incorporate the effect on the motion into the first-order perturbation. In this sense, the perturbed worldline itself is quasi-invariant under a gauge transformation on long timescales; it may only be shifted by a small amount of order $\e$, while the shift due to the self-force is a much larger effect, of order $\sim 1/\e$ (or $\sim \e^0$, depending on whether one is looking at orbital phase or orbital radius in an inspiral, for example). 

For this same reason, if one is interested in long timescales, one need not include the second-order perturbation in order to obtain gauge-invariant information about an inspiral. For any given short patch of that inspiral, the effects of the self-force will be pure gauge; but the \emph{accumulation} of those effects over long timescales is invariant within the class of gauges that are well behaved on these timescales.

\subsection{Future directions}

Although the foundations seem in place and the second-order results seem sufficient for the practical purposes we can presently imagine, several open avenues remain to be explored. 

First, at present we have obtained second-order equations only for objects whose spin and quadrupole moments vanish at leading order. It would be worthwhile, and astrophysically relevant, to obtain second-order equations of motion for more generic compact objects. 
As is well known~\cite{Dixon:74}, the object's leading order quadrupole moment would generate a force by coupling to the external curvature. Additionally, there may be an effect due the correction $\delta S^{i}$ to the object's spin. 

The mention of this subleading spin raises another question to consider: what is the physical content of the perturbed multipole moments? I have discussed above how the monopole correction $\delta m_{\mu\nu}$ is pure gauge (at least on short timescales). 
Will higher corrections, or corrections to higher moments, contain more physical information than this? We can expect that they will: tidal heating and torquing, for example, should create physical corrections to the mass and spin~\cite{Poisson:04b,Chatziioannou-Poisson-Yunes:12}, and at least for a material body, the external tidal fields should correct the higher moments by physically deforming the object~\cite{Damour-Nagar:09,Binnington-Poisson:09,Landry-Poisson:14}. Ultimately, to model the motion of a particular class of objects, such as realistic neutron stars, one will need to explicitly match an inner expansion to the outer expansion I have discussed; the matching procedure will uniquely identify the multipole moments of the objects as they appear in the metric in the buffer region. To this end, one might consider metrics of tidally perturbed neutron stars, such as those described in Refs.~\cite{Damour-Nagar:09,Binnington-Poisson:09}.

One might also try to describe less compact astrophysical objects, or more exotic objects, by broadening the scope of the perturbative expansion. Rather than assuming the object is compact, such that its linear dimension $d$ is of the same order as its mass $m$, one might consider a two-parameter family corresponding to $m$ and $d$. This would alter the orders at which various multipole moments appear in the buffer region, since they scale as $md^\ell$. One could even try to describe approximately string-like objects by examining a limit in which two of the object's linear dimensions go to zero while one linear dimension remains finite.

To gain further insight into the physical content of the multipole moments, one could also relate them to the moments appearing in other formalisms, such as the non-perturbative formalism of Harte~\cite{Harte:12}. Another point of comparison with Harte would be the definition of the effectively external (i.e., the ``regular'') field. As I have discussed, the choice of effectively external field is far from unique, and in fact an infinite number of choices could be made that would still guarantee the center-of-mass motion is geodesic in the effective metric. Here, I have taken as my guideline that the effective metric should be a vacuum solution that is causal on the worldline and in which the motion is geodesic; these conditions still do not uniquely identify an effective field, but they narrow the range of options, and I have shown how a choice satisfying these conditions arises naturally in the process of solving the Einstein equation in the buffer region outside the object. However, Harte makes a different choice, which does not satisfy the vacuum Einstein equation. How are the two related? The answer is not obvious. One way of establishing a stronger relationship would be to find perturbative equations of motion for objects with higher multipole moments: because moments of higher $\ell$ couple to higher derivatives of the metric, they feel much more of the field than does a monopole. With Harte's definition of effective field, the equation of motion is precisely that of a test body in the effective metric. Would the same be true in the effective metric I have defined here?

In addition to comparison with Harte, the results I have described should be related to Gralla's more closely related, perturbative results~\cite{Gralla:12}. At second order, Gralla makes a choice of regular field that \emph{is} a vacuum solution (at least to the order in $r$ to which he defines it), but in which the center-of-mass motion is \emph{not} geodesic. 

Although I have emphasized the idea of splitting the physical metric into a self-field and an ``effectively external" metric, and although I have linked that split to a generalized equivalence principle, the array of choices of regular field should make clear that there is a danger of over-interpreting the physical meaning of any particular choice, no matter how nice its properties. This risk is also present because in general, a regular field satisfying nice properties on the object's worldline is acausal when evaluated away from the worldline. As a practical matter, the freedom to define different effective metrics may make future comparisons of self-force results to post-Newtonian results more hairy; given the vast freedom---even while maintaining the property that the motion is geodesic in the effective metric, for example---it is remarkable and fortuitous that agreement has so far been found for so many different effects that appear to rely on particular choices of this field~\cite{Detweiler:08,Dolan-etal:13,Dolan-etal:14}.

\section*{Acknowledgements}
I thank Leor Barack, Eric Poisson, and Abraham Harte for thought-provoking discussions that helped shape my thinking on self-force theory. This work received funding from the European Research Council under the European Union's Seventh Framework Programme (FP7/2007-2013)/ERC Grant No. 304978.

\appendix
\section{Expansions of the geodesic equation}\label{geodesic_equation_expansion}
\subsection{Expansion in powers of a metric perturbation}\label{geodesic_expansion_in_h}
In this appendix, I examine the expansion of the geodesic equation in any sufficiently smooth metric ${\sf g}_{\mu\nu}$; the treatment is generic, not specialized to a spacetime containing a small object. I expand only in powers of a sufficiently smooth metric perturbation; I do not expand the worldline itself. Hence, the analysis is meant to apply to the self-consistent representation of motion, not to the Gralla-Wald representation. 

The geodesic equation reads 
\beq
\frac{d\dot z^\mu}{ds}+{}^{\sf g}\Gamma^\mu_{\nu\rho}\dot z^\nu \dot z^\rho=\kappa \dot z^\mu,
\eeq
where $s$ is a potentially non-affine parameter on the curve $z^\mu(s)$, $\dot z^\mu\equiv\frac{dz^\mu}{ds}$ is its tangent vector field, ${}^{\sf g}\Gamma^\mu_{\nu\rho}$ is the Christoffel symbol corresponding to ${\sf g}_{\mu\nu}$, and $\kappa=\frac{d}{ds}\ln\sqrt{-{\sf g}_{\mu\nu}\dot z^\mu \dot z^\nu}$. 

If we now write the metric as the sum of two pieces, ${\sf g}_{\mu\nu}=g_{\mu\nu}+h_{\mu\nu}$, and if we take $s=\tau$, the proper time on $z^\mu$ in $g_{\mu\nu}$, and if we rewrite the geodesic equation in terms of covariant derivatives compatible with $g_{\mu\nu}$, we find
\begin{equation}
a^\mu = -\Delta \Gamma^\mu_{\nu\rho}u^\nu u^\rho+\kappa u^\mu,\label{exact_geodesic}
\end{equation}
where $a^\mu\equiv \frac{D^2z^\mu}{d\tau^2}$, $u^\mu\equiv\frac{dz^\mu}{d\tau}$, and 
\begin{align}
\Delta \Gamma^\alpha_{\beta\gamma} &\equiv {}^{\sf g}\Gamma^\alpha_{\beta\gamma}-\Gamma^\alpha_{\beta\gamma}
			= \frac{1}{2}{\sf g}^{\alpha\delta}\left(2h_{\delta(\beta;\gamma)}-h_{\beta\gamma;\delta}\right)
\end{align}
is the difference between the Christoffel symbol associated with the full metric ${\sf g}_{\mu\nu}$ and that associated with the background metric $g_{\mu\nu}$. With $\tau$ as a parameter, $\kappa$ becomes
\begin{equation}
\kappa = \frac{\frac{d}{d\tau}\sqrt{1-h_{\mu\nu}u^\mu u^\nu}}{\sqrt{1-h_{\mu\nu}u^\mu u^\nu}}.
\end{equation}

So far no approximation has been made; Eq.~\eqref{exact_geodesic} is exact. If we now expand $\Delta\Gamma^\mu_{\nu\rho}$ and $\kappa$ in powers of $h_{\mu\nu}$, we find
\begin{align}
a^\alpha &= -\frac{1}{2}(g^{\alpha\delta}-h^{\alpha\delta})\!\left(2h_{\delta(\beta;\gamma)}-h_{\beta\gamma;\delta}\right)\!u^\beta u^\gamma
	-\frac{1}{2}h_{\beta\gamma;\delta}u^\alpha u^\beta u^\gamma u^\delta\nonumber\\
		&\quad	-\frac{1}{2}h_{\mu\nu}h_{\beta\gamma;\delta}u^\alpha u^\beta u^\gamma u^\delta u^\mu u^\nu 
	-h_{\beta\gamma}u^\alpha a^\beta u^\gamma+\O(h^3). \label{expanded_geodesic}
\end{align}
This equation is complicated by the fact that the acceleration appears on both sides in a nontrivial way. To disentangle the acceleration from the perturbation, I assume that $a^\mu$, too, has an expansion in powers of $h_{\mu\nu}$,
\begin{equation}
a^\mu = a^\mu_{\rm lin}+a^\mu_{\rm quad}+\O(h^3),
\end{equation}
where $a^\mu_{\rm lin}$ is linear in $h_{\mu\nu}$ and $a^\mu_{\rm quad}$ is quadratic in it. Substituting this expansion into Eq.~\eqref{expanded_geodesic}, one finds
\begin{align}
a^\alpha_{\rm lin} &= -\frac{1}{2}P^{\alpha\delta}\!\left(2h_{\delta(\beta;\gamma)}-h_{\beta\gamma;\delta}\right)\!u^\beta u^\gamma,\\
a^\alpha_{\rm quad} &= -\frac{1}{2}P^{\alpha\mu}h^\delta{}_\mu\!\left(2h_{\delta(\beta;\gamma)}-h_{\beta\gamma;\delta}\right)\!u^\beta u^\gamma,
\end{align}
where $P^{\alpha\mu}\equiv g^{\alpha\mu}+u^\alpha u^\mu$. Summing these, we have
\begin{align}
\frac{D^2z^\mu}{d\tau^2} =  -\frac{1}{2}P^{\alpha\mu}
		(g^\delta{}_\mu-h^\delta{}_\mu)\!\left(2h_{\delta(\beta;\gamma)}-h_{\beta\gamma;\delta}\right)\!u^\beta u^\gamma +\O(h^3).\label{2nd-geo}
\end{align}

As applied to the case of the effective metric $\tilde g_{\mu\nu}=g_{\mu\nu}+h^\R_{\mu\nu}$ (i.e., replacing $h_{\mu\nu}$ with $h^\R_{\mu\nu}$), Eq.~\eqref{2nd-geo} agrees with the second-order self-forced equation of motion derived in the body of the paper.

\subsection{Expansion in powers of a metric perturbation and a worldline deviation}\label{geodesic_expansion_in_h_and_dz}
In the last section, I expanded the geodesic equation while holding the solution $z^\mu_\e(s)$ to that equation fixed. I now expand $z^\mu_\e(s)$ as well. This procedure yields a sequence of equations for the terms in the expansion of $z^\mu_\e(s)$, suitable for a Gralla-Wald approximation. My approach to the expansion closely follows the treatment of geodesic deviation in Sec. 1.10 in Ref.~\cite{Poisson:04a}

I first describe the geometry of the situation. Consider a family of worldlines $z^\mu(\tau,\e)$, with each member $z_\e^\mu(\tau)=z^\mu(\tau,\e)$ governed by the equation of motion~\eqref{exact_geodesic}. Each member satisfies 
\begin{equation}\label{family_acceleration}
\frac{D^2 z_\e^\mu}{d\tau^2}=F^\mu(\tau,\e),
\end{equation}
where $\tau$ is proper time on $z_\e^\mu$, and $F^\mu$ is given by the right-hand side of Eq.~\eqref{exact_geodesic}. The family generates a two dimensional surface $\mathcal{S}$ with a tangent bundle spanned by $u^\mu\equiv\frac{\partial x^\mu}{\partial\tau}$ and $v^\mu\equiv\frac{\partial x^\mu}{\partial\e}$. An important relation between these vector fields can be found from $\frac{\partial^2 x^\mu}{\partial\tau\partial\e}=\frac{\partial^2 x^\mu}{\partial\e\partial\tau}$, which implies $\mathcal{L}_u v^\mu=0=\mathcal{L}_v u^\mu$, and from there,
\begin{equation}\label{commutation}
v^\mu{}_{;\nu}u^\nu = u^\mu{}_{;\nu}v^\nu.
\end{equation}


Now, we seek to describe the deviation of an accelerated worldline $z^\mu_\e(\tau)$ from the zeroth order, geodesic worldline $z^\mu_0(\tau)\equiv z^\mu(\tau,0)$. The first step is to expand the worldline in the power series
\begin{equation}\label{z_expansion_fixedtau}
z^\mu(\tau,\e) = z^\mu_0(\tau) +\e z^\mu_1(\tau) + \e^2 z_2^\mu(\tau)+ O(\e^3),
\end{equation}
where
\begin{equation}
z^\mu_n(\tau) = \frac{1}{n!}\e^n\frac{\partial^n z^\mu}{\partial\e^n}(\tau,0).
\end{equation}
We may also write the expansion as $z^\mu(\tau,\e)=\sum\frac{1}{n!}\e^n\Lie^n_v z^\mu|_{z_0(\tau)}$. Note that here $z^\mu$ is a scalar field equal to the $\mu$th coordinate field on the surface $\mathcal{S}$. The leading-order term is the family member $z_0^\mu(\tau)\equiv z^\mu(\tau,0)$. The second term is  $z_1^\mu(\tau)=\Lie_v z^\mu|_{z_0}=v^\mu(z_0(\tau))$, a vector on $z^\mu_0$. But at second order and beyond, a subtlety arises: unlike the first derivative along a curve, which is a tangent vector, second and higher derivatives are not immediately vectorial quantities. The function $z^\mu(\tau,\e)$ describes a curve in a particular set of coordinates, and the corrections $z^\mu_n$ depend on the coordinate system in which one defines $z^\mu(\tau,\e)$. Since my notion of an object's center is established with reference to a comoving normal coordinate system, I wish my covariant measure of the second-order deviation to agree, component by component, with $\frac{1}{2}\frac{\partial^n z^\mu}{\partial\e^n}(\tau,0)$ when evaluated in a normal coordinate system centered on $z^\mu_0$; Sec.~\ref{worldline} describes the utility of this choice when re-expanding a self-consistent approximation into Gralla-Wald or osculating-geodesics form. With that in mind, I define the vector 
\beq
w^\alpha\equiv\frac{1}{2}\frac{Dv^\alpha}{d\e} = \frac{1}{2}v^\beta\nabla_\beta v^\alpha
\eeq
and I seek an evolution equation for its restriction to $z_0^\mu$, 
\beq
z^\alpha_{2{\rm F}}(\tau)\equiv w^\alpha|_{z_0(\tau)}.
\eeq
$z^\alpha_{2{\rm F}}$ is the second-order term in the expansion~\eqref{z_expansion_fixedtau} when that expansion is performed in Fermi normal coordinates centered on $z_0^\mu$.

In addition to the choice of coordinates, the expansion~\eqref{z_expansion_fixedtau} depends on the particular choice of parametrization $(\tau,\e)$ of the surface $\mathcal{S}$. A change of parametrization alters the direction of expansion away from $z_0^\mu$. Here, the parametrization is chosen such that $\tau$ is proper time along each curve $z^\mu_\e(\tau)$, and a flow line generated by $v^\mu$ links points on different curves $z^\mu_\e(\tau)$ at the same value of $\tau$. When restricted to $z_0^\mu$, the parameter $\tau$ is $\tau_0$, the proper time on $z_0^\mu$.

With all preliminaries established, I now proceed to find the evolution equations for $z_0^\mu$, $z_1^\mu$, and $z_2^\mu$. The leading term clearly satisfies
\beq
\frac{D^2z_0^\mu}{d\tau_0^2}=F^\mu(\tau,0)=0. 
\eeq
For the others, I first find evolution equations for $v^\mu$ and $w^\mu$ and then evaluate the results on $z^\mu_0$. At first order, using Eqs.~\eqref{family_acceleration} and \eqref{commutation}, we have 
\begin{align}
\frac{D^2v^\alpha}{d\tau^2} &= \left(v^\alpha{}_{;\beta}u^\beta\right)_{;\gamma}u^\gamma\\
							&= \left(u^\alpha{}_{;\beta}v^\beta\right)_{;\gamma}u^\gamma\\
							&= F^\alpha_{;\gamma}v^\gamma-R^\alpha{}_{\mu\beta\nu}u^\mu v^\beta u^\nu,
\end{align}
where the second line follows from Eq.~\eqref{commutation} and the third line follows from the Ricci identity and Eq.~\eqref{family_acceleration}. Evaluating on $z_0^\mu$, I write this as
\begin{equation}\label{z1_generic}
\frac{D^2z^\alpha_1}{d\tau_0^2} = \check{F}_1^\alpha(\tau_0)-R^\alpha{}_{\mu\beta\nu}u_0^\mu z^\beta_1 u_0^\nu,
\end{equation}
where $\check{F}^\alpha_1\equiv \frac{DF^\alpha}{d\e}|_{\gamma_0}$. This is a generalization from the usual geodesic deviation equation to the deviation between neighbouring accelerating worldlines; it is valid even if $F^\mu(\tau,0)\neq0$.

At second order, repeated use of Eq.~\eqref{commutation} and Ricci's identity leads to
\begin{align}
\frac{D^2w^\alpha}{d\tau^2} &= \frac{1}{2}\left[\left(v^\alpha{}_{;\beta}v^\beta\right)_{;\mu}u^\mu\right]_{;\nu}u^\nu \\
								&= \frac{1}{2}\left[\left(u^\alpha{}_{;\beta}u^\beta\right)_{;\gamma}v^\gamma\right]_{;\delta}v^\delta
		 +\frac{1}{2}R^\alpha{}_{\mu\beta\nu;\gamma}\left(v^\mu u^\beta v^\nu u^\gamma-u^\mu v^\beta u^\nu v^\gamma\right)\nonumber\\
&\quad -R^\alpha{}_{\mu\beta\nu}\left(u^\mu w^\beta u^\nu+2u^\mu{}_{;\gamma}v^\gamma v^\beta u^\nu +\tfrac{1}{2}v^\mu v^\beta u^\nu_{;\gamma}u^\gamma\right).
\end{align}
Evaluating on $z^\mu_0$ and using Eq.~\eqref{family_acceleration}, we can write this as
\begin{align}\label{z2_generic}
\frac{D^2z_{2{\rm F}}^\alpha}{d\tau_0^2} &= \check{F}^\alpha_2(\tau_0)  -R^\alpha{}_{\mu\beta\nu}\left(u_0^\mu z_{2{\rm F}}^\beta u_0^\nu+2u_1^\mu z_1^\beta u_0^\nu\right)
							+2R^\alpha{}_{\mu\beta\nu;\gamma}z_1^{(\mu} u_0^{\beta)} z_1^{[\nu} u_0^{\gamma]}
\end{align}
where $\check{F}^\alpha_2\equiv\frac{1}{2}\frac{D^2F^\alpha}{d\e^2}|_{\gamma_0}$ and $u_1^\mu\equiv \frac{Dz_1^\mu}{d\tau}$. Equation~\eqref{z2_generic} describes the second deviation between neighbouring accelerating worldlines. 
In the case of neighbouring geodesics, it agrees with ``Bazanski's equation" in the form given in Eq.~(5.9) of Ref.~\cite{Vines:14}.

The quantities $\check{F}^\mu_n$ appearing in Eqs.~\eqref{z1_generic} and \eqref{z2_generic} can be straightforwardly evaluated by performing the expansion $h_{\mu\nu}(x,\e)=\e\check{h}^1_{\mu\nu}(x)+\e^2\check{h}^2_{\mu\nu}(x)+\O(\e^3)$ in Eq.~\eqref{2nd-geo} and then taking covariant derivatives with respect to $v^\mu$. The results are
\begin{equation}\label{F1_generic}
\check{F}_1^\mu = \frac{1}{2}P^{\mu\nu}_0\left(\check{h}^1_{\sigma\lambda;\rho}-2\check{h}^1_{\rho\sigma;\lambda}\right)u^\sigma_0 u^\lambda_0
\end{equation}
and 
\begin{align}
\check{F}_2^\mu &= -\frac{1}{2}P_0^{\mu\nu}\left(2\check{h}^2_{\nu\sigma;\lambda}-\check{h}^2_{\sigma\lambda;\nu}\right)u^\sigma_0 u^\lambda_0 
		-\frac{1}{2}P_0^{\mu\nu}\left(2\check{h}^1_{\nu\sigma;\lambda\delta}-\check{h}^1_{\sigma\lambda;\nu\delta}\right)u^\sigma_0 u^\lambda_0z_1^\delta\nonumber\\
		&\quad 	 -\left(2\check{h}^1_{\nu\sigma;\lambda}-\check{h}^1_{\sigma\lambda;\nu}\right)
		\left(u^{(\mu}_1 u^{\nu)}_0u^\sigma_0u_0^\lambda+P_0^{\mu\nu}u^{(\sigma}_1 u^{\lambda)}_0\right)\nonumber\\
		&\quad +P_0^{\mu\nu}\check{h}^1_\nu{}^\rho\left(2\check{h}^1_{\rho\sigma;\lambda}-\check{h}^1_{\sigma\lambda;\rho}\right)u^\sigma_0 u^\lambda_0.
		\label{F2_generic}
\end{align}

As applied to the case of the effective metric $\tilde g_{\mu\nu}=g_{\mu\nu}+h^\R_{\mu\nu}$ (i.e., replacing $\check{h}^n_{\mu\nu}$ with $\check{h}^{\R n}_{\mu\nu}$),  Eqs.~\eqref{z1_generic} and \eqref{z2_generic}, together with Eqs.~\eqref{F1_generic} and \eqref{F2_generic}, are the second-order expansion of the motion that apply in  a Gralla-Wald approximation.


\section{Expansion of point-particle fields in powers of a worldline deviation}\label{pp_quantities_expansion}
In this appendix, I derive the linear terms in expansions of the point particle stress-energy $T_1^{\mu\nu}(x;z)$ and the Lorenz-gauge retarded field $h^1_{\mu\nu}(x;z)$ when the worldline is expanded as $z^\mu(s,\e)=z_0^\mu(s)+\e z_1^\mu(s)+\O(\e^2)$, where $s$ is an arbitrary parameter. I also establish the identification of these linear terms with the mass dipole moment terms found from the local analysis in Sec.~\ref{algorithm}.

\subsection{Stress-energy}\label{T1_expansion}
I write the stress-energy in the parametrization-invariant form~\cite{Poisson-Pound-Vega:11}
\begin{equation}
T^{\alpha\beta}_1(x;z) = m\int_\gamma g^\alpha_{\alpha'}(x,z)g^\beta_{\beta'}(x,z) \dot z^{\alpha'}\dot z^{\beta'} \delta(x,z)
									\frac{ds}{\sqrt{-g_{\mu'\nu'}(z)\dot z^{\mu'}\dot z^{\nu'}}},
\end{equation}
where $g^\alpha_{\alpha'}(x;z)$ is a parallel propagator from the source point $x'=z(s,\e)$ to the field point $x$, and $\dot z^\mu\equiv \frac{dz^\mu}{ds}$.  

Substituting the expansion~\eqref{z_expansion} into  this stress-energy tensor, we obtain
\begin{align}
T_1^{\alpha\beta}(x;z) &= m\int_{\gamma_0} \left[g^\alpha_{\alpha'}(x,z_0)g^\beta_{\beta'}(x,z_0) \dot z_0^{\alpha'}\dot z_0^{\beta'}
				+\epsilon z_1^{\mu'}\nabla_{\mu'}(g^\alpha_{\alpha'}g^\beta_{\beta'} \dot z^{\alpha'}\dot z^{\beta'})|_{\epsilon=0}\right]\nonumber\\
				&\quad \times\left[\delta(x,z_0)+\epsilon z^{\nu'}_1\nabla_{\nu'}\delta(x,z)|_{\epsilon=0}\right]
				\left[1-\epsilon \frac{\dot z_{0\delta'}z_1^{\gamma'}\nabla_{\gamma'}\dot z_0^{\delta'}}{\dot z^{\kappa'}_0\dot z_{0\kappa'}}\right] 
				\frac{ds}{\sqrt{-\dot z^{\rho'}_0\dot z_{0\rho'}}}\nonumber\\
				&\quad +O(\epsilon^2).
\end{align}
In each instance, the evaluation at $\epsilon=0$ occurs \emph{after} taking the derivative.

I simplify this expression using the distributional identities $\nabla_{\mu'}\delta(x,z)=-g^\mu_{\mu'}\nabla_\mu\delta(x,z)$ and $g^\alpha_{\alpha';\beta'}\delta(x,z)=0$~\cite{Poisson-Pound-Vega:11}. I also use the identity
\begin{equation}
z_1^{\mu'}(\nabla_{\mu'}\dot z^{\alpha'})\big|_{\epsilon=0}=\frac{Dz_1^{\alpha'}}{ds}\equiv \dot z_1^{\alpha'},\label{z1dot}
\end{equation}
which follows in the same manner as Eq.~\eqref{commutation}.

The result of those simplifications is
\begin{equation}\label{T_expansion}
\e T_1^{\alpha\beta}(x;z) = \epsilon T^{\alpha\beta}_1(x;z_0)+\epsilon^2 \delta T_1^{\alpha\beta}(x;z_0,z_1)+O(\epsilon^3),
\end{equation}
with
\begin{align}
T^{\alpha\beta}_1(x;z_0) &= m\int_{\gamma_0} g^\alpha_{\alpha'}g^\beta_{\beta'} u_0^{\alpha'}u_0^{\beta'}\delta(x,z_0)d\tau_0',\label{T1_z0}\\
\delta T^{\alpha\beta}_1(x;z_0,z_1) &= m\int_{\gamma_0} g^\alpha_{\alpha'}g^\beta_{\beta'}
	\Bigl[\left(2u_0^{(\alpha'}u_{1}^{\beta')}-u_0^{\alpha'}u_0^{\beta'}u_{0\gamma'}u_{1}^{\gamma'}\right)\delta(x,z_0)\nonumber\\
	&\quad -u_0^{\alpha'}u_0^{\beta'}z_{1}^{\gamma'}g^\gamma_{\gamma'}\nabla_{\gamma}\delta(x,z_0)\Bigr]d\tau_0',\label{dT}
\end{align}
where $u_1^\mu(\tau_0)\equiv\frac{Dz_{1}^{\mu}}{d\tau_0}$, and the parallel propagators are evaluated at $(x,z_0(\tau_0'))$. I have simplified these expressions by reparametrizing $z_0^\mu$ in terms of $\tau_0$, the proper time on $\gamma_0$, but note that this does \emph{not} correspond to choosing the original parameter $ s=\tau_0$. Equations~\eqref{T_expansion}--\eqref{dT} are valid for any choice of parameter $ s$, and $z^\mu_1(\tau_0)$ actually \emph{depends} on the original choice of $ s$: a change of parametrization $ s\to s'( s,\e)$ will change the direction of $z^\mu_1$, in particular changing whether or not $z^\mu_1$ is orthogonal to $u_0^\mu$.

To eliminate this dependence on the initial choice of parametrization, I rewrite $\delta T^{\alpha\beta}_1$ in terms of the orthogonal part of $z^\mu_1$, $z^\mu_{1\perp}\equiv(\delta^\mu_\nu+u^\mu_0u_{0\nu})z^\nu_1$. The result is
\begin{equation}\label{dT_zperp}
\delta T^{\alpha\beta}_1 = m\!\int_{\gamma_0}\! g^\alpha_{\alpha'}g^\beta_{\beta'}
					\left[2u_0^{(\alpha'}u_{1\perp}^{\beta')}\delta(x,z_0)
					-u_0^{\alpha'}u_0^{\beta'}z_{1\perp}^{\gamma'}g^\gamma_{\gamma'}\nabla_{\gamma}\delta(x,z_0)\right]d\tau_0'.
\end{equation}
where $u_{1\perp}^\mu\equiv\frac{Dz_{1\perp}^{\mu}}{d\tau_0}$. Note that the part of $z^\mu_1$ parallel to $u_0^\mu$ does not appear in this expression. Again, this result does not depend on the initial choice of parametrization. One need not choose a parametrization by hand that enforces $z_1^\mu u_{0\mu}=0$; no matter the choice, only the perpendicular part plays a role in the field equations. 

The quantity $\delta T^{\mu\nu}_1(x;z_0)$ is equal to $\varLie_v T^{\mu\nu}(x;z_0)$, where $v^\mu=\frac{\partial z^\mu(s,\e)}{\partial\e}$ is introduced in Appendix~\ref{geodesic_expansion_in_h_and_dz}, and $\varLie_v$ is the Lie derivative introduced in Sec.~\ref{worldline}. In fact, for any vector $\xi^\mu$, $\varLie_\xi T^{\mu\nu}(x;z)$ is given by Eq.~\eqref{dT_zperp} with the replacement $z_1^\mu\to\xi^\mu$ and $u^\mu_0\to u^\mu$. This quantity is useful when considering gauge transformations in the self-consistent approximation. Also useful is the ordinary Lie derivative of $T^{\mu\nu}_1$; taking similar steps as above, one finds
\begin{align}
\Lie_\xi T^{\alpha\beta}_1(x;z) &= -m\!\int_\gamma\! g^\alpha_{\alpha'}g^\beta_{\beta'}
					\bigg\{\left[2u^{(\alpha'}\frac{D\xi_{\perp}^{\beta')}}{d\tau'} +u^{\alpha'}u^{\beta'}\left(\frac{d\xi_{\parallel}}{d\tau}+\xi^{\rho'}{}_{;\rho'}\right)\right] \delta(x,z)\nonumber\\
					&\quad-u^{\alpha'}u^{\beta'}\xi_{\perp}^{\gamma'}g^\gamma_{\gamma'}\nabla_{\gamma}\delta(x,z)\bigg\}d\tau',\label{LieT}
\end{align}
where $\xi_{\perp}^{\beta'}\equiv P^{\beta'}{}_{\alpha'}\xi^{\alpha'}$ and $\xi_{\parallel}\equiv u_{\mu'}\xi^{\mu'}$. The sum of the two Lie derivatives yields the simple result
\beq\label{LieT+varLieT}
(\Lie_\xi+\varLie_\xi)T^{\mu\nu}_1(x;z) = -m\int g^\mu_{\mu'}g^\nu_{\nu'}u^{\mu'}u^{\nu'}\left(\frac{d\xi_{\parallel}}{d\tau}+\xi^{\rho'}{}_{;\rho'}\right)\delta(x,z)d\tau'.
\eeq
A $\xi^\mu\nabla_\mu\delta(x,z)$ term signals that the mass $m$ is displaced from $z^\mu$ by an amount $\xi^\mu$; the lack of any $\nabla_\mu\delta(x,z)$ term in Eq.~\eqref{LieT+varLieT} signals that the displacements due to the two derivatives cancel one another, leaving the mass $m$ moving on $z^\mu$.

\subsection{Metric perturbation}
In the Lorenz gauge, the first-order self-consistent field, given some global boundary conditions, is given by 
\beq\label{h1_z}
h^1_{\mu\nu}(x;z) = 4m\int \bar G_{\mu\nu\mu'\nu'}u^{\mu'}u^{\nu'}d\tau',
\eeq
where $G_{\mu\nu\mu'\nu'}$ is the Green's function that comports with the global boundary conditions. I wish to expand this about $z^\mu=z_0^\mu$ to obtain something of the form
\beq
\e h^1_{\mu\nu}(x;z) = \e h^1_{\mu\nu}(x;z_0)+\e^2 \delta h^1_{\mu\nu}(x;z_0,z_1)+\O(\e^3).
\eeq
There are two methods available for achieving this: directly, following steps analogous to those in the previous section; or by making use of the result of the previous section. 

Here I adopt the second method. Noting that $\delta h^1_{\mu\nu}(x;z_0,z_1)=\varLie_v h^1_{\mu\nu}(x;z_0)$, that $\delta T^{\alpha\beta}_1=\varLie_v T_1^{\mu\nu}(x;z_0)$, and that $\varLie_v$ commutes with derivatives acting at $x$, we have
\beq
E_{\mu\nu}[\varLie_v \bar h^1](x;z_0) = \varLie_v E_{\mu\nu}[\bar h^1](x;z_0) = -16\pi \varLie_v T^1_{\mu\nu}(x;z_0).
\eeq
Hence, 
\beq
\delta h^1_{\mu\nu}(x;z_0,z_1)=\varLie_v h^1_{\mu\nu}(x;z_0) = 4\int \bar G_{\mu\nu\mu'\nu'}\delta T^{\mu'\nu'}(x';z_0,z_1)dV'.
\eeq
From Eq.~\eqref{dT_zperp}, this evaluates to
\beq\label{dh_zperp}
\delta h^1_{\mu\nu}(x;z_0,z_1) = 4m\!\int_{\gamma_0}\!\! \left(2\bar G_{\mu\nu\mu'\nu'}u_0^{(\mu'}u_{1\perp}^{\nu')}
			+ \bar G_{\mu\nu\mu'\nu';\gamma'}u^{\mu'}_0u^{\nu'}_0z_{1\perp}^{\gamma'}\right)d\tau_0'.
\eeq


\subsubsection{Gauge condition}
It is worth examining how $h^1_{\alpha\beta}(x;z)$ and $\delta h^1_{\alpha\beta}(x;z_0,z_1)$ contribute to the Lorenz gauge condition. Those contributions are easily found by invoking the identity $\nabla^\nu G_{\mu\nu\mu'\nu'}=-G_{\mu(\mu';\nu')}$~\cite{Poisson-Pound-Vega:11}, where $G_{\mu\mu'}$ is a Green's function for the vector wave equation $\Box V_\mu=S_\mu$, and both Green's functions must satisfy the same boundary conditions. Performing a trace-reversal on Eq.~\eqref{h1_z}, taking the divergence of the result, using the Green's-function identity, and integrating by parts yields
\beq\label{gauge_condition_SC}
\nabla^\beta \bar h^1_{\alpha\beta}(x;z) = 4\int G_{\alpha\alpha'}\frac{D}{d\tau'}(mu^{\alpha'})d\tau'.
\eeq
(Earlier results in this section assumed constant $m$, but here I momentarily leave it arbitrary for generality.)
The contribution to the gauge condition is determined entirely by $\frac{dm}{d\tau}$ and the acceleration of $z^\mu$. In a Gralla-Wald expansion, one has $\nabla^\beta \check{\bar h}^1_{\alpha\beta}(x;z_0)=0$, from which one can read off $\frac{dm}{d\tau}=0$ and $a_0^\mu=0$. In a self-consistent expansion, one instead has  $\nabla^\beta [\e\bar h^1_{\alpha\beta}(x;z)+\e^2\bar h^2_{\alpha\beta}(x;z)]=\O(\e^3)$ [or more precisely, Eq.~\eqref{gauge_conditions_SC}], which determines Eq.~\eqref{Detweiler-Whiting-form1}.

Doing the same with Eq.~\eqref{dh_zperp} yields
\beq\label{gauge_condition_GW}
\nabla^\beta \delta\bar h^1_{\alpha\beta}(x;z_0,z_1) = 
	4m\int G_{\alpha\alpha'}\left(\frac{D^2z^{\alpha'}_{1\perp}}{d\tau'^2_0}+R^{\alpha'}{}_{\beta'\gamma'\delta'}u_0^{\beta'}z_{1\perp}^{\gamma'}u_0^{\delta'}\right)d\tau'_0.
\eeq
The contribution to the gauge condition is determined entirely by the acceleration of the deviation from $z_0^\mu$ (together with the geodesic-deviation term). In a Gralla-Wald expansion, $\delta h^1_{\alpha\beta}(x;z_0,z_1)$ is included in $\check{h}^2_{\alpha\beta}(x;z_0)$, and $\nabla^\beta \check{\bar h}^2_{\alpha\beta}(x;z_0)=0$ determines Eq.~\eqref{GW_ddotz1}.

\subsection{Local expansion and identification of mass dipole moment}\label{M-dz-identification}
In Sec.~\ref{skeleton}, I showed that the mass dipole moment of the object creates a term~\eqref{TM} (and contributes to the term~\eqref{Tdm}) in the object's skeletal stress-energy. By comparing that result to Eq.~\eqref{dT_zperp}, we can make the identification $M^i=mz^i_1$, exactly as concluded elsewhere in the paper. Since the two stress-energy tensors are the same, it follows that  $\delta h^1_{\mu\nu}(x;z_0,z_1)$ includes the entire contribution to $\check{h}^2_{\mu\nu}$ coming from the mass dipole moment.

Here, I complete the circle by performing a local expansion of Eq.~\eqref{dh_zperp} near $z_0^\mu$ and showing $\delta h^1_{\mu\nu}(x;z_0,z_1)$ reproduces the order-$1/r^2$ and $1/r$ terms for the mass dipole seed solution in Fermi normal coordinates. The method of local expansion is elaborated in Ref.~\cite{Poisson-Pound-Vega:11}. In particular, I follow steps analogous to those in Sec. 23.2 of that reference. To avoid belabouring the details, here I provide only the barest sketch; I leave it to the interested reader to fill in the gaps using the tools of Ref.~\cite{Poisson-Pound-Vega:11}.

The starting point is the Hadamard decomposition of the retarded Green's function, $G_{\mu\nu\mu'\nu'}=U_{\mu\nu\mu'\nu'}\delta_+(\sigma) + V_{\mu\nu\mu'\nu'}\theta_+(-\sigma)$. Here $\delta_+(\sigma)$ is a Dirac delta function supported on the past light cone of $x$, and $\theta_+(-\sigma)$ is a Heaviside step function supported in the interior of that light cone. $\sigma(x,x')$ is one-half the squared geodesic distance from $x'$ to $x$, such that it vanishes when a null geodesic connects the two points. 

From this starting point, the final result is obtained by a two-step process: (i) evaluating the integrals over $\tau_0'$ by changing the integration variable to $\sigma$, and (ii) writing the retarded distance from $x'$ to $x$ in terms of the Fermi radial coordinate $r$. The results for the two terms in Eq.~\eqref{dh_zperp} are
\beq\label{integral1}
8m\int \bar G_{\mu\nu\mu'\nu'}u_0^{(\mu'}\frac{Dz_{1\perp}^{\nu')}}{d\tau'_0} d\tau'_0 = \frac{8m}{r}u_{0(\mu}e^i_{\nu)}\dot z_{1i} +\O(r^0),
\eeq
where $\dot z^i_1 = \frac{dz^i_1}{dt}$, and
\beq\label{integral2}
4m\int \bar G_{\mu\nu\mu'\nu';\gamma'}u^{\mu'}_0u^{\nu'}_0z_{1\perp}^{\gamma'}d\tau_0' = \frac{2m}{r^2}z_1^an_a\delta_{\mu\nu}+\O(r^0).
\eeq
The components in Fermi coordinates are
\begin{subequations}\label{dh1}%
\begin{align}
\delta h^1_{tt} &= \frac{2mz^i_1n_i}{r^2}+\O(r^0),\\
\delta h^1_{ta} &= \frac{4m \dot z_{1a}}{r}+\O(r^0),\\
\delta h^1_{ab} &= \frac{2mz^i_1n_i}{r^2}\delta_{ab}+\O(r^0). 
\end{align}
\end{subequations}
Here we see that with the identification $M^i=mz_1^i$, the $tt$ and $ab$ components are precisely those in Eq.~\eqref{M_seed}, and the $ta$ component is precisely the term proportional to $\dot M^i$ in Eq.~\eqref{GW_dm_hR}. In other words, the integral~\eqref{integral1} corresponds to the mass dipole moment's contribution to the monopole seed $h^\seed_{\mu\nu}(x;z_0,\delta m)$, while the integral~\eqref{integral2} corresponds to the mass-dipole seed $h^\seed_{\mu\nu}(x;z_0,M)$.

\section{Identities for gauge transformations of curvature tensors}\label{gauge-identities}
Let $A[g]$ be a tensor of any rank constructed from a metric $g$. (To streamline the presentation, I adopt index-free notation throughout most of this appendix.) Now define
\beq\label{dnA-gen}
\delta^n A[f_1,\ldots,f_n] \equiv \frac{1}{n!}\frac{d^n}{d\lambda_1\cdots d\lambda_n}A[g+\lambda_1 f_1 +\cdots + \lambda_n f_n]\big|_{\lambda_1=\cdots=\lambda_n=0}.
\eeq
This tensor is linear in each of its arguments $f_1,\ldots,f_n$; it is also symmetric in them.  In the case that all the arguments are the same, we have $\delta^n A[h,\ldots,h] = \frac{1}{n!}\frac{d^n}{d\lambda^n}A[g+\lambda h]\big|_{\lambda=0}$, the piece of $A[g+h]$ containing precisely $n$ factors of $h$ and its derivatives. 

The following identities are easily proved by writing Lie derivatives as ordinary derivatives: 
\begin{align}
\Lie_\xi A[g] &=\delta A[\Lie_\xi g] , \label{Lie A}\\
\tfrac{1}{2}\Lie^2_\xi A[g] &=\tfrac{1}{2}\delta A[\Lie^2_\xi g] + \delta^2A[\Lie_\xi g,\Lie_\xi g] ,\label{Lie2 A}\\
\Lie_\xi\delta A[h] &= \delta A[\Lie_\xi h] + 2\delta^2 A[\Lie_\xi g, h] .\label{Lie dA}
\end{align}
Note that $\delta^2 A[\Lie_\xi g,h]= \delta^2 A[h,\Lie_\xi g]=\frac{1}{2}\left(\delta^2 A[h,\Lie_\xi g]+\delta^2 A[\Lie_\xi g,h]\right)$. As an example, if $A$ is the Ricci tensor, then
\begin{align}
\Lie_\xi R_{\mu\nu}[g] &=\delta R_{\mu\nu}[\Lie_\xi g], \label{Lie R}\\
\tfrac{1}{2}\Lie^2_\xi R_{\mu\nu}[g] &=\tfrac{1}{2}\delta R_{\mu\nu}[\Lie^2_\xi g] + \delta^2R_{\mu\nu}[\Lie_\xi g,\Lie_\xi g] ,\label{Lie2 R}\\
\Lie_\xi\delta R_{\mu\nu}[h] &= \delta R_{\mu\nu}[\Lie_\xi h] + 2\delta^2 R_{\mu\nu}[h,\Lie_\xi g],\label{Lie dR}
\end{align}
where I have restored indices to avoid confusion with the Ricci scalar.

To establish Eq.~\eqref{Lie A}, one can write the metric as a function of a parameter $\lambda$ along the flow generated by $\xi$ and then perform a Taylor expansion:
\beq
\Lie_\xi A[g] = \frac{d}{d\lambda}A\!\left[g(0)+\lambda\frac{dg}{d\lambda}\Big|_{\lambda=0}\right]\!\!\bigg|_{\lambda=0} 
= \delta A\left[\frac{dg}{d\lambda}\big|_{\lambda=0}\right] = \delta A[\Lie_\xi g].
\eeq
Similarly, to establish Eq.~\eqref{Lie2 A}, one can write
\begin{subequations}
\begin{align}
\tfrac{1}{2}\Lie^2_\xi A[g] &= \tfrac{1}{2}\frac{d^2}{d\lambda^2} A\!\left[g(0)+\lambda\frac{dg}{d\lambda}\Big|_{\lambda=0}+\tfrac{1}{2}\lambda^2\frac{dg}{d\lambda^2}\Big|_{\lambda=0}\right]\!\!\bigg|_{\lambda=0}\\
&= \tfrac{1}{2}\delta A[\Lie^2_\xi g] + \delta^2A[\Lie_\xi g,\Lie_\xi g] ,
\end{align}
\end{subequations}
and to establish Eq.~\eqref{Lie dA}, one can write $g$ as a function of parameters $(\lambda,\e)$, where $h\equiv \frac{dg}{d\e}\big|_{\e=0}$, and then write
\begin{subequations}
\begin{align}
\Lie_\xi\delta A[h] &= \frac{d^2}{d\lambda d\e}A\!\left[g(\lambda,0)+\e \frac{d g}{d\epsilon}(\lambda,0)\right]\!\!\bigg|_{\lambda=\e=0}\\
&=  \frac{d^2}{d\lambda d\e}A\!\left[g(0,0)+\lambda \frac{d g}{d\lambda}(0,0) + \e \frac{d g}{d\e}(0,0)+\lambda \e \frac{d^2 g}{d\lambda d\e}(0,0)\right]\!\!\bigg|_{\lambda=\e=0}\nonumber\\
&= \delta A[\Lie_\xi h] + 2\delta^2 A[h,\Lie_\xi g].
\end{align}
\end{subequations}

\bibliographystyle{apsrev4-1}


\bibliography{../bibfile}

\end{document}